\documentclass[11pt,a4paper]{article}  
\usepackage{comment}
\usepackage{graphicx}
\usepackage{xcolor}
\usepackage{amsmath}

\usepackage{txfonts}

\usepackage{makecell}
\usepackage[switch]{lineno}
\usepackage[utf8]{inputenc}
\usepackage{natbib}
\usepackage[hidelinks]{hyperref}

\usepackage{geometry}
\providecommand{\keywords}[1]{\textbf{\textbf{Key words.}} #1}

\begin{document} 

\title{Multiwavelength study of the galactic PeVatron candidate LHAASO J2108+5157}

\author{
S.~Abe$^{1}$ \and
A.~Aguasca-Cabot$^{2}$ \and
I.~Agudo$^{3}$ \and
N.~Alvarez~Crespo$^{4}$ \and
L.~A.~Antonelli$^{5}$ \and
C.~Aramo$^{6}$ \and
A.~Arbet-Engels$^{7}$ \and
M.~Artero$^{8}$ \and
K.~Asano$^{1}$ \and
P.~Aubert$^{9}$
% ...
}

\date{
$^1$Institute for Cosmic Ray Research, University of Tokyo, 5-1-5, Kashiwa-no-ha, Kashiwa, Chiba 277-8582, Japan
; $^2$Departament de Física Quàntica i Astrofísica, Institut de Ciències del Cosmos, Universitat de Barcelona, IEEC-UB, Martí i Franquès, 1, 08028, Barcelona, Spain
; $^3$Instituto de Astrofísica de Andalucía-CSIC, Glorieta de la Astronomía s/n, 18008, Granada, Spain
; $^4$Grupo de Electronica, Universidad Complutense de Madrid, Av. Complutense s/n, 28040 Madrid, Spain
; $^5$INAF - Osservatorio Astronomico di Roma, Via di Frascati 33, 00040, Monteporzio Catone, Italy
; $^6$INFN Sezione di Napoli, Via Cintia, ed. G, 80126 Napoli, Italy
; $^7$Max-Planck-Institut für Physik, Föhringer Ring 6, 80805 München, Germany
; $^8$Institut de Fisica d'Altes Energies (IFAE), The Barcelona Institute of Science and Technology, Campus UAB, 08193 Bellaterra (Barcelona), Spain
; $^9$Univ. Savoie Mont Blanc, CNRS, Laboratoire d'Annecy de Physique des Particules - IN2P3, 74000 Annecy, France
; $^{10}$Universität Hamburg, Institut für Experimentalphysik, Luruper Chaussee 149, 22761 Hamburg, Germany
; $^{11}$Graduate School of Science, University of Tokyo, 7-3-1 Hongo, Bunkyo-ku, Tokyo 113-0033, Japan
; $^{12}$EMFTEL department and IPARCOS, Universidad Complutense de Madrid, 28040 Madrid, Spain
; $^{13}$INAF - Osservatorio di Astrofisica e Scienza dello spazio di Bologna, Via Piero Gobetti 93/3, 40129 Bologna, Italy
; $^{14}$Centro Brasileiro de Pesquisas Físicas, Rua Xavier Sigaud 150, RJ 22290-180, Rio de Janeiro, Brazil
; $^{15}$INFN Sezione di Padova and Università degli Studi di Padova, Via Marzolo 8, 35131 Padova, Italy
; $^{16}$Instituto de Astrofísica de Canarias and Departamento de Astrofísica, Universidad de La Laguna, La Laguna, Tenerife, Spain
; $^{17}$CIEMAT, Avda. Complutense 40, 28040 Madrid, Spain
; $^{18}$Department of Physics, TU Dortmund University, Otto-Hahn-Str. 4, 44227 Dortmund, Germany
; $^{19}$INFN Sezione di Bari and Politecnico di Bari, via Orabona 4, 70124 Bari, Italy
; $^{20}$INFN Sezione di Trieste and Università degli studi di Udine, via delle scienze 206, 33100 Udine, Italy.
; $^{21}$INFN Sezione di Catania, Via S. Sofia 64, 95123 Catania, Italy
; $^{22}$INAF - Istituto di Astrofisica e Planetologia Spaziali (IAPS), Via del Fosso del Cavaliere 100, 00133 Roma, Italy
; $^{23}$Aix Marseille Univ, CNRS/IN2P3, CPPM, Marseille, France
; $^{24}$INFN Sezione di Torino, Via P. Giuria 1, 10125 Torino, Italy
; $^{25}$Palacky University Olomouc, Faculty of Science, 17. listopadu 1192/12, 771 46 Olomouc, Czech Republic
; $^{26}$Department of Physics and Technology, University of Bergen, Museplass 1, 5007 Bergen, Norway
; $^{27}$University of Geneva - Département de physique nucléaire et corpusculaire, 24 Quai Ernest Ansernet, 1211 Genève 4, Switzerland
; $^{28}$Port d'Informació Científica, Edifici D, Carrer de l'Albareda, 08193 Bellaterrra (Cerdanyola del Vallès), Spain
; $^{29}$INFN Sezione di Bari and Università di Bari, via Orabona 4, 70126 Bari, Italy
; $^{30}$University of Rijeka, Department of Physics, Radmile Matejcic 2, 51000 Rijeka, Croatia
; $^{31}$Institute for Theoretical Physics and Astrophysics, Universität Würzburg, Campus Hubland Nord, Emil-Fischer-Str. 31, 97074 Würzburg, Germany
; $^{32}$Institut für Theoretische Physik, Lehrstuhl IV: Plasma-Astroteilchenphysik, Ruhr-Universität Bochum, Universitätsstraße 150, 44801 Bochum, Germany
; $^{33}$ILANCE, CNRS - University of Tokyo International Research Laboratory, Kashiwa, Chiba 277-8582, Japan
; $^{34}$Physics Program, Graduate School of Advanced Science and Engineering, Hiroshima University, 739-8526 Hiroshima, Japan
; $^{35}$INFN Sezione di Roma Tor Vergata, Via della Ricerca Scientifica 1, 00133 Rome, Italy
; $^{36}$Faculty of Physics and Applied Informatics, University of Lodz, ul. Pomorska 149-153, 90-236 Lodz, Poland
; $^{37}$University of Split, FESB, R. Boškovića 32, 21000 Split, Croatia
; $^{38}$Department of Physics, Yamagata University, Yamagata, Yamagata 990-8560, Japan
; $^{39}$Tohoku University, Astronomical Institute, Aobaku, Sendai 980-8578, Japan
; $^{40}$Josip Juraj Strossmayer University of Osijek, Department of Physics, Trg Ljudevita Gaja 6, 31000 Osijek, Croatia
; $^{41}$Kitashirakawa Oiwakecho, Sakyo Ward, Kyoto, 606-8502, Japan
; $^{42}$INFN Sezione di Roma La Sapienza, P.le Aldo Moro, 2 - 00185 Rome, Italy
; $^{43}$Department of Astronomy, University of Geneva, Chemin d'Ecogia 16, CH-1290 Versoix, Switzerland
; $^{44}$Astronomical Institute of the Czech Academy of Sciences, Bocni II 1401 - 14100 Prague, Czech Republic
; $^{45}$Faculty of Science, Ibaraki University, Mito, Ibaraki, 310-8512, Japan
; $^{46}$Faculty of Science and Engineering, Waseda University, Shinjuku, Tokyo 169-8555, Japan
; $^{47}$Department of Physics, Tokai University, 4-1-1, Kita-Kaname, Hiratsuka, Kanagawa 259-1292, Japan
; $^{48}$INFN Sezione di Trieste and Università degli Studi di Trieste, Via Valerio 2 I, 34127 Trieste, Italy
; $^{49}$INFN and Università degli Studi di Siena, Dipartimento di Scienze Fisiche, della Terra e dell'Ambiente (DSFTA), Sezione di Fisica, Via Roma 56, 53100 Siena, Italy
; $^{50}$Escuela Politécnica Superior de Jaén, Universidad de Jaén, Campus Las Lagunillas s/n, Edif. A3, 23071 Jaén, Spain
; $^{51}$Saha Institute of Nuclear Physics, Bidhannagar, Kolkata-700 064, India
; $^{52}$Institute for Nuclear Research and Nuclear Energy, Bulgarian Academy of Sciences, 72 boul. Tsarigradsko chaussee, 1784 Sofia, Bulgaria
; $^{53}$FZU - Institute of Physics of the Czech Academy of Sciences, Na Slovance 1999/2, 182 21 Praha 8, Czech Republic
; $^{54}$Dipartimento di Fisica e Chimica 'E. Segrè' Università degli Studi di Palermo, via delle Scienze, 90128 Palermo
; $^{55}$Department of Applied Physics, University of Zagreb, Horvatovac 102a, 10000 Zagreb, Croatia
; $^{56}$Institute of Space Sciences (ICE, CSIC), and Institut d'Estudis Espacials de Catalunya (IEEC), and Institució Catalana de Recerca I Estudis Avançats (ICREA), Campus UAB, Carrer de Can Magrans, s/n 08193 Bellatera, Spain
; $^{57}$Hiroshima Astrophysical Science Center, Hiroshima University, Higashi-Hiroshima, Hiroshima 739-8526, Japan
; $^{58}$Charles University, Institute of Particle and Nuclear Physics, V Holešovičkách 2, 180 00 Prague 8, Czech Republic
; $^{59}$Institute for Space-Earth Environmental Research, Nagoya University, Chikusa-ku, Nagoya 464-8601, Japan
; $^{60}$Kobayashi-Maskawa Institute (KMI) for the Origin of Particles and the Universe, Nagoya University, Chikusa-ku, Nagoya 464-8602, Japan
; $^{61}$Graduate School of Technology, Industrial and Social Sciences, Tokushima University, Tokushima 770-8506, Japan
; $^{62}$INFN Dipartimento di Scienze Fisiche e Chimiche - Università degli Studi dell'Aquila and Gran Sasso Science Institute, Via Vetoio 1, Viale Crispi 7, 67100 L'Aquila, Italy
; $^{63}$IRFU, CEA, Université Paris-Saclay, Bât 141, 91191 Gif-sur-Yvette, France
; $^{64}$Graduate School of Science and Engineering, Saitama University, 255 Simo-Ohkubo, Sakura-ku, Saitama city, Saitama 338-8570, Japan
; $^{65}$Department of Physical Sciences, Aoyama Gakuin University, Fuchinobe, Sagamihara, Kanagawa, 252-5258, Japan
; $^{66}$Division of Physics and Astronomy, Graduate School of Science, Kyoto University, Sakyo-ku, Kyoto, 606-8502, Japan
; $^{67}$Dipartimento di Fisica - Universitá degli Studi di Torino, Via Pietro Giuria 1 - 10125 Torino, Italy
; $^{68}$Department of Physics, Konan University, Kobe, Hyogo, 658-8501, Japan;
(\today)\\
}

\maketitle

\begin{abstract}

   \textit{Context:} Several new ultrahigh-energy (UHE) gamma-ray sources have recently been discovered by the Large High Altitude Air Shower Observatory (LHAASO) collaboration. These represent a step forward in the search for the so-called Galactic PeVatrons, the enigmatic sources of the Galactic cosmic rays up to PeV energies. However, it has been shown that multi-TeV gamma-ray emission does not necessarily prove the existence of a hadronic accelerator in the source;  indeed this emission could also be explained as inverse Compton scattering from electrons in a radiation-dominated environment. A clear distinction between the two major emission mechanisms would only be made possible by taking into account multi-wavelength data and detailed morphology of the source.  

   \textit{Aims:} We aim to understand the nature of the unidentified source LHAASO J2108+5157, which is one of the few known UHE sources with no very high-energy (VHE) counterpart.

   \textit{Methods:} We observed LHAASO J2108+5157 in the X-ray band with \emph{XMM-Newton} in 2021 for a total of 3.8 hours and at TeV energies with the Large-Sized Telescope prototype (LST-1), yielding 49 hours of good-quality data. In addition, we analyzed 12 years of \emph{Fermi}-LAT data, to better constrain emission of its high-energy (HE) counterpart 4FGL J2108.0+5155. We used \texttt{naima} and \texttt{jetset} software packages to examine the leptonic and hadronic scenario of the multi-wavelength emission of the source.

   \textit{Results:} We found an excess ($3.7\sigma$) in the LST-1 data at energies $E>3\, \mathrm{TeV}$. Further analysis of the  whole LST-1 energy range, assuming a point-like source, resulted in a hint ($2.2\sigma$) of hard emission, which can be described with a single power law with a photon index of $\Gamma = 1.6 \pm 0.2$ the range of $0.3 - 100$ TeV. We did not find any significant extended emission that could be related to a supernova remnant (SNR) or pulsar wind nebula (PWN) in the \emph{XMM-Newton} data, which puts strong constraints on possible synchrotron emission of relativistic electrons. We revealed a new potential hard source in \emph{Fermi}-LAT data with a significance of $4\sigma$ and a photon index of $\Gamma = 1.9 \pm 0.2$, which is not spatially correlated with LHAASO J2108+5157, but including it in the source model we were able to improve spectral representation of the HE counterpart 4FGL J2108.0+5155. 

   \textit{Conclusions:} The LST-1 and LHAASO observations can be explained as inverse Compton-dominated leptonic emission of relativistic electrons with a cutoff energy of $100^{+70}_{-30}$ TeV. The low magnetic field in the source imposed by the X-ray upper limits on synchrotron emission is compatible with a hypothesis of a PWN or a TeV halo. Furthermore, the spectral properties of the HE counterpart are consistent with a Geminga-like pulsar, which would be able to power the VHE-UHE emission. Nevertheless, the lack of a pulsar in the neighborhood of the UHE source is a challenge to the PWN/TeV-halo scenario. The UHE gamma rays can also be explained as $\pi^0$ decay-dominated hadronic emission due to interaction of relativistic protons with one of the two known molecular clouds in the direction of the source. Indeed, the hard spectrum in the LST-1 band is compatible with protons escaping a shock around a middle-aged SNR because of their high low-energy cut-off, but the origin of the HE gamma-ray emission remains an open question.
\end{abstract}


%
\section{Introduction}

Cosmic rays (CRs) with energies up to the knee ($\sim$1 PeV) are believed to be produced in hadronic PeVatrons, cosmic accelerators located in our Galaxy \citep[for a review see e.g.,][]{2019IJMPD..2830022G}. Despite substantial observational efforts in the last decade, the origin of the highest-energy galactic CRs remains unknown, mainly because of the difficulty in reconstructing the direction of their origin, as they are subject to deflection in the Galactic magnetic field. When accelerated protons interact with ambient matter, they emit gamma rays through $\pi^0$ decay. Similarly, electrons and positrons produce gamma rays via inverse Compton (IC) scattering on low-energy photon fields via bremsstrahlung when colliding with atomic nuclei of ambient matter, or via synchrotron radiation when interacting with magnetic fields. Studying very high-energy (VHE, $0.1 < E < 100 \, \mathrm{TeV}$) and ultrahigh-energy (UHE, $E>0.1 \, \mathrm{PeV}$) cosmic gamma rays and disentangling the different origins of the radiation are therefore essential in order to search for cosmic PeVatrons (\citealt[]{2019scta.book.....C}, CTA Consortium, in prep.).

Diffusive shock acceleration (DSA; \citet{1978MNRAS.182..147B}) taking place in supernova remnants (SNRs) and pulsar wind nebulae (PWNe) has been proposed as a possible mechanism to accelerate CRs \citep{2013APh....43...56B}. In several SNRs, a characteristic spectral feature known as a "pion-decay bump"  has been detected, providing evidence that proton acceleration takes place in these sources \citep{2013Sci...339..807A, 2016ApJ...816..100J, 2018A&A...612A...5H, 2019A&A...623A..86A, 2022ApJ...933..204A}. However, none of the gamma-ray spectra of the sources firmly identified as SNRs extend beyond 100 TeV \citep{2019NatAs...3..561A, 2019ApJ...874...50Z}, which suggests that these sources are probably not capable of proton acceleration up to PeV energies.

The search for PeVatrons continues, and in the last few years several new candidates showing gamma-ray emission above 100 TeV have been discovered by the Tibet Air Shower (AS) collaboration \citep{2019PhRvL.123e1101A, 2021NatAs...5..460T} and the High Altitude Water Cherenkov (HAWC) observatory \citep{2020PhRvL.124b1102A}. Recently, the Large High Altitude Air Shower Observatory (LHAASO) collaboration, exploiting the unprecedented sensitivity of the LHAASO-KM2A instrument in the UHE range, reported the discovery of 12 UHE gamma-ray sources reaching energies up to $1.4 \, \mathrm{PeV}$ \citep{2021Natur.594...33C}. Among these, there is only one unidentified source with an as-of-yet undetected TeV counterpart, namely LHAASO J2108+5157.

LHAASO J2108+5157 is the first gamma-ray source directly discovered in the UHE band, and was detected with a post-trial significance of $6.4\sigma$ above 100 TeV \citep{2021ApJ...919L..22C}. The position of the source is $\mathrm{R.A.} = 317.22^\circ \pm 0.07^\circ, \, \mathrm{Dec} =  51.95^\circ \pm 0.05^\circ$. The source is reported to be point-like with a $95\%$ confidence level upper limit on its extension of $0.26^\circ$ with a two-dimensional symmetrical Gaussian shape assumption. The spectrum of LHAASO J2108+5157 above 25 TeV can be described by a single power law (PL) with a photon index of $2.83\pm0.18$. There is no VHE or X-ray counterpart to the source, but \citet{2021ApJ...919L..22C} identified a close high-energy (HE) point source, 4FGL J2108.0+5155 \citep{2020ApJS..247...33A}, at an angular distance of $0.13^\circ$. A dedicated analysis suggested that the HE source might be spatially extended (4FGL J2108.0+5155e) with extension of $0.48^\circ$. Its spectrum can be described with a single PL with a photon index of $2.3$ between 1 GeV and 1 TeV. However, the physical connection between the spectral energy distributions (SEDs) of the HE and UHE sources is not particularly clear because of the very different spectral indices. \citet{2021ApJ...919L..22C} found the source to be coincident with the position of a molecular cloud [MML2017]4607 \citep{2017ApJ...834...57M}, which would support the hypothesis that the emission  has a  hadronic origin, if CR protons collide with the ambient gas. The authors suggested that the extended emission of 4FGL J2108.0+5155e could be related to an old SNR, while the point-like UHE emission could be due to interaction of the escaping CRs from the SNR with the molecular cloud. Alternatively, the authors proposed that the relativistic CRs could be accelerated in one of the nearby open stellar clusters, but confirmation of these hypotheses is complicated because of the unknown distance of the source. A lepto-hadronic emission scenario was also proposed by \citet{Kar_2022}, whereby shock-accelerated electrons and protons were injected in the molecular cloud several thousand years ago during an explosion.

TeV halos in the vicinity of pulsars were recently established as a class of extended VHE sources \citep{2017PhRvD..96j3016L, 2017Sci...358..911A, 2022NatAs...6..199L}, featuring bright TeV emission and a hard spectrum \citep{2019PhRvD.100d3016S}. Gamma-ray emission in such sources can be produced in IC scattering of ambient photons by VHE electrons and positrons accelerated by the pulsar-wind termination shock \citep{2019PhRvD.100d3016S}. Even though IC gamma-ray emission beyond 100 TeV is suppressed because of the Klein-Nishina effect, it has been shown that IC can still dominate UHE emission in radiation-dominated environments \citep[e.g.,][]{2009A&A...497...17V, 2021ApJ...908L..49B}. This mechanism was used to explain UHE gamma-ray emission of extended sources detected by HAWC, and three LHAASO sources associated with pulsars \citep{2021ApJ...908L..49B, 2022A&A...660A...8B}. In addition, the study of \citet{2021ApJ...911L..27A} suggests that UHE gamma-ray emission may be a generic feature in the vicinity of pulsars with high spin-down powers $\dot{E} > 10^{36} \, \mathrm{erg s^{-1}}$. The same limit on $\dot{E}$ was also derived from first principles, showing that only very energetic pulsars can power PeV gamma-ray emission \citep{2022ApJ...930L...2D}.

According to the ATNF database\footnote{https://www.atnf.csiro.au/people/pulsar/psrcat/} \citep{Manchester_2005} there is no detected pulsar within a $1^\circ$ radius around LHAASO J2108+5157. This does not a priory exclude the PWN/TeV halo scenario, as the pulsar might remain undetected if its beam is not pointing towards us. The spectral analysis of \citet{2021ApJ...919L..22C} showed that a PWN scenario can also explain the observed UHE emission of LHAASO J2108+5157. With a lack of other observational data, and especially the missing VHE counterpart, the nature of the source remains unknown.

In this paper, we present the results of a dedicated observation of the source region with the first Large-Sized Telescope (LST-1) and \emph{XMM-Newton}\footnote{Proposed as Target of Opportunity observation, PI: R. Walter.}, and results of a dedicated analysis of \emph{Fermi}-LAT data, providing strong constraints on LHAASO J2108+5157 gamma-ray and X-ray emission and the physical nature of the source. We also use all these available data to carry out a detailed modeling of the multiwavelength emission of the source, including a discussion of possible emission scenarios.

The paper is structured as follows: in Section~\ref{sec_data_analysis} we describe our detailed analysis of LST-1 and \emph{Fermi}-LAT data, followed by a dedicated analysis of the \emph{XMM-Newton} data, and an analysis of $^{12}$CO(1-0) emission lines in the direction of the source that were collected in a composite survey by \citet{2001ApJ...547..792D}. In Section~\ref{sec_sed_modeling}, we present and discuss the results of multiwavelength spectral modeling of the source. Finally, a summary and conclusions can be found in Section~\ref{sec_conclusions}.

\section{Observations and data analysis}\label{sec_data_analysis}

\subsection{LST-1}
LHAASO J2108+5157 was observed with LST-1 (\citealt[]{CTALSTproject:2021mfp}, LST Collaboration, in prep.) for 91 hours over 49 nights from June to September 2021. The data were taken in Wobble mode ---which allows  the background to be evaluated from the same observations \citep{FOMIN1994137}--- using four positions centered at $\mathrm{R.A.}=317.15^\circ$, $\mathrm{Dec}=51.95^\circ$, that is, at coordinates that lie between the LHAASO source and the possible \emph{Fermi}-LAT counterpart. The offset of each Wobble with respect to the center of the field of view (FoV) was $0.5^\circ$, instead of the standard $4^\circ$, in order to decrease the number of excess events leaking into the background regions in case the source was found to be extended. The observations presented were taken up to $> 55^\circ$ zenith angle, and with the Moon below horizon. We applied quality cuts based on the stability of the trigger rate, the atmospheric transmission (using MAGIC LIDAR measurements \citep{Fruck:2022igg}), and the rate of CR events, resulting in 49.3 hours of good-quality data (dead-time-corrected) used for the analysis.

The data calibration and shower reconstruction was carried out using the standard pipeline implemented in \texttt{lstchain v0.9} \citep{ruben_lopez_coto_2022_6458862}. To separate pixels containing Cherenkov light emitted by the atmospheric shower particles from background noise, we carried out image cleaning, taking into account the pixel-wise noise fluctuations  (LST Collaboration, in prep.). This method exploits background level information provided by dedicated interleaved pedestal events (containing only noise) acquired during observations at a rate of 100 Hz to reduce the effect of increased pixel noise due to stars in the FoV  for example (LST Collaboration, in prep.). The cleaned shower images were parameterized with Hillas parameters \citep{1985ICRC....3..445H}. We then applied a random forest (RF) algorithm  trained on Monte Carlo (MC) simulated images of gamma and proton events in order to separate the gamma rays from the hadrons (resulting in an estimation of the so-called Gammaness parameter for each event)\footnote{Gammaness parameter indicates how likely it is that the event is created by a gamma ray, rather than a proton or other cosmic-ray particle.} and to reconstruct the energy and arrival direction of each event, using the Hillas parameters as inputs to the RF. As shower development for a primary particle or photon of given energy depends on the specific orientation of each shower relative to the local magnetic field, and the amount of Cherenkov light collected with the telescope depends on the zenith angle of the shower, we trained the RFs on MC events simulated along a declination line that follows the approximate path of the source in zenith and azimuth during the night (LST Collaboration, in prep.). While a constant low night-sky background (NSB) level was assumed in MC simulations, the real observations are performed in a wide range of NSB conditions. To reach the best possible performance in this particular analysis, we tuned the NSB levels in the training sample of MC-simulated images to that of real data, adopting the so-called "noise padding" method (for further details see LST Collaboration, in prep.).

Global selection cuts on Gammaness\footnote{\texttt{lstchain} currently does not allow to apply energy dependent Gammaness cut optimized on a source detection significance to create data files in a format needed for \texttt{Gammapy} spectral analysis (DL3), and thus global cut was used in this case.} and the squared angular distance between the reconstructed event direction and the source ($\theta^2$) ---assuming a point-like nature for the source--- were optimized on 36 high-quality runs of Crab Nebula observations taken in 2021, applying the same selection criteria as in the LHAASO J2108+5157 source analysis. Crab Nebula is the brightest persistent gamma-ray source in the sky, which makes it an ideal target for LST-1 calibration and validation of new data-reconstruction methods \citep[e.g.,][]{CTALSTproject:2021mfp, 2022icrc.confE.716A, Jurysek:2021iig}. The Crab Nebula detection significance was evaluated on a grid of Gammaness $\in (0.5, 0.98)$ with a step of $0.02$ and $\theta^2$ $\in (0.01, 0.1) \, \mathrm{deg^2}$ with a step of $0.002 \, \mathrm{deg^2}$, resulting in the best global selection cuts of Gammaness > 0.84 and $\theta^2 < 0.04 \, \mathrm{deg^2}$ used in the spectral analysis. In order to reach the best possible performance for a potential source detection, we also optimized the Gammaness cut on Crab detection significance in individual energy bins (five bins per decade) keeping the $\theta^2$ cut fixed.

Figure 1 shows a $\theta^2$ plot for three OFF regions reflected with respect to the center of the FOV after selection cuts optimized on Crab detection significance. There is no significant source detection in either of the four energy bins, but at the highest energies (3 - 100 TeV), we see a hint of a source with a detection
significance of $3.67 \sigma$ \citep{1983ApJ...272..317L} and a signal-to-noise ratio (S/N) of $46\,\%$ under the point-like source assumption. This signal is insufficient to claim a detection of the source in the TeV energy range. However, if the excess above $3 \, \mathrm{TeV}$ proves to be significant in future observations, then no excess seen at lower energies may suggest a hard spectral index of the potential VHE source.

\begin{figure*}
\centering
        {
        \includegraphics[width=0.24\hsize]{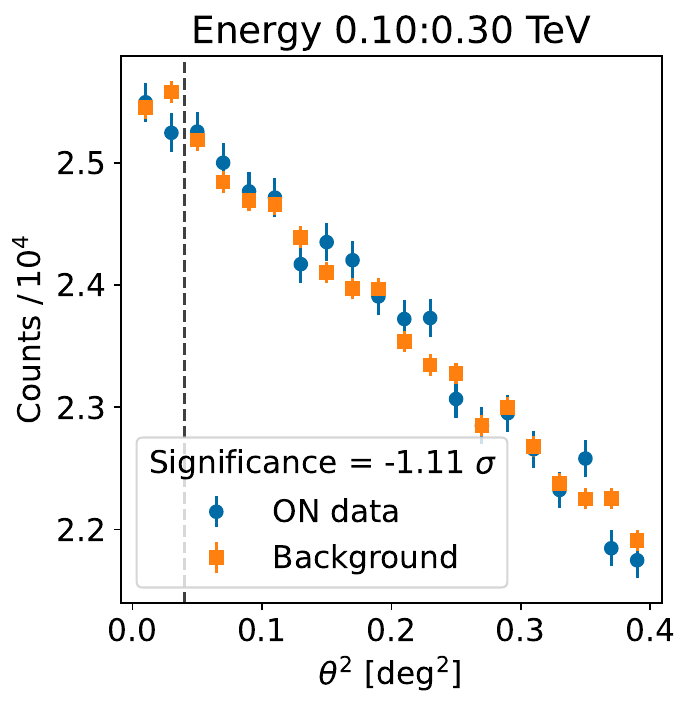}
        \includegraphics[width=0.24\hsize]{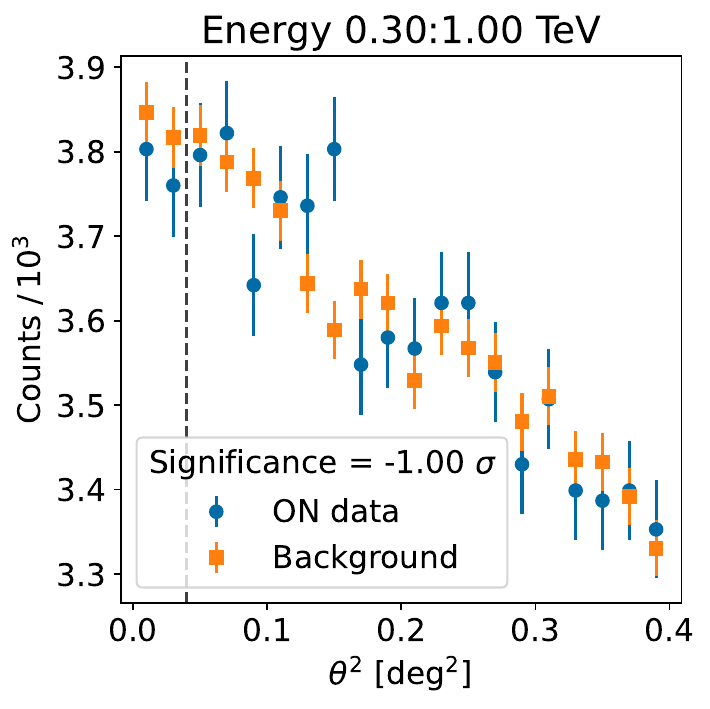}
        \includegraphics[width=0.24\hsize]{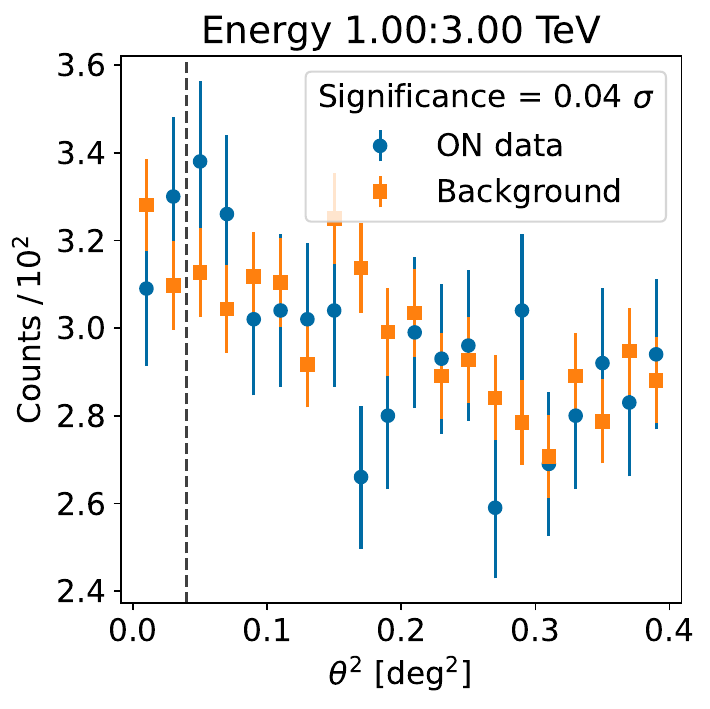}
        \includegraphics[width=0.24\hsize]{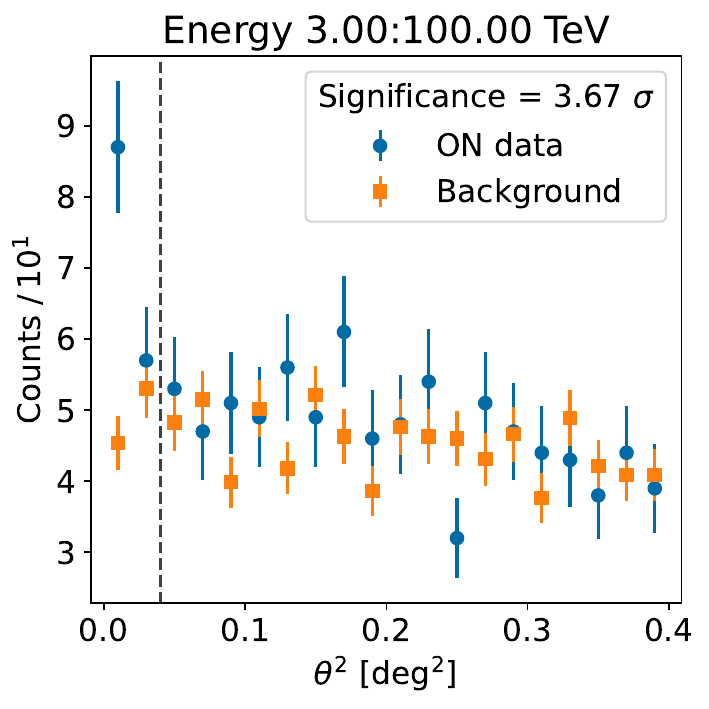}
        }
  \caption{ON (blue) and OFF (orange) counts detected by the LST-1 telescope after selection cuts in 49.3 hours of effective observation time in four blindly selected energy bins. Number of excess events in the first two $\theta^2$ bins for the highest energies is $45\pm13$ with a Li and Ma detection significance of $3.67 \sigma$.}
     \label{fig_lhaaso_theta2_onoff}
\end{figure*}

\subsubsection{Spectral analysis}

We performed a 1D spectral analysis using the \texttt{Gammapy} \citep{2017ICRC...35..766D} package, adopting the source coordinates reported by \citet{2021ApJ...919L..22C}. As LST-1 is in the commissioning phase and the analysis software and methods are still under intensive development, we only performed a preliminary 2D analysis using an acceptance model taken from real LST data, which was then used to correct for radial dependencies in the background models in \texttt{Gammapy}. The results of the 2D analysis can be found in Appendix~\ref{sec_appendix}, but a detailed analysis of the source morphology is left for future studies.

The 1D spectral analysis was performed in the energy range of 100 GeV to 100 TeV using the `reflected regions background' method to estimate the number of background events in the signal region. We assumed the source spectrum to follow a single PL defined as $\mathrm{d}N/\mathrm{d}E = N_0 (E/E_0)^{-\Gamma}$, with fixed reference energy $E_0 = 1 \, \mathrm{TeV}$. Having no morphological information and no sign of a potential source extension in the LST-1 data, the best-fit spectral parameters were found under the  assumption of a point-like source. The resulting spectral parameters are listed in Table~\ref{tab_sp_fit}. Despite the large statistical uncertainties, the resulting photon indices suggest a relatively hard spectrum in the TeV range. We estimated the systematic uncertainty due to the energy resolution on the fitted photon index $\Gamma$ to be $\sim 2 \%$; this is negligible compared to the statistical uncertainty.

\begin{table*}[]
\centering
\begin{tabular}{cccccc}
\hline
Data & Spectral & $N_0$ & $\Gamma$ & $E_\mathrm{cutoff}$ & $-2 \log \mathcal{L}$\\ & model & $[\times 10^{-14} \mathrm{cm^{-2} s^{-1} TeV^{-1}}]$ & & [TeV] & \\
\hline
LST-1 & PL & $8.0 \pm 5.4$ & $1.62 \pm 0.23$ & \ldots & 5.17 \\
LST-1 + LHAASO & ECPL & $7.6 \pm 4.8$ & $1.37 \pm 0.22$ & $50 \pm 14$ & 7.30 \\
\hline
\end{tabular}
\vspace{0.1cm}
\caption{Best-fit parameters for the spectral analysis performed on the LST-1 data alone using a PL model of the spectrum, and for the joint fit to LST-1 and LHAASO data using ECPL.}
\label{tab_sp_fit}
\end{table*}

We also performed a joint likelihood fit of the LST-1 data and LHAASO flux points to find the spectrum of the source in the multi-GeV to multi-TeV range. Provided the hard TeV and soft multi-TeV spectrum, for the spectral shape, we considered a PL with an exponential cutoff (ECPL) defined as $\mathrm{d}N/\mathrm{d}E = N_0 (E/E_0)^{-\Gamma} \exp(-E/E_\mathrm{cutoff})$. The best-fit parameters are listed in Table~\ref{tab_sp_fit}. 

\begin{figure}
\centering
\resizebox{0.5\hsize}{!}
        {\includegraphics[width=0.48\hsize]{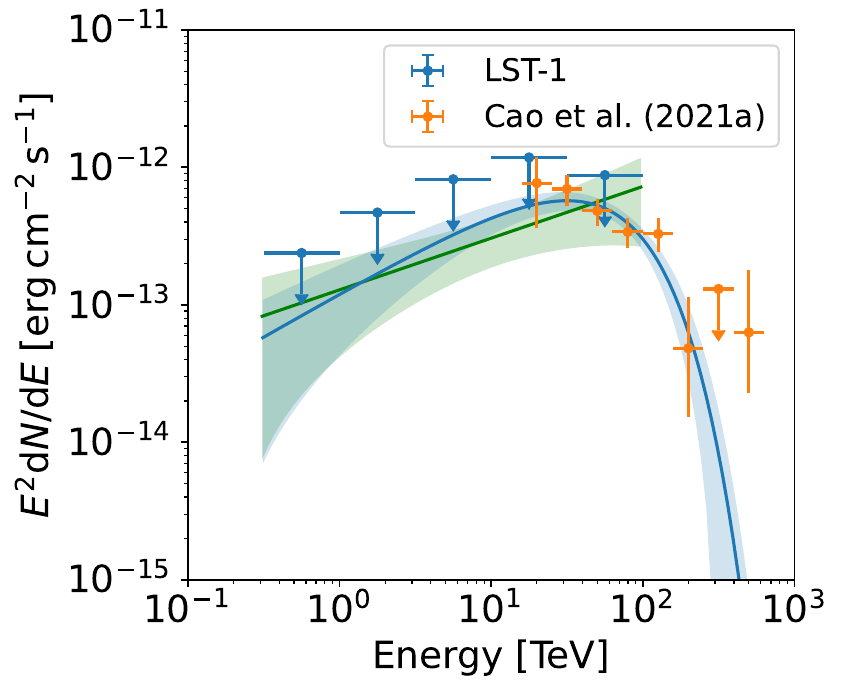}}
  \caption{Spectral energy distribution of the LHAASO J2108+5157 source observed with LST-1. The green confidence band represents the best-fitting PL spectral model of LST-1 data and its statistical uncertainties.  The blue confidence band shows a joint likelihood fit of the LST-1 data and LHAASO flux points with an ECPL spectral model. The ECPL spectral model was used to estimate the 95\% confidence level ULs on the differential fluxes shown in all energy bins.}
     \label{fig_lst_sed_pl_ecpl}
\end{figure}

As a second step, we performed a maximum-likelihood estimation of the source flux in six logarithmically spaced energy bins between 100 GeV and 100 TeV, using ECPL spectral parameters fitted in the previous step. We did not reach a significant source detection in the TeV range, and therefore calculated $95\%$ confidence level differential flux upper limits (ULs) in individual energy bins, which are shown in Figure~\ref{fig_lst_sed_pl_ecpl}. In the first energy bin (0.1-0.316 TeV), the telescope effective area drops below $10\%$, which we used as a safe threshold in the analysis, and therefore the \texttt{Gammapy} flux point estimator only calculated the flux in five energy bins. Table~\ref{tab_lst_fluxes} summarizes the LST-1-measured flux ULs and the corresponding TS in each energy bin.

Although no significant point-like source was detected at energies above 300 GeV ($2.2\sigma$ for PL spectral model), the LST-1 ULs provide strong constraints on the source emission, which is further discussed in Section~\ref{sec_sed_modeling}. The resulting integral $95\%$ UL on the flux of the source is $F(E>300\, \mathrm{GeV}) < 5.0 \times 10^{-13} \, \mathrm{ph\,cm^{-2}\, s^{-1}}$.

\begin{table}[]
\centering
\begin{tabular}{cc|cc}
\hline
E min & E max & Flux ULs & TS \\
 $[\mathrm{TeV}]$ & $[\mathrm{TeV}]$  &
    \Bigg[$
    \makecell{ \times 10^{-14}\\ \mathrm{cm^{-2} s^{-1}}}$\Bigg] &
     \\
\hline
\hline
0.32 & 1.00 & 30.8       & 0.85  \\
1.00 & 3.16 & 19.2       & 0.23 \\
3.16 & 10.00 & 10.6 & 4.19 \\
10.00 & 31.62 & 4.86 & 7.07  \\
31.62 & 100.00 & 1.20 & 0.15 \\
\hline
\end{tabular}
\vspace{0.1cm}
\caption{LST-1 flux ULs ($95\%$ confidence level) assuming a point-like source with an ECPL spectral model.}
\label{tab_lst_fluxes}
\end{table}

\subsection{Fermi-LAT}
We performed a dedicated binned analysis of the region around  LHAASO J2108+5157 with the publicly available \emph{Fermi}-LAT science tools\footnote{https://fermi.gsfc.nasa.gov/ssc/data/analysis/software/}. We followed the standard recommendations from the \emph{Fermi}-LAT team to analyze data from 1 GeV up to 500 GeV within $11^\circ$ of the LHAASO source in order to take into account the broad instrument point spread function (PSF). We specifically selected data with \texttt{evclass=128} and \texttt{evtype=3} and a \texttt{zenith angle $<90^\circ$}. The selected data cover the time interval from $4^\mathrm{}$ August 2008 until $31^\mathrm{}$ January 2022, and we filtered out bad time intervals using \texttt{(DATA\_QUAL>0)\&\&(LAT\_CONFIG==1)}. The data were reprocessed with the most recent \texttt{P8R3\_SOURCE\_V3} instrument response function (IRF), and we used \texttt{gll\_iem\_v07}\footnote{https://fermi.gsfc.nasa.gov/ssc/data/analysis/software/aux/4fgl/\\Galactic\_Diffuse\_Emission\_Model\_for\_the\_4FGL\_Catalog\_Analysis.pdf} for the most updated Galactic diffuse emission model and \texttt{iso\_P8R3\_SOURCE\_V3\_v1} for the isotropic emission. We used a spatial binning of $0.1^\circ$/pixel and eight logarithmically uniform energy bins per decade. As opposed to the LHAASO Collaboration, who used the 10-year 4FGL-DR2 catalog \citep{2020arXiv200511208B, 2020ApJS..247...33A}, we used the more recent 12-year 4FGL-DR3 catalog \citep{2022arXiv220111184F} in our analysis to create the source model, which includes more sources. More specifically, the LHAASO team manually added two new $\gamma$-ray sources in their source model to describe the region of interest (ROI) found with the "find source" tool of \texttt{fermipy} \citep{2017ICRC...35..824W}. One of the two sources visible in their plot (see bottom left square in the left of Figure~6 of \cite{2021ApJ...919L..22C}) is now confirmed in the new DR3 catalog as an interstellar gas clump, but with slightly shifted coordinates (4FGL J2115.2+5219c). The other source does not appear in the new \emph{Fermi}-LAT catalog, and is also not detected in our analysis. As recommended, we included sources in our model up to $5^\circ$ beyond our ROI, keeping their parameters frozen in the likelihood fit.

\begin{figure}
\centering
\includegraphics[width=0.5\hsize]{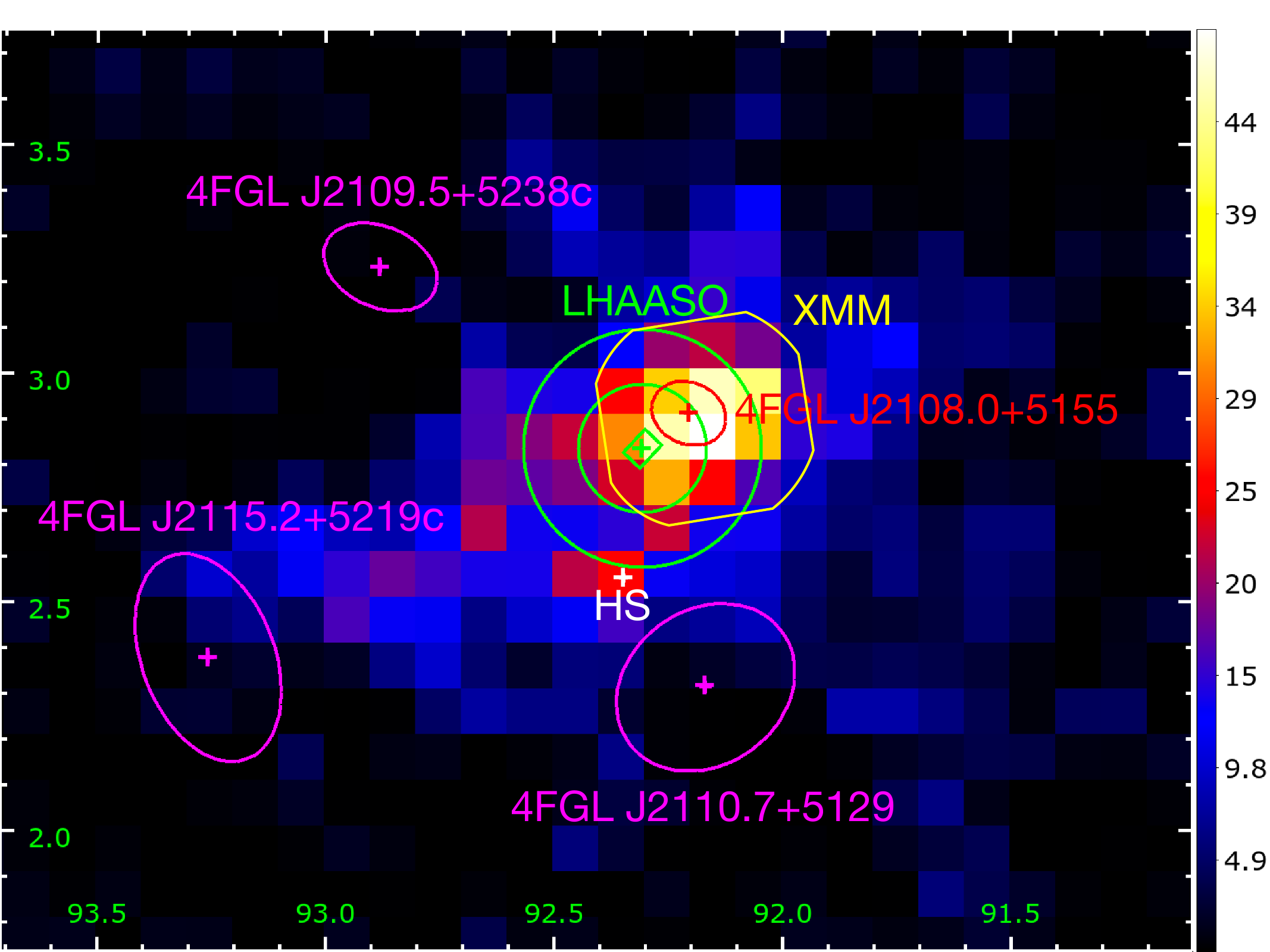}
  \caption{Fermi-LAT TS map in Galactic coordinate above 2 GeV, which shows the sources present in the 4FGL-DR3 catalog with their 95\% positional errors (magenta and red ellipses). The small green rectangle indicates the position of the LHAASO source with statistical uncertainty on R.A. and Dec derived from a two-dimensional Gaussian model, while the smaller green circle represents 95\% position uncertainty of $0.14^\circ$ reported by \citet{2021ApJ...919L..22C}. The larger green circle indicates the $95\%$ UL on the source extension ($0.26^\circ$). The white cross highlights the position of a new potential  hard source, whereas the yellow contour indicates the FoV of the previously discussed XMM observation.}
     \label{fig_fermi_TS_map}
\end{figure}

Any source in the catalog with a nonzero \texttt{flags} parameter is affected by systematic errors. Sources with these indicators should be used with great care. These correspond to significant excesses of photons, but such excesses can result from residual extended emission or confused source pile-up. In the 4FGL-DR3 catalog, three sources within less than $1.5^\circ$ of LHAASO J2108+5157 are flagged with "c", which stands for "interstellar gas clump". More than 50\% of the neighboring sources within less than $4^\circ$ of LHAASO J2108+5157 present at least one nonzero flag \citep[for more details see][]{2022arXiv220111184F}, which make the low-energy \emph{Fermi}-LAT analysis nontrivial.

Looking at all sources present in the 4FGL-DR3 catalog and their 95\% positional errors, there are no counterparts overlapping with the uncertainty position provided by LHAASO (see Figure~\ref{fig_fermi_TS_map}). The closest source is 4FGL J2108.0+5155, which lies $0.13^\circ$ away. Above 1 GeV, the source spectrum is well fitted by a log parabola with a normalization of $(9.8\pm0.9)\times10^{-13}$ ph/cm$^2$/s/MeV, $\alpha=2.5\pm0.2$ and $\beta=0.37\pm0.18$, assuming the same $E_b$ value as that of the catalog (i.e., $E_b = 1580.67$~MeV). The other three 4FGL sources visible in Figure~\ref{fig_fermi_TS_map} are fainter and present a softer log-parabolic spectrum with a turnover at lower energies, with the only exception being J2109.5+5238c, whose spectrum is a PL with a photon index of 2.6, which locally overtakes the flux of 4FGL J2108.0+5155 above a few tens of GeV.

The spectrum of the closest source to LHAASO J2108+5157, namely 4FGL J2108.0+5155, presents a steep decrease above a few GeVs, which is not compatible with the UHE LHAASO points. Therefore, its physical relation to the UHE source is challenging (see the discussion in the following sections). By rerunning the analysis, extending the low-energy threshold to 500~MeV and to 300~MeV, and properly increasing and adapting the selected ROI, the fitted spectra that we obtain present some scatter at low energy, which is due to the large instrument PSF. Although it depends on how much freedom we allow in the fit to the neighboring sources and to the Galactic diffuse emission, in all cases the trend converges toward a unique and consistent behavior above a few GeVs.

\begin{figure*}
\centering
\includegraphics[width=0.99\hsize]{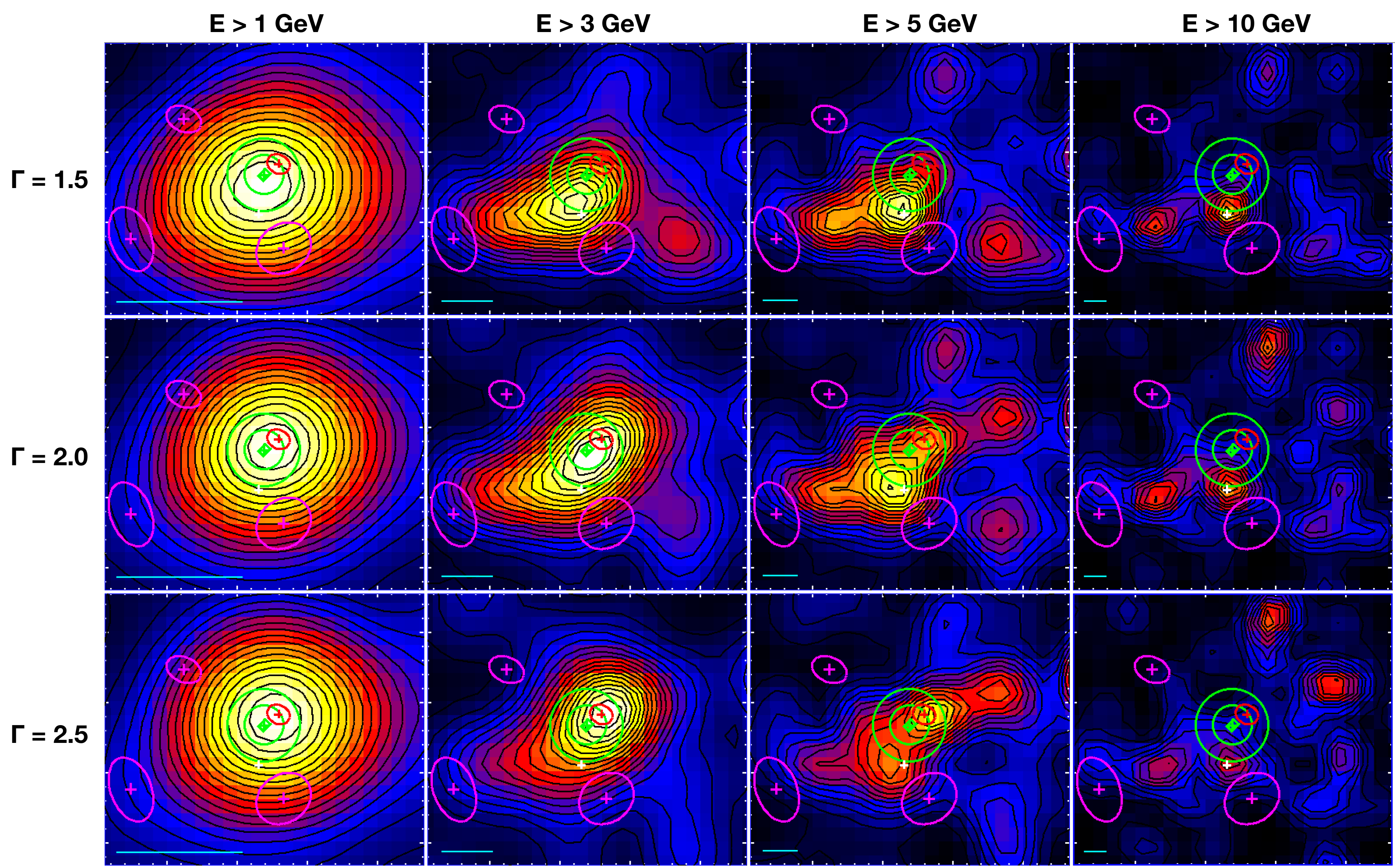}
  \caption{\emph{Fermi}-LAT TS maps computed assuming various photon indexes for the putative source, and above different threshold energies. Each TS map is smoothed with a Gaussian whose sigma is equal to 68\% of the \emph{Fermi}-LAT PSF containment radius measured at each corresponding threshold energy. This size is reported with a cyan segment in the bottom left corner of each plot. Black contours are overplotted with a linear scale to better localize the position of the TS peaks. Green, red, magenta, and white elements are the same as those used and described in Figure~\ref{fig_fermi_TS_map}. The small white ticks on both axes are in units of $0.1^\circ$. Each subplot has been renormalized to its own maximum value to make its color-scale and isocontours comparable.
  }
     \label{fig_fermi_TS_map_table}
\end{figure*}

In order to verify the goodness of the used source model at high energies, we constructed a $15^\circ \times 15^\circ$ TS map centered on the LHAASO source, removing the source 4FGL J2108.0+5155 from the model. We computed the TS map above different threshold energies, from 1~GeV to 10~GeV, and we used a PL spectrum for the putative source, assuming different gamma indices (from -1.5 to -3). Some of these TS maps are reported in Figure~\ref{fig_fermi_TS_map_table}. Each TS map has been smoothed with a Gaussian with a standard deviation equal to 68\% of the \emph{Fermi}-LAT containment angles at each different threshold energy. From this analysis, we can clearly see that, assuming a very soft photon index, above 1 GeV the peak of the TS map coincides with the position of 4FGL J2108.0+5155, whereas using a harder photon index moves the peak toward the southeast (in Galactic coordinates). This trend becomes even more evident when we move towards higher energies. Already above 2~GeV, the excess of the TS maps assumes an elongated shape toward the southeast, and can no longer be considered as point-like, nor can it be reproduced by an extended symmetric Gaussian. These TS maps (Figure~\ref{fig_fermi_TS_map_table}) confirm the very soft spectral behavior of 4FGL J2108.0+5155, whose flux steeply drops above a few GeVs, and suggest the presence of two different sources with clearly distinct spectra, located at two different positions separated by $\sim0.4^\circ$. One of these sources is 4FGL J2108.0+5155, which is already included in the 4FGL-DR3 catalog, whereas the other is a new hard source (hereafter HS), approximately located at $l=92.35^\circ$ and $b=2.56^\circ$, not included in the catalog. Such sources are difficult to distinguish from one another at low energies because of the relatively large PSF of the \emph{Fermi}-LAT instrument, and it is not trivial to spatially disentangle them. On the contrary, they are clearly distinguishable above a few GeVs, where the PSF becomes smaller than the two source separations\footnote{As a matter of reference, $0.4^\circ$ corresponds to more than 68\% containment angle above 3~GeV for the \emph{Fermi}-LAT instrument. At higher energies, the PSF decreases, reaching $0.2^\circ$ and better above 10 GeV. For a detailed \emph{Fermi}-LAT PSF dependence on energy see: \url{https://www.slac.stanford.edu/exp/glast/groups/canda/lat_Performance.htm}}. The existence of two distinct peaks is also evident in the nonsmoothed TS maps. Assuming a flat spectrum, the excess at the position of HS dominates over that of 4FGL J2108.0+5155 above $\sim4$~GeV. If instead we assume a harder spectrum, a similar transition occurs at even lower energies. It is important to mention that the new HS source does not spatially correlate with the local structure of the diffuse Galactic emission model.

Adding the new HS source in the original source model and rerunning the likelihood fit analysis provides slightly different results for the spectral shape of 4FGL J2108.0+5155, which is now fitted with a log parabola with a normalization of $(9.9\pm0.9)\times10^{-13}$ ph/cm$^2$/s/MeV, $\alpha=2.7\pm0.2$ and $\beta=0.32\pm0.16$, assuming the same fixed value for $E_b$. The new HS source is detected with a significance of $\sim4\sigma$, and its spectrum can be fitted with a PL with a normalization of $(1.5\pm0.9)\times10^{13}$ ph/cm$^2$/s/MeV and a photon index of $\Gamma=1.9\pm0.2$, using an energy scale of $E_0=1$~GeV. If we fix the photon index, the normalization accuracy of HS improves to $(1.5\pm0.5)\times10^{-13}$ ph/cm$^2$/s/MeV. Due to the HS small flux at low energies, its inclusion in the model does not significantly affect the spectral results of the neighboring sources, in particular at low energies. Using a different model to represent the HS source, such as a log parabola or ECPL does not improve the likelihood fitting, and so the simple PL is preferred, which presents fewer degrees of freedom. The angular separation of this HS from the LHAASO J2108+5157 source is $0.27^\circ$, which is larger than the $95\%$ upper limit of the extension provided in \citet{2021ApJ...919L..22C}, and  is therefore unlikely to be its counterpart.

\begin{figure}
\centering
\includegraphics[width=0.5\hsize]{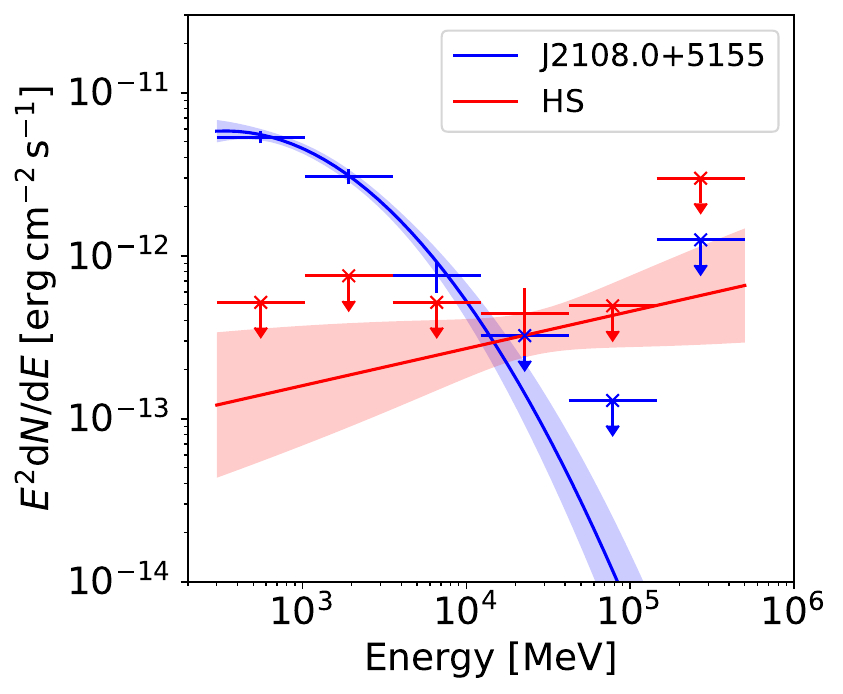}
  \caption{\emph{Fermi}-LAT SED for J2108.0+5155 (blue) and HS (red) analyzed with the energy threshold of 300 MeV (see the text for details). Fluxes in energy bins with TS>10 are drawn as flux points, and for lower TS $95\%$ confidence level ULs are shown.}
     \label{fig_fermi_SED}
\end{figure}

The SED points of J2108.0+5155 and HS shown in Figure~\ref{fig_fermi_SED} were computed by running a separate independent likelihood analysis in each smaller energy band, replacing the source of interest with a simple PL spectrum. The normalization of this spectrum was let free to vary in the fit, whereas its photon index was fixed to the local slope ($\alpha$) of the log parabola in the case of  J2108.0+5155, and to the previous obtained photon index $\Gamma$ in the case of the HS source. The error bar represents $1~\sigma$ statistical error. The confidence band represents the $1~\sigma$ error obtained from the covariance matrix of the fit.

The discrepancy between our flux and that provided by \citet{2021ApJ...919L..22C} can arise from the several differences present between the two analyses, which we highlight in this article. In particular, we used a more recent IRF, a more recent source catalog, and a more recent isotropic diffuse emission component. Furthermore, \citet{2021ApJ...919L..22C} provided the integral flux value assuming a symmetric Gaussian extended source with a radius of $0.48^\circ$, and our TS map results suggest this is not a correct assumption (see Figure~\ref{fig_fermi_TS_map_table}).

\subsection{XMM-Newton}

The field surrounding LHAASO J2108+5157 was observed by \emph{XMM-Newton} on June 11, 2021, for a total of 13.6 ks. The observation was centered on R.A.(J2000)=$317.0170^\circ$, Dec(J2000)=$+51.9275^\circ$. We reduced the data from the European Photon Imaging Camera (EPIC)\footnote{The EPIC is made of three co-aligned detectors: MOS1, MOS2 and pn.} of \emph{XMM-Newton} using XMMSAS v19.1 and the X-COP data-analysis pipeline \citep{eckert17,ghirardini19}. After screening the data and creating calibrated event files using the standard chains, we used the XMMSAS tasks \texttt{pn-filter} and \texttt{mos-filter} to filter out time periods affected by strong soft proton flares. After excising the flaring time periods, the clean exposure time is 4.7 ks (MOS1), 4.9 ks (MOS2), and 3.0 ks (pn). From the clean event files, we extracted images in the soft (0.5-2 keV) and hard (2-7 keV) bands, and used the \texttt{eexpmap} task to create effective exposure maps accounting for vignetting, bad pixels, and chip gaps. To estimate the nonX-ray CR-induced background (NXB), we made use of the unexposed corners of the detectors to rescale the filter-wheel-closed event files available in the calibration database. We then reprojected the filter-wheel-closed data to match the attitude file of our observation and extracted model-particle background images from the rescaled filter-wheel-closed event files. Finally, we summed up the images of the three EPIC instruments as well as the NXB maps. We also created a total EPIC exposure map by summing up the exposure maps of the three instruments, weighted by their respective effective area.

We ran the XMMSAS task \texttt{ewavelet} to detect X-ray sources in the field. The resulting catalog contains 11 sources in the soft band and 2 in the hard band. The brightest source in the field is associated with the eclipsing binary star V* V1061 Cyg (RX J2107.3+5202). We extracted the spectrum of the source and fitted it in XSPEC. We find that the eclipsing binary exhibits a very soft spectrum, which appears better represented by a thermal model with $kT\sim0.8$ keV than with a PL. The total model flux of the source is $3.7\times10^{-13} \, \mathrm{erg\, cm^{-2} \, s^{-1}}$ (0.5-2 keV) and $2.0\times10^{-14} \, \mathrm{erg\, cm^{-2} \, s^{-1}}$ (2-7 keV). Assuming a thermal nature of the spectrum, this is an unlikely counterpart for the LHAASO source. The other sources in the field are substantially fainter than the fluxes reported here. 

If  the source were associated with leptonic emission from a SNR or PWN, we would expect the corresponding X-ray source to be extended. We therefore attempted to place upper limits on the possible X-ray flux of an extended source with varying radius and centered on the LHAASO source position. To this end, we used the public code \texttt{pyproffit} \citep{eckert20} to extract the brightness profile of unresolved emission across the field. The NXB-subtracted, vignetting-corrected profiles are consistent with a flat brightness distribution, which is compatible with the expected sky background emission made of a combination of the local hot bubble, the Galactic halo, and the cosmic X-ray background. If the source is smaller than the \emph{XMM-Newton} FoV ($30'$ in diameter), we can use the measured count rates in the outer regions of the field to estimate the local sky background emission and set the maximum allowed number of source counts on top of the background. For aperture photometry performed in the Poisson regime, the S/N is given by

\begin{equation}
S/N = \frac{N_{\rm src}}{\sqrt{N_{\rm src} + N_{\rm bkg}}}
,\end{equation}

where $N_{\rm src}$ is the number of (background-subtracted) source counts and $N_{\rm bkg}$ is the total number of background counts (sky background + NXB). A $2\sigma$ upper limit can therefore be set by computing the number of source counts for which S/N=2, that is

\begin{equation}
    N_{2\sigma} = 2 + 2\sqrt{1+N_{\rm bkg}}.
\end{equation}

For a uniform, circular source, the corresponding upper limit depends on the assumed source radius because the background expectation is proportional to the area. We computed such upper limits for three possible apertures: a radius of 0.5 arcmin (point-like source) and extended sources with radii of 3 and 6 arcmin, respectively, which was the maximum possible radius considering the relatively small FoV of the EPIC camera and that the \emph{XMM-Newton} observation is not centered at the position of the UHE source. The estimated $2\sigma$ upper limits on the allowed number of counts and count rates are given in Table \ref{tab:xmm_uls}.

To convert the upper limits on the count rates into fluxes, a model spectrum needs to be assumed; it is also especially important that we understand whether or not the source is affected by Galactic extinction. In the case where the source is located at the opposite side of the Galaxy, the measured X-ray fluxes will be attenuated by the Galactic $N_H$, which is substantial along the Galactic plane. The Galactic HI column density at the position of the LHAASO source is $1.05\times10^{22}$ cm$^{-2}$ \citep{2016A&A...594A.116H}, which implies that the majority of the low-energy photons will be absorbed along the way, such that the actual emitted flux can be substantially higher. Conversely, if the source is local, the source spectrum will be mostly unabsorbed. Here, we provide upper limits for the two extreme cases of a completely absorbed and a completely unabsorbed source to bracket all possible scenarios. We simulated \emph{XMM-Newton} spectra assuming that the source spectrum is a PL with a photon index of 2.0 and computed the conversion between count rate and flux in the two energy bands and for the absorbed and unabsorbed scenarios alike. The conversion factors were then used to determine the corresponding flux upper limits. Table \ref{tab:xmm_uls} provides a list of upper limits for a wide range of scenarios. The allowed values are in the range $10^{-15}-10^{-13} \, \mathrm{erg\, cm^{-2} \, s^{-1}}$ depending on the assumed source size and spectrum. 

In order to estimate a flux upper limit in the full region of a possible extension of the UHE source from the data available (95\% extension upper limits has a radius of 16 arcmin), we scaled the 6 arcmin upper limits by the ratio of the two radii. However, we note that there is no surface brightness gradient across the \emph{XMM-Newton} FoV, which would be expected for a very extended source centered at  the UHE source coordinates. This means that any low-surface-brightness extended source would need to be not only very extended but also uniform across the \emph{XMM-Newton} FoV, which is not very likely.

\begin{table*}[]
    \centering
    \footnotesize

    \begin{tabular}{ccccccccc}
    \hline
Radius & $N_{\rm src,0.5-2}$ & $CR_{0.5-2}$ & $F_{\rm 0.5-2, abs}$ & $F_{\rm 0.5-2, unabs}$ & $N_{\rm src,2-7}$ & $CR_{2-7}$ & $F_{\rm 2-7, abs}$ & $F_{\rm 2-7, unabs}$\\
$[\textrm{arcmin}]$ & [cts] & $[\mathrm{cts\, s^{-1}}]$ & $[\mathrm{erg\, cm^{-2} \, s^{-1}}]$ & $[\mathrm{erg\, cm^{-2} \, s^{-1}}]$ & [cts] & $[\mathrm{cts\, s^{-1}}]$ & $[\mathrm{erg\, cm^{-2} \, s^{-1}}]$ & $[\mathrm{erg\, cm^{-2} \, s^{-1}}]$\\
\hline
\hline
0.5 & 10.214 & 5.940e-4 & 1.69e-14 & 2.36e-15 & 10.014 & 6.391e-4 & 1.28e-14 & 1.04e-14 \\
3 & 47.415 & 3.070e-3 & 8.76e-14 & 1.22e-14 & 47.892 & 3.380e-3 & 6.77e-14 & 5.51e-14\\
6 & 88.527 & 6.464e-3 & 1.84e-13 & 2.57e-14 & 90.055 & 7.221e-3 & 1.45e-13 & 1.18e-13\\
16 & \ldots & \ldots & 4.9e-13 & 6.9e-14 & \ldots & \ldots & 3.8e-13 & 3.15e-13 \\
\hline
\end{tabular}
\vspace{0.1cm}
\caption{ \label{tab:xmm_uls} \emph{XMM-Newton} $2\sigma$ upper limits on the number of counts, count rate (CR), and the X-ray flux of the LHAASO source for varying source apertures, energy bands, and source emission models. The fluxes are estimated assuming that the source spectrum can be described by a PL with a photon index of 2.0, either on the other side of the Galaxy (i.e., absorbed by the local Galactic column density) or local (unabsorbed). In the absorbed case, the corresponding flux is the equivalent unabsorbed source flux. Upper limits for the radius of 16 arcmin are 6 arcmin scaled to the angular size of the UHE source extension.}
\end{table*}

\subsection{Molecular clouds}\label{sec_molecular_clouds}

LHAASO J2108+5157 is located in the direction of relatively dense molecular clouds. \citet{2021ApJ...919L..22C} searched the database of molecular clouds in the Galactic plane \citep{2017ApJ...834...57M} and found its position close to the direction of two molecular clouds [MML2017]2870 and [MML2017]4607 with distances of 1.43 kpc and 3.28 kpc, and masses of $3.5 \times 10^{4} \, \mathrm{M_\odot}$ and $8.5 \times 10^3 \, \mathrm{M_\odot}$, respectively. 

To estimate the total gas density in the direction of LHAASO J2108+5157, we considered a contribution of $\mathrm{H_2}$ molecular density and of neutral hydrogen $\mathrm{HI}$ density. We used $^{12}$CO(1-0) line-emission observations collected during the composite survey by \citet{2001ApJ...547..792D} to estimate the density and distance of the molecular clouds. We note that, given the Galactic longitude of the source, namely $l>90^\circ$, the kinematic distance of each molecular cloud has only a single solution \citep[see e.g.,][]{2009ApJ...699.1153R}. In order to get a mean radial-velocity spectrum of brightness temperature $T_\mathrm{B}(v)$ in the UHE source region, we averaged $T_\mathrm{B}(v)$ in all pixels within $0.26^\circ$, which is $95\%$ ULs on the UHE source extension reported by \citet{2021ApJ...919L..22C}. Consistently with \citet{2021ApJ...919L..22C}, we identified three peaks in the $T_\mathrm{B(H_2, v)}$ spectrum, which can be well described by three Gaussians at $v_1 \approx -11.8 \, \mathrm{km \, s^{-1}}$, $v_2 \approx -2.7 \, \mathrm{km \, s^{-1}}$ , and $v_3 \approx 8.4 \, \mathrm{km \, s^{-1}}$. Two of these correspond to the centroid velocities of the molecular clouds, of namely, $-1.1 \, \mathrm{km \, s^{-1}}$ and $-13.7 \, \mathrm{km \, s^{-1}}$, identified by \citep{2017ApJ...834...57M}, but the origin of the last peak remains unknown, and we further consider only the clouds with negative radial velocities. 

\begin{figure*}
\resizebox{\hsize}{!}
        {\includegraphics[width=0.48\hsize]{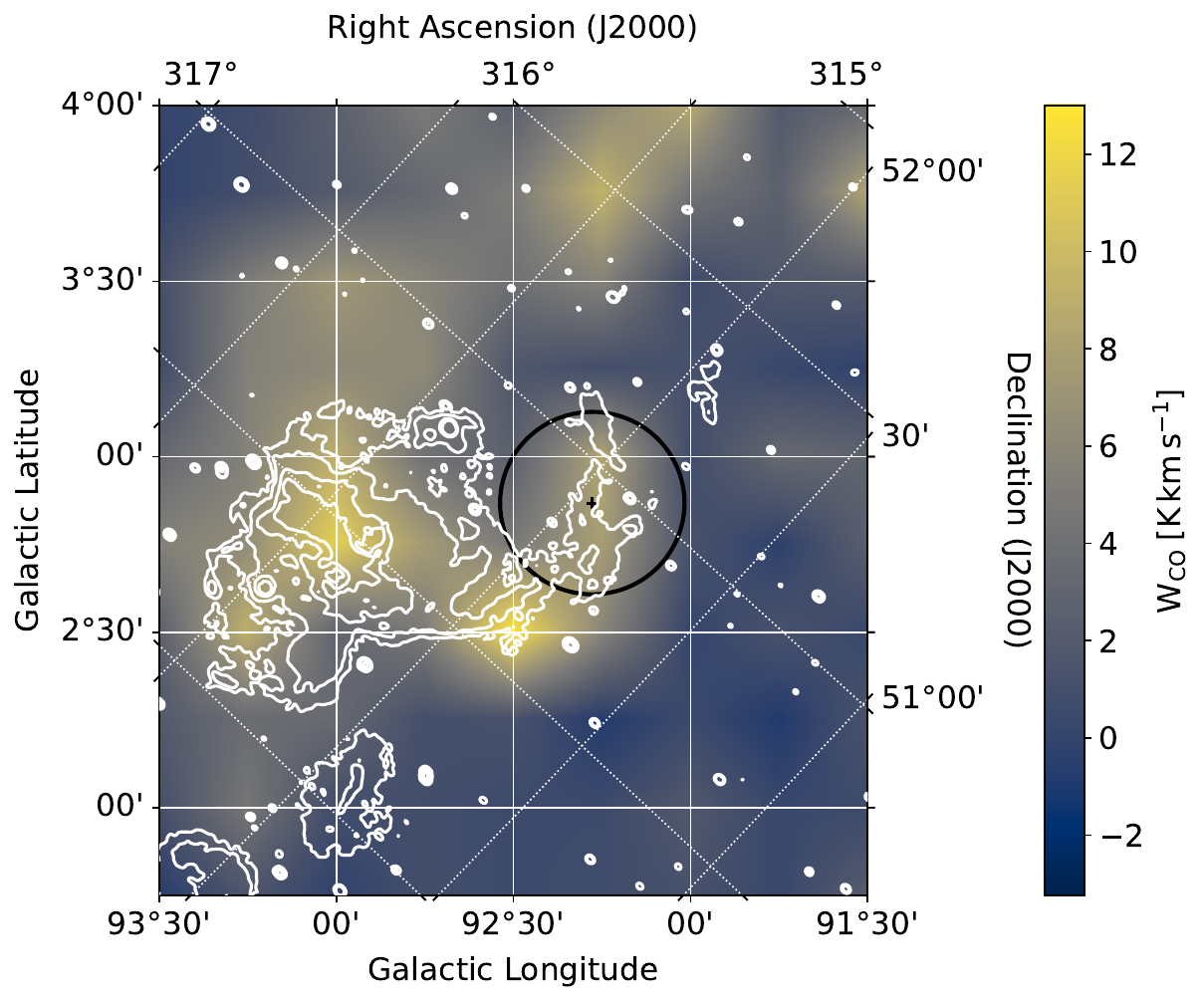}
        \includegraphics[width=0.48\hsize]{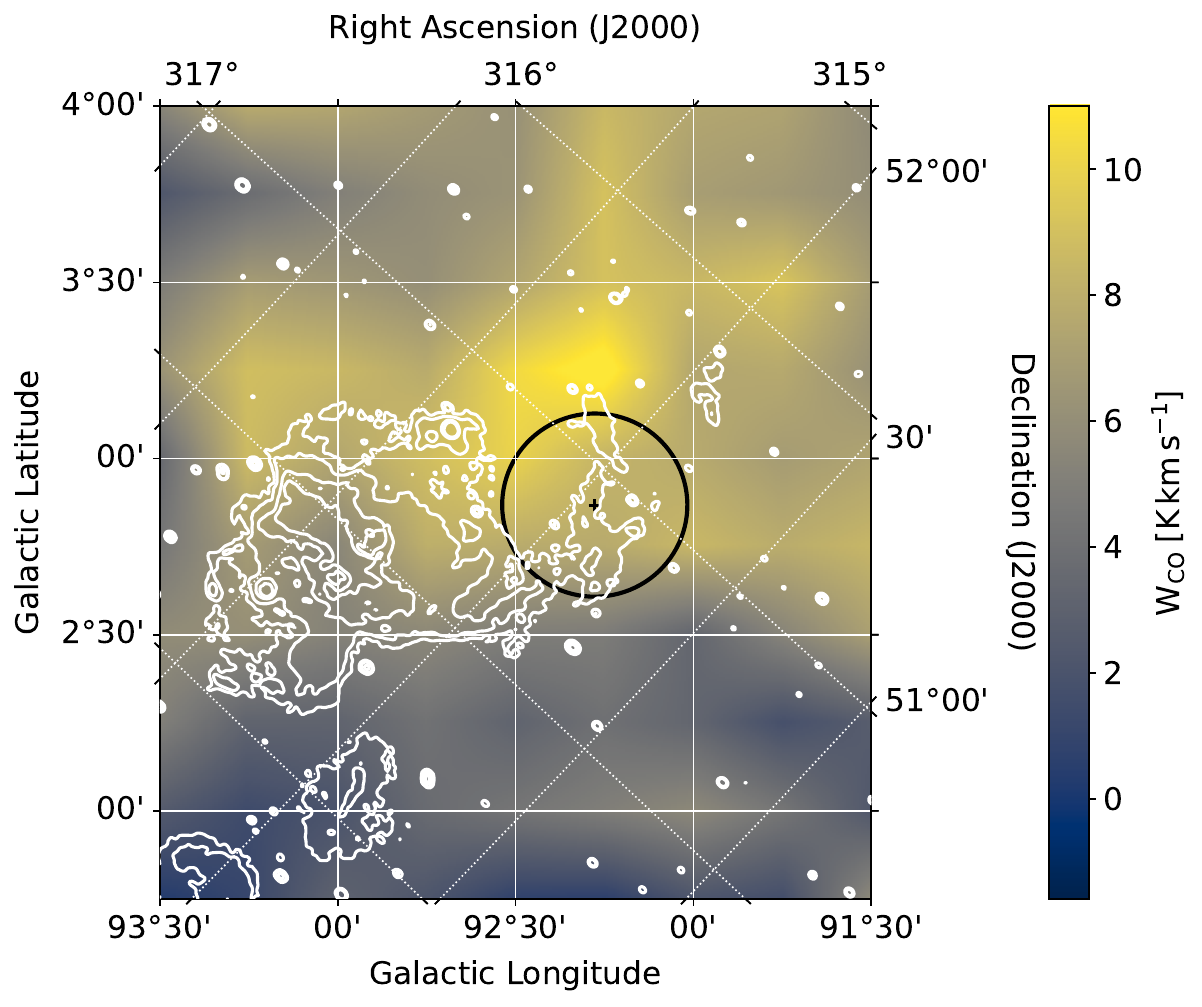}}
  \caption{Velocity-integrated $^{12}$CO intensity ($W_\mathrm{CO}$) of two molecular clouds spatially coincident with the direction of LHAASO J2108+5157. \textit{Left:} Integrated velocity of the first Gaussian component peaking at $v_1 \approx -11.8 \, \mathrm{km \, s^{-1}}$, with corresponding distance of $d_1 \approx 3.1 \, \mathrm{kpc}$. \textit{Right:} Integral of the second Gaussian component at $v_2 \approx -2.7 \, \mathrm{km \, s^{-1}}$ and $d_1 \approx 2.0 \, \mathrm{kpc}$. The white contour represents 1420 MHz continuum emission from the Canadian Galactic
Plane Survey \citep{2003AJ....125.3145T}. The position of  LHAASO J2108+5157 is marked with a black cross, and $95\%$ UL on its extension ($0.26^\circ$) is indicated with a black circle \citep{2021ApJ...919L..22C}. Bilinear interpolation is used to smooth out the contributions from individual pixels.}
     \label{fig_molecular_clouds}
\end{figure*}

For determination of the kinematic distances $d$ to the molecular clouds, we adopted a PL rotation curve \citep{2017A&A...601L...5R} with a galactocentric radius of $R_0 = 8.34 \, \mathrm{kpc}$ and with the same orbital velocity as the Sun, $V_0 = 240 \, \mathrm{km \, s^{-1}}$, leading to $d_1 \approx 3.1 \, \mathrm{kpc}$ and $d_2 \approx 2.0 \, \mathrm{kpc}$. The angular radius of the UHE emission can then be converted to a physical radius of the source, which gives $r_1 = 7.1 \, \mathrm{pc}$ and $r_2 = 4.5  \, \mathrm{pc}$, depending on its distance, if spatially coincident with the molecular clouds. 

In order to obtain the average densities of the molecular clouds in the source region, we integrated the individual Gaussian contributions $W_\mathrm{CO}$ over the velocity ranges $v_1 \in (-23, -1)  \, \mathrm{km \, s^{-1}}$ and $v_2 \in (-9, 4)  \, \mathrm{km \, s^{-1}}$, given by $3\sigma$ ranges of the two Gaussian contributions. $W_\mathrm{CO}$ can then be converted into the column density as $N(\mathrm{H_2}) =  X_\mathrm{CO} W_\mathrm{CO} / N_\mathrm{px}$, where $X_\mathrm{CO} = 2 \times 10^{20} \, \mathrm{cm^{-2} \, s \, K^{-1} \, km^{-1}}$ \citep{2013ARA&A..51..207B}, and $N_\mathrm{px}$ is the number of pixels on the sky within the source extension. Assuming a spherically symmetrical UHE source emission region, we estimate the number density of molecular hydrogen to be $n_1(\mathrm{H_2}) = N_1(\mathrm{H_2}) / 2r_1 = 51 \, \mathrm{cm^{-3}}$ and $n_2(\mathrm{H_2}) = 170 \, \mathrm{cm^{-3}}$, for the distant and the close molecular cloud, respectively. A sky map of the velocity-integrated $^{12}$CO intensity is shown in Figure~\ref{fig_molecular_clouds} for both molecular clouds.

To estimate the density of neutral hydrogen $\mathrm{HI}$, we used the brightness-temperature velocity spectrum $T_\mathrm{B(HI,v)}$ in the direction of the source averaged over its angular extension from the HI4$\pi$ survey \citep{2016A&A...594A.116H}. The column density of $\mathrm{HI}$ can be estimated under the optically thin limit assumption as $N(\mathrm{HI}) = 1.823 \int T_\mathrm{B (HI,v)}\,\mathrm{d}v \, \mathrm{cm^{-2}}$ \citep{1990ARA&A..28..215D}. Assuming the same distance for both gas phases, we integrated over the same velocity ranges as for the $\mathrm{H_2}$ molecular clouds. Following the same geometrical assumptions as for $\mathrm{H_2}$, we estimate the number density of neutral hydrogen to be $n_1(\mathrm{HI}) = 64 \, \mathrm{cm^{-3}}$ and $n_2(\mathrm{HI}) = 70 \, \mathrm{cm^{-3}}$. The total number density of the gas in the source region can therefore be estimated for both clouds as $n_1 = n_1(\mathrm{HI}) + n_1(\mathrm{H_2}) = 115 \, \mathrm{cm^{-3}}$ and $n_2 = 240 \, \mathrm{cm^{-3}}$.

\section{Spectral modeling and discussion}\label{sec_sed_modeling}

\subsection{Leptonic scenario of emission}

As most of the VHE sources in our Galaxy have been identified as PWNe \citep{2018A&A...612A...1H}, we first examine this possible scenario of emission. Here, we provide quantitative estimates of and limits on the leptonic scenario of emission derived from the data, and compare them with the physical properties of known PWNe/TeV halos. 

We used the \texttt{naima}\footnote{https://naima.readthedocs.io/} package \citep{2015ICRC...34..922Z} to derive a parent electron distribution reproducing IC dominated \citep{2014ApJ...783..100K} VHE to UHE emission of LHAASO J2108+5157. For the sake of simplicity, we assume a single electron spectrum in the form of an ECPL $f(E) \sim E^{-\alpha} \exp(-E/E_\mathrm{c})$. As the source was not detected in the X-ray range, we only considered IC emission using the LST-1 and LHAASO flux points \citep{2021ApJ...919L..22C}. The target photon field for IC is expected to consist of cosmic microwave background (CMB) and far-infrared (FIR) radiation of the dust, with temperatures of 2.83 K and 20 K, and energy densities of $0.26 \; \mathrm{eV \, cm^{-3}}$ and $0.3 \; \mathrm{eV \, cm^{-3}}$, respectively, which are typical values at the galactocentric radius of the Sun \citep{2009ARA&A..47..523H, 2017MNRAS.470.2539P}.

\begin{figure}
        {\includegraphics[width=0.99\hsize]{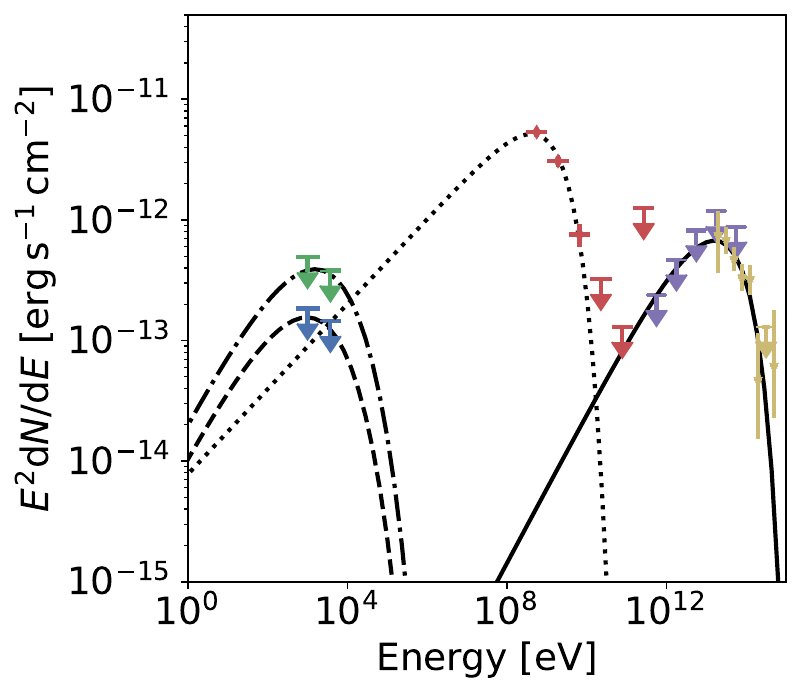}}
  \caption{Multiwavelength SED of LHAASO J2108+5157 showing a leptonic scenario of emission. Observations with different instruments are represented by data points of different colors: \emph{XMM-Newton} $r=6'$ (blue), $r=16'$ (green), \emph{Fermi}-LAT (red), LST-1 (purple), and LHAASO-KM2A (yellow). The black solid line represents the best-fitting IC-dominated emission of LST-1 and LHAASO data. The corresponding synchrotron radiation of the same population of electrons is represented with dashed and dash-dotted lines for $B = 1.2 \, \mathrm{\mu G}$ and $B = 1.9 \, \mathrm{\mu G}$, respectively. The dotted line represents a phenomenological model of a tentative pulsar: the best-fit PL with a subexponential cutoff on the \emph{Fermi}-LAT data. }
     \label{fig_leptonic_sed}
\end{figure}

The electron distribution that best describes the observations is shown in Figure~\ref{fig_leptonic_sed}; it has a cutoff at $E_c = 100^{+70}_{-30} \, \mathrm{TeV}$ and a spectral index of $\alpha = 1.5 \pm 0.4$. All parameters of the model are summarized in Table~\ref{tab:naima_leptonic}. If there is magnetic field, the same population of electrons must emit synchrotron radiation \citep{2010PhRvD..82d3002A}, which together with the \emph{XMM-Newton} upper limits on the X-ray emission allow us to put constraints on the strength of the magnetic field  in the PWN. Figure~\ref{fig_leptonic_sed} shows $95\%$ \emph{XMM-Newton} upper limits on the absorbed emission in a circular region with $r = 6'$, which is a reasonable upper limit on the angular size of a Galactic PWN emitting X-rays at a distance $d \leq 10 \, \mathrm{kpc}$ \citep{Bamba_2010}. The comparison with the $95\%$ CL of synchrotron emission constrains the magnetic field to $B \lesssim 1.2 \, \mathrm{\mu G}$, which is lower than a typical value of Galactic magnetic field $B_\mathrm{G} \approx 3 \, \mu G$. We note that for unabsorbed X-ray upper limits, which are relevant if the source is relatively close, the constraints on the magnetic field would be even stronger, namely $B \lesssim 0.5 \, \mathrm{\mu G}$. Given its Galactic latitude of $b \approx 3^\circ$, the source is close to the Galactic plane if it is not too distant from the Sun, and one should not expect a background magnetic field strength significantly below the typical level; therefore the absorbed case is favored. The possibility of greater extension of the undetected PWN ---which would potentially lead to more relaxed constraints on its magnetic field--- cannot be excluded. However, we note that even the approximate absorbed X-ray flux ULs scaled on the full UHE source extension lead to a relatively low $B \lesssim 1.9 \, \mathrm{\mu G}$ compared to the average Galactic magnetic field (also shown in Figure~\ref{fig_leptonic_sed} for reference).

\begin{table}[]
    \centering
    \begin{tabular}{rcc}
    \hline
Parameter &  Best fit value & Frozen \\
\hline
$E_0 \, [\mathrm{TeV}]$ & $1$ & True \\
$E_\mathrm{e,min}  \, [\mathrm{GeV}] $ &        $0.1$ & True \\
$E_\mathrm{e,max}  \, [\mathrm{TeV}] $ &        $1000$ & True \\
$N_0 \, [\times 10^{43} \, \mathrm{TeV^{-1}}]$ & $ 1.7^{+4}_{-1.4}$ & False \\
$E_\mathrm{c} \, [\mathrm{TeV}]$ & $100^{+70}_{-30}$ & False \\
$\alpha$ & $1.5 \pm 0.4$ & False \\
\hline
\end{tabular}
\vspace{0.1cm}
\caption{Best-fit parameters of an ECPL electron distribution in the form $\mathrm{d}N/\mathrm{d}E = N_0 (E/E_0)^{-\alpha} \exp(-(E/E_\mathrm{c}))$, where $E_0$ is the energy scale, $\alpha$ the spectral index, and $E_\mathrm{c}$ the cutoff energy. Normalization of the spectrum $N_0$ is calculated for the source distance of 1 kpc. VHE-UHE emission of LHAASO J2108+5157 is assumed to be dominated by emission due to IC scattering of electrons on CMB ($T = 2.83 \, \mathrm{K}$, $u = 0.26 \, \mathrm{eV cm^{-3}}$) and FIR ($T = 20 \, \mathrm{K}$, $u = 0.3 \, \mathrm{eV cm^{-3}}$) seed photon fields.}
\label{tab:naima_leptonic}
\end{table}

Such a weak magnetic field, needed to suppress the synchrotron emission of LHAASO J2108+5157, is on the lower end of the typical range seen for $B_\mathrm{PWN}$, which is, $1 - 100 \, \mathrm{\mu G}$ \citep{2014MNRAS.443..138M, 2018A&A...609A.110Z}. However, we note that a relatively weak magnetic field is needed to explain a leptonic UHE emission, which is only possible in radiation-dominated environments \citep{2009A&A...497...17V, 2021ApJ...908L..49B, 2022A&A...660A...8B}. MILAGRO \citep{Abdo:2009ku} and HAWC \citep{2017Sci...358..911A} detected an extended $2^\circ$ TeV emission surrounding the pulsar Geminga, leading to the recent establishment of a new class of TeV-halo sources \citep{2017PhRvD..96j3016L, 2019PhRvD.100d3016S}. Resulting from propagation of relativistic electrons that already left the PWN in the interstellar medium (ISM), magnetic field in the TeV halos can be expected to follow the level of the magnetic field in the ISM. However, \citet{Liu_2019} obtained an upper limit on the magnetic field in the  halo of Geminga of   $B < 1 \, \mathrm{\mu G}$, and therefore the TeV halo scenario for LHAASO J2108+5157 is also feasible. The TeV nebula surrounding Geminga has a large angular extension, but this pulsar is also relatively close ($d=250$ pc \citep{2007Ap&SS.308..225F}). In the Geminga-like scenario, the lower limit on the distance of LHAASO J2108+5157 is approximately 2 kpc in order not to violate the source-extension UL of $0.26^\circ$ provided by \citet{2021ApJ...919L..22C}.

Inverse-Compton-dominated radiation of a single electron population cannot explain the soft GeV emission of 4FGL J2108.0+5155, which is spatially coincident with LHAASO J2108+5157. There are 117 gamma-ray pulsars identified in the \emph{Fermi}-LAT data showing similar spectral properties to 4FGL J2108.0+5155 \citep{2013ApJS..208...17A}. We therefore put forward the hypothesis that the GeV emission is the signature of a gamma-ray pulsar. \citet{2016ApJ...820....8S} applied machine learning methods to classify sources in the Third \emph{Fermi}-LAT catalog in two major classes: AGNs and pulsars. 3FGL J2108.1+5202, which is the Third \emph{Fermi}-LAT general catalog \citep{2015ApJS..218...23A} counterpart of 4FGL J2108.0+5155, was classified consistently with logistic regression (LR) and RF classifiers as a pulsar, which provides support for our hypothesis. However, we note that the resulting LR and FR probabilities are relatively low, that is, only about $30\, \%$, and therefore we cannot exclude the possibility of misclassification of the source, and an extragalactic origin of the HE emission cannot be excluded \citep[for further details see][]{2016ApJ...820....8S}.

Gamma-ray pulsars are characterized by soft spectra, with the flux steeply falling above a few GeV \citep[e.g.,][]{2020A&A...643L..14M}. In the \emph{Fermi}-LAT energy band, the typical differential spectrum can be described with a PL with a subexponential cutoff $\mathrm{d}N/\mathrm{d}E = N_0 (E/E_0)^{-\Gamma} \exp(-(E/E_\mathrm{cutoff})^b)$, where $E_0$ is the energy scale, $\Gamma$ the photon index, $E_\mathrm{cutoff}$ the cutoff energy, and $b$ the cut-off strength \citep{Leung_2014, 2020A&A...643L..14M}. In order to reduce the degeneracy of the model parameters, considering that there are only three significant \emph{Fermi}-LAT flux points, we fixed $b = 0.7$, which is the cut-off strength of the PL with a subexponential cutoff model of the Geminga pulsar SED in the GeV band \citep{2020A&A...643L..14M}. The best fit of the \emph{Fermi}-LAT data consistent with \emph{XMM-Newton} ULs shown in Figure~\ref{fig_leptonic_sed} has $\Gamma = 1.5^{+0.1}_{-0.2}$ and $E_\mathrm{cutoff} = 0.9 \pm 0.2 $ GeV. Despite the large uncertainty, the photon index is consistent with that of gamma-ray pulsars with a spin-down power of $\dot{E} = 10^{34} - 10^{37} \, \mathrm{erg \, s^{-1}}$ \citep{2013ApJS..208...17A}. The gamma-ray luminosity of 4FGL J2108.0+5155 of between 1 and 100 GeV is $L_{1-100 \, \mathrm{GeV}} = 2 \times 10^{33} (d/\mathrm{1 \, kpc})^2 \, \mathrm{erg \, s^{-1}}$. One should note that, assuming a Galactic origin for the source, the Galactic longitude of the source, of namely $l = 92.2^\circ$, implies an UL on the source distance of $d \lesssim 8 \, \mathrm{kpc}$,
because of the geometry between the position of the Sun and the edge of the Galaxy at the Galactic longitude of the source. Such UL on the source distance leads to an UL on the source luminosity of $L_{1-100 \, \mathrm{GeV}} \lesssim 1.3 \times 10^{35} \, \mathrm{erg \, s^{-1}}$, which is consistent with the population of  gamma-ray pulsars for any possible distance to the source.

The allowed spin-down power range of the tentative pulsar also implies limits on the luminosity in the TeV band for the nebula powered by the pulsar of $L_{1-10 \, \mathrm{TeV}} \approx 10^{31} - 10^{35} \, \mathrm{erg \, s^{-1}}$ when compared with the sample of known TeV PWNe \citep[see Figure 7 in][]{2018A&A...612A...2H}. The TeV luminosity of LHAASO J2108+5157 is $L_{1-10 \, \mathrm{TeV}} \approx 6 \times 10^{32} (d/\mathrm{1 \, kpc})^2 \, \mathrm{erg \, s^{-1}}$, which makes the pulsar scenario valid for any possible distance of the source within the Galaxy. In the Geminga-like scenario, assuming the source extension of the order of 10 pc, the minimum possible distance of 2 kpc implies a TeV PWN luminosity of $L_{1-10 \, \mathrm{TeV}} > 2 \times 10^{33} \, \mathrm{erg \, s^{-1}}$  and therefore a pulsar spin-down power of $\dot{E} > 10^{35} \, \mathrm{erg \, s^{-1}}$, which is consistent with the estimates derived from the SED of the \emph{Fermi}-LAT counterpart.

The total energy in electrons dominating the TeV emission is $E(>1 \, \mathrm{GeV}) \approx 1 \times 10^{45} (d / 1 \, \mathrm{kpc)^2 \, erg}$, which can be compared with the total energy released by the pulsar during relativistic electron cooling time, given by 
\begin{align}
t_\mathrm{cool, yr} &= (1/t_\mathrm{IC, yr} + 1/t_\mathrm{syn, yr})^{-1},\\
t_\mathrm{IC, yr} &\approx 3.1 \times 10^5 (1 + 40 E_\mathrm{e,TeV} (kT)_\mathrm{eV})^{1.5} u_\mathrm{rad, eV cm^{-3}}^{-1} E_\mathrm{e,TeV}^{-1},\\
t_\mathrm{syn, yr} &\approx 1.3 \times 10^7 E_\mathrm{e,TeV}^{-1} B_\mathrm{\mu G}^{-2},
\end{align}
where $E_\mathrm{e}$ is electron energy, and $u_\mathrm{rad}$ and $T$ are the energy density and temperature of the radiation field, respectively \citep{2005MNRAS.363..954M, 2009ARA&A..47..523H}. For the cutoff-energy electrons $E_\mathrm{c} = 100 \, \mathrm{TeV}$, CMB radiation field, and $B = 1.9 \, \mathrm{\mu G}$, we get $t_\mathrm{cool} \approx 20$ kyr. The total energy released by the pulsar would be in the range of $E \approx 6 \times 10^{46} - 6 \times 10^{48} \, \mathrm{erg}$, for the spin-down power in the range of $\dot{E} \approx 10^{35} - 10^{37} \, \mathrm{erg \, s^{-1}}$. Provided that the distance to the source is $d \leq 5 \, \mathrm{kpc}$, only a small fraction of the total energy released by the pulsar during its lifetime needs to be invested in acceleration of the HE electrons, even for $\dot{E}$ on the lower end of the allowed spin-down powers. For $\dot{E} \gtrsim 10^{36} \, \mathrm{erg \, s^{-1}}$, the pulsar could power the IC emission for any possible distance of the source in the Galaxy.

We also note that there is an unidentified point-like radio source, NVSS 210803+515255 \citep{1998AJ....115.1693C} or WENSS B2106.4+5140 \citep{2000yCat.8062....0D}, well within the $95\%$ error ellipse of the source 4FGL J2108.0+5155. We did not use the radio flux points to further constrain the HE emission of the tentative pulsar as,  in general, it does not follow PL from radio to sub-GeV. 

\subsection{Hadronic scenario of emission}

The absence of an X-ray counterpart supports a hadronic emission mechanism, where UHE gamma rays are produced through inelastic collisions of hadronic relativistic CR with thermal protons followed by $\pi^0$ decay (see \citet{2014PhRvD..90l3014K}). There is an open stellar cluster, Kronberger 80, and an open cluster candidate, Kronberger 82 \citep{2006A&A...447..921K}, in the close vicinity of LHAASO J2108+5157. These open clusters may potentially act as sources of accelerated PeV protons, as young massive stars seem to contribute significantly to the Galactic CR \citep{2019NatAs...3..561A}. Interaction of the PeV protons with ambient gas leads to gamma-ray emission via $\pi^0$ decay, and therefore may be responsible for the observed emission of LHAASO J2108+5157, which is spatially coincident with two relatively dense molecular clouds. However, the $95\%$ confidence interval for the distance to Kronberger 80 derived from Gaia DR2 data is 7.9-13.7 kpc \citep{2020A&A...633A..99C}, which puts the mutual distance of the cluster and the more distant cloud ($d_1 \approx 3.1$ kpc) at more than $4.8$ kpc, excluding the possibility of protons accelerated in this stellar cluster interacting with both molecular clouds. We note that the same can be concluded for the previously reported distance of Kronberger 80, which is 4.98 kpc \citep{2016A&A...585A.101K}. On the other hand, the distance to Kronberger 82 remains unknown and one cannot exclude that this open cluster is close to one of those molecular clouds. Another source of CR protons are SNRs, which have been shown to accelerate protons emitting gamma rays of multi-TeV energies \citep{2013Sci...339..807A}, which diffuse away from their acceleration sites and interact with ISM and molecular clouds, as observed in several cases (i.e., W28 \citep{2008A&A...481..401A} or IC443 \citep{2007ApJ...664L..87A}).

\begin{figure}
        {\includegraphics[width=0.99\hsize]{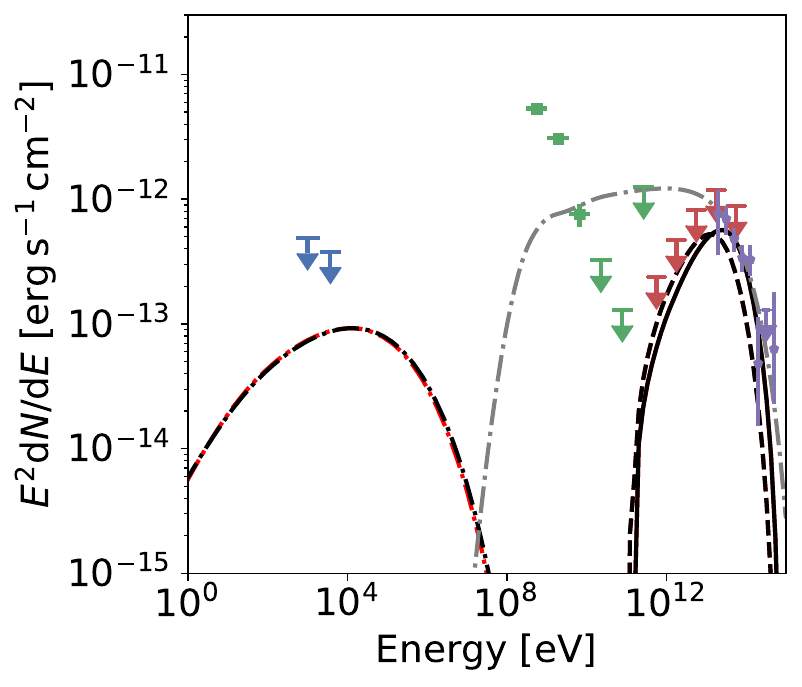}}
  \caption{Multiwavelength SED of LHAASO J2108+5157 with hadronic models of emission. Observations with different instruments are represented by datapoints of different colors: \emph{XMM-Newton} $r=16'$ (blue), \emph{Fermi}-LAT (green), LST-1 (red), and LHAASO-KM2A (purple). The best-fitting hadronic model of VHE-UHE emission (solid line) with fixed spectral index of the proton PL distribution $\alpha = 2.75$ has $\gamma_\mathrm{min} = 1.6 \times 10^5$ for both clouds. Dashed line represents the total neutrino flux for both clouds. Black and red dash-dotted lines represent the synchrotron emission of secondary particles for cloud 1 and cloud 2, respectively. The gray dash-dotted line represents the $\pi^0$ decay emission model with $\alpha = 2$ and $\gamma_\mathrm{min} = 1$ shown for reference \citep{2021ApJ...919L..22C}.}
     \label{fig_hadronic_sed}
\end{figure}

Assuming a $\pi^0$ decay-dominated origin of the UHE emission, we used the \texttt{jetset} v1.2.2\footnote{\url{https://jetset.readthedocs.io/}} \citep{Tramacere2009,Tramacere2011, 2020ascl.soft09001T} package and its implemented frequentist fitting routines (based on \texttt{iminuit}) to fit the LST-1 and LHAASO. The hadronic proton-proton (pp) model implemented in \texttt{jetset} is based on the parametrization of \cite{kelner_2006}, and takes into account the $\gamma$-ray emission  from $\pi^0$ decay, radiation (synchrotron, IC, and bremsstrahlung) from the secondaries (evolved to equilibrium) of charged pions, and neutrinos. This means that, for each step of the minimization, secondaries pairs temporarily evolve to equilibrium state, taking into account synchrotron radiation and escape timescales. In Figure~\ref{fig_hadronic_sed}, we show for reference the $\pi^0$ decay emission model with $\alpha = 2.$ and $\gamma_{\rm min}=1$ proposed by \citet{2021ApJ...919L..22C}, which we find to be ruled out by the \emph{Fermi}-LAT and LST-1 ULs, 
suggesting a very hard spectrum or a high value of the low-energy cut-off in the proton distribution responsible for the pp-induced $\gamma-$ray emission.
Based on this observational evidence, we frame our model in the phenomenological scenario proposed by \cite{celli2019}, that is, we assume  that the hard spectrum (dictated by the LST-1 upper limits) is generated by protons escaping a shock around a middle-aged SNR, and illuminating the molecular cloud. \cite{celli2019} showed that  the escaped proton distribution can be very narrow, with a low-energy cut-off of up to $\approx 10^5$ GeV.
In order to test such a scenario, we model the proton distribution with a simple PL, we leave the value of $\gamma_{\rm min}$  free during  the fit, and we freeze the value of the spectral index to $ \alpha=2.75$, which is similar to that of the galactic CR in this energy range \citep{Acero_2016}. In this simplified scenario,  $\gamma_{\rm min}$ is mimicking the energy break investigated in \cite{celli2019}, and $\alpha$ represents the index of the protons escaping from the acceleration region. This index is expected to soften with respect to the index of the  confined accelerated proton  ($\alpha_{\rm acc}$) \citep{celli2019}.  Table~\ref{tab:jetset_hadronic} shows the best-fit model parameters shown in Figure~\ref{fig_hadronic_sed} for both molecular clouds in the direction of the source, where the distances, total number densities of the target protons, and sizes of the emitting regions were fixed on the values reported in Section~\ref{sec_molecular_clouds},
In particular, we notice that the value of $\gamma_{\rm min}$  and  $\alpha$ fit well within the parameter space investigated in \cite{celli2019} for the case of $\alpha_{\rm acc} \in [2,2.3]$, and are in agreement with the expectations from standard DSA theory \citep[e.g.,][]{1978MNRAS.182..147B}.
The total required energy of all relativistic protons interacting with molecular cloud is $E_\mathrm{T,1} = 7.5 \times 10^{46} \, \mathrm{erg}$ and $E_\mathrm{T,2} = 1.5 \times 10^{46} \, \mathrm{erg}$, assuming the interaction of protons with the more distant and closer molecular cloud, respectively. This is well below the energy content of CR protons interacting with molecular clouds in the vicinity of W28 and IC 443, which is $1\%-10 \, \%$ of the total energy of a typical SN, which is  $E_\mathrm{SN} \approx 10^{51} \, \mathrm{erg}$ \citep{2013Sci...339..807A, 2018ApJ...860...69C}.

The total neutrino flux resulting from $\pi^{+/-}$ decay is comparable with the gamma-ray flux in the TeV range (see Figure~\ref{fig_hadronic_sed}), which makes this source an interesting candidate for a follow-up analysis of data from a neutrino detector in this region. However, we note that the sensitivity of current neutrino detectors is about an order of magnitude lower than the predicted neutrino flux, and only future instruments will have the potential to definitively confirm or reject the hadronic emission hypothesis \citep[e.g.,][]{2019BAAS...51g.288G}.

\begin{table}[]
    \centering
    \begin{tabular}{rccc}
    \hline
Parameter       &  \multicolumn{2}{c}{Best fit value} & Frozen \\
& Cloud 1 & Cloud 2 & \\
\hline
$n \, [\mathrm{cm^{-3}}]$ & 115 & 240 & True \\
$d \, [\mathrm{kpc}]$ & 3.1 & 2.0 & True \\
R [pc] & 7.1 & 4.5 & True \\
$\gamma_\mathrm{min}$ [$\times 10^5$] & $1.6\pm0.5$ & $1.6\pm5$ & False \\
$\gamma_\mathrm{max}$ [$\times 10^6$] & 1.0 & 1.0 &     True \\
B [mG] & $9 \pm 5$ & $\leq 8$ & False \\
$N \, [\times 10^{-15} \, \mathrm{cm^{-3}}]$ & $4 \pm 1$ & $3 \pm 1$ & False \\
$\alpha$ & $2.75$ & $2.75$ & True \\
\hline
\end{tabular}
\vspace{0.1cm}
\caption{Best-fit parameters of $\pi^0$ decay-dominated VHE-UHE emission of LHAASO J2108+5157 for both molecular clouds in the direction of the source. The injected protons are assumed to be distributed according to ECPL with $\gamma$-factor in the range ($\gamma_\mathrm{min}$, $\gamma_\mathrm{max}$), cutoff at $\gamma_\mathrm{cut}$, spectral index $\alpha$, and total numeric density $N$.}
\label{tab:jetset_hadronic}
\end{table}

The HE gamma-ray emission cannot be explained in a single-component hadronic scenario together with VHE-UHE emission. \citet{2021ApJ...919L..22C} suggested that the spectrum of the extended source 4FGL J2108.0+5155e may be associated with an old SNR, which usually features a soft spectrum above 1 GeV \citep{2016ApJS..224....8A, 2021ApJ...913L..33L}. However, our dedicated analysis of \emph{Fermi}-LAT data shows that the Gaussian extended-source assumption is not correct. Fitting the SED of 4FGL J2108.0+5155 in the \emph{Fermi}-LAT energy band above 1 GeV with a single PL provides a photon index of $\Gamma = 3.2 \pm 0.2$, which in turn tends to be too soft compared to the observations of old SNRs interacting with dense molecular clouds \citep[see][]{2012ApJ...761..133Y}. One might also consider a significant contribution of HE emission from the sea of CRs. The energy density of CRs at Galactocentric radii $>8 \, \mathrm{kpc}$ is $u_\mathrm{CR}(E>1 \, \mathrm{GeV}) \approx 0.5 \, \mathrm{eV \, cm^{-3}}$ \citep{PhysRevD.93.123007}, which is lower than observed in the Solar System. Considering the angular resolution of \emph{Fermi}-LAT at 1 GeV, which is about $0.9^\circ$, the UL on the radius of 4FGL J2108.0+5155 to still appear as a point-like source can be written as $R < d \tan{(0.9^\circ)}$. This results in rather weak limits on the proton energy density $u_\mathrm{p,1}(E>1 \, \mathrm{GeV}) > 0.14 \, \mathrm{eV \, cm^{-3}}$ and $u_\mathrm{p,2}(E>1 \, \mathrm{GeV}) > 0.10 \, \mathrm{eV \, cm^{-3}}$, for the more distant and closer molecular cloud, respectively, and a hadronic origin for the HE emission  therefore cannot be excluded.

\section{Summary and conclusions}\label{sec_conclusions}

In this work, we present a multiwavelength study of the unidentified UHE gamma-ray source LHAASO J2108+5157, which has not yet been found to be associated with any PWN, SNR, or pulsar. Dedicated observations of the source with LST-1 ---yielding 49 hours of good-quality data--- resulted in a hint ($2.2\sigma$) of hard-spectrum emission at energies between 300 GeV and 100 TeV, which can be described with a single PL with a photon index of $\Gamma = 1.62 \pm 0.23$. Our data analysis with selection cuts optimized for source detection show a possible excess ($3.7\sigma$) at energies $E > 3 \, \textrm{TeV}$. Although a confirmed detection of the VHE emission would require deeper observations, the LST-1 data provide important constraints on the source emission in the TeV range.

The VHE-UHE gamma-ray emission can be well described with IC-dominated emission of relativistic electrons with a spectral index of $\alpha = 1.5 \pm 0.4$ and a cut-off energy of $E_\mathrm{c} = 100^{+70}_{-30}$ TeV, favouring the PWN scenario. However, there is no sign of any X-ray source in 13.6 ks of dedicated observation with \emph{XMM-Newton}, which puts strong constraints on the magnetic field in the emission region, $B \lesssim 1.2 - 1.9 \, \mathrm{\mu G}$, depending on the angular extension of the X-ray-emitting region. Such a weak magnetic field is on the lower end of a typical magnetic field in PWNe, and also compatible with a magnetic field in the TeV halo around Geminga. A detailed morphological study of the region with a high-resolution instrument, such as the future completed CTA observatory, or a deeper X-ray observation would shed more light on the nature of the source and help to distinguish between the PWN and TeV-halo hypotheses. 

The lack of detection of a pulsar associated with the UHE source presents another challenge for the PWN/TeV-halo scenario. Our dedicated analysis of the 12 years of \emph{Fermi}-LAT data allowed us to precisely determine the spectral properties of the HE source 4FGL J2108.0+5155, which is spatially consistent with LHAASO J2108+5157. The spectral analysis shows that the HE emission is compatible with the spectral properties of the population of known gamma-ray pulsars, which are characterized with soft spectra. Limits on the total energy released by the tentative pulsar derived from the luminosity of the source and relativistic electron cooling time significantly exceed the total energy in relativistic electrons, and such an object would therefore be able to power the VHE-UHE emission.

The presence of two molecular clouds in the direction of LHAASO J2108+5157 supports a hadronic scenario of emission, with the UHE gamma rays being produced by the interaction of relativistic protons accelerated in a close stellar cluster or SNR with one of the two molecular clouds. In such a scenario, X-ray ULs on the synchrotron radiation of secondaries would allow significantly higher levels of magnetic field in the source compared to the leptonic scenario of emission. We performed a detailed spectral modeling under the assumption of $\pi^0$ decay-dominated emission based on the phenomenological scenario proposed by \cite{celli2019}, that is, assuming  that the hard spectrum in the LST spectral window is due to protons escaping a shock around a middle-aged SNR and illuminating the molecular cloud. The best-fit value of $\gamma_\mathrm{min} \approx 1.5\times 10^5$ agrees with one of the scenarios propose by \cite{celli2019}, which predicts a  narrow  distribution of escaped protons, with a low-energy cut-off up to $\approx 10^5$ GeV.
The proposed model reproduces the observed broad-band SED reasonably well, but a followup analysis is still needed to get a better constraint on the spectrum of the escaped proton illuminating the molecular cloud, and will be presented in a future publication. For example, a narrower proton distribution and/or a softer spectrum in the escaped protons can still account for the observed SED, with implications as to the age of the possible SNR. Nonetheless, the current analysis shows the potential of the LST in providing robust observational constraints useful for testing theoretical frameworks, and reducing the degeneracy in the parameter space.

The HE gamma-ray emission in the hadronic scenario cannot be explained by a single-component model together with VHE-UHE emission. While the old SNR scenario seems to be unlikely due to the very soft spectral index of the emission, the lower limit on the relativistic proton energy density is compatible with an interaction between the  sea of CRs and the molecular clouds.

The total neutrino flux predicted in our model is comparable with the gamma-ray flux in the TeV range, which makes LHAASO J2108+5157 an interesting candidate for future neutrino experiments  of sufficient sensitivity to either confirm or reject the hadronic scenario of emission.

\section*{Author Contribution}

M. Balbo: \emph{Fermi}-LAT data analysis, paper drafting and edition; D. Eckert: \emph{XMM-Newton} data analysis, paper drafting and edition; J. Jury\v{s}ek: project leadership, LST data analysis, paper drafting and edition, theoretical modelling, theoretical interpretation; A. Tramacere: theoretical modelling, theoretical interpretation; G. Pirola: LST analysis cross-check; R. Walter: project initiator \& leadership, theoretical interpretation, XMM ToO. The rest of the authors have contributed in one or several of the following ways: design, construction, maintenance and operation of the instrument(s) used to acquire the data; preparation and/or evaluation of the observation proposals; data acquisition, processing, calibration and/or reduction; production of analysis tools and/or related Monte Carlo simulations; discussion and approval of the contents of the draft.

\section*{Acknowledgement}

We gratefully acknowledge financial support from the following agencies and organisations:

\bigskip

Conselho Nacional de Desenvolvimento Cient\'{\i}fico e Tecnol\'{o}gico (CNPq), Funda\c{c}\~{a}o de Amparo \`{a} Pesquisa do Estado do Rio de Janeiro (FAPERJ), Funda\c{c}\~{a}o de Amparo \`{a} Pesquisa do Estado de S\~{a}o Paulo (FAPESP), Funda\c{c}\~{a}o de Apoio \`{a} Ci\^encia, Tecnologia e Inova\c{c}\~{a}o do Paran\'a - Funda\c{c}\~{a}o Arauc\'aria, Ministry of Science, Technology, Innovations and Communications (MCTIC), Brasil; Ministry of Education and Science, National RI Roadmap Project DO1-153/28.08.2018, Bulgaria; Croatian Science Foundation, Rudjer Boskovic Institute, University of Osijek, University of Rijeka, University of Split, Faculty of Electrical Engineering, Mechanical Engineering and Naval Architecture, University of Zagreb, Faculty of Electrical Engineering and Computing, Croatia; Ministry of Education, Youth and Sports, MEYS  LM2015046, LM2018105, LTT17006, EU/MEYS CZ.02.1.01/0.0/0.0/16\_013/0001403, CZ.02.1.01/0.0/0.0/18\_046/0016007 and CZ.02.1.01/0.0/0.0/16\_019/0000754, Czech Republic;  CNRS-IN2P3, France; Max Planck Society, German Bundesministerium für Bildung und Forschung (Verbundforschung / ErUM), Deutsche Forschungsgemeinschaft (SFBs 876 and 1491), Germany; Istituto Nazionale di Astrofisica (INAF), Istituto Nazionale di Fisica Nucleare (INFN), Italian Ministry for University and Research (MUR); ICRR, University of Tokyo, JSPS, MEXT, Japan; JST SPRING - JPMJSP2108; Narodowe Centrum Nauki, grant number 2019/34/E/ST9/00224, Poland; The Spanish groups acknowledge the Spanish Ministry of Science and Innovation and the Spanish Research State Agency (AEI) through the government budget lines PGE2021/28.06.000X.411.01, PGE2022/28.06.000X.411.01 and PGE2022/28.06.000X.711.04, and grants PGC2018-095512-B-I00, PID2019-104114RB-C31, PID2019-107847RB-C44, PID2019-104114RB-C32, PID2019-105510GB-C31, PID2019-104114RB-C33, PID2019-107847RB-C41, PID2019-107847RB-C43, PID2019-107988GB-C22; the “Centro de Excelencia Severo Ochoa” program through grants no. SEV-2017-0709, CEX2019-000920-S; the “Unidad de Excelencia Mar\'ia de Maeztu” program through grants no. CEX2019-000918-M, CEX2020-001058-M; the “Ram\'on y Cajal” program through grant RYC-2017-22665; the “Juan de la Cierva-Incorporaci\'on” program through grant no. IJC2019-040315-I. They also acknowledge the the “Programa Operativo” FEDER 2014-2020, Consejer\'ia de Econom\'ia y Conocimiento de la Junta de Andaluc\'ia (Ref. 1257737), PAIDI 2020 (Ref. P18-FR-1580) and Universidad de Ja\'en; “Programa Operativo de Crecimiento Inteligente” FEDER 2014-2020 (Ref.~ESFRI-2017-IAC-12), Ministerio de Ciencia e Innovaci\'on, 15\% co-financed by Consejer\'ia de Econom\'ia, Industria, Comercio y Conocimiento del Gobierno de Canarias; the “CERCA” program of the Generalitat de Catalunya; and the European Union’s “Horizon 2020” GA:824064 and NextGenerationEU; We acknowledge the Ramon y Cajal program through grant RYC-2020-028639-I; State Secretariat for Education, Research and Innovation (SERI) and Swiss National Science Foundation (SNSF), Switzerland;

The research leading to these results has received funding from the European Union's Seventh Framework Programme (FP7/2007-2013) under grant agreements No~262053 and No~317446.
This project is receiving funding from the European Union's Horizon 2020 research and innovation programs under agreement No~676134.

This research has made use of the observations obtained with \emph{XMM-Newton}, an ESA science mission with instruments and contributions directly funded by ESA Member States and NASA. This research has made use of the SIMBAD database, operated at CDS, Strasbourg, France.

\bibliographystyle{aa} 
\bibliography{lhaaso}

\begin{thebibliography}{92}
\expandafter\ifx\csname natexlab\endcsname\relax\def\natexlab#1{#1}\fi

\bibitem[{{Abdo} {et~al.}(2013){Abdo}, {Ajello}, {Allafort}, {Baldini},
  {Ballet}, {Barbiellini}, {Baring}, {Bastieri}, {Belfiore}, {Bellazzini},
  {Bhattacharyya}, {Bissaldi}, {Bloom}, {Bonamente}, {Bottacini}, {Brandt},
  {Bregeon}, {Brigida}, {Bruel}, {Buehler}, {Burgay}, {Burnett}, {Busetto},
  {Buson}, {Caliandro}, {Cameron}, {Camilo}, {Caraveo}, {Casandjian}, {Cecchi},
  {{\c{C}}elik}, {Charles}, {Chaty}, {Chaves}, {Chekhtman}, {Chen}, {Chiang},
  {Chiaro}, {Ciprini}, {Claus}, {Cognard}, {Cohen-Tanugi}, {Cominsky},
  {Conrad}, {Cutini}, {D'Ammando}, {de Angelis}, {DeCesar}, {De Luca}, {den
  Hartog}, {de Palma}, {Dermer}, {Desvignes}, {Digel}, {Di Venere}, {Drell},
  {Drlica-Wagner}, {Dubois}, {Dumora}, {Espinoza}, {Falletti}, {Favuzzi},
  {Ferrara}, {Focke}, {Franckowiak}, {Freire}, {Funk}, {Fusco}, {Gargano},
  {Gasparrini}, {Germani}, {Giglietto}, {Giommi}, {Giordano}, {Giroletti},
  {Glanzman}, {Godfrey}, {Gotthelf}, {Grenier}, {Grondin}, {Grove},
  {Guillemot}, {Guiriec}, {Hadasch}, {Hanabata}, {Harding}, {Hayashida},
  {Hays}, {Hessels}, {Hewitt}, {Hill}, {Horan}, {Hou}, {Hughes}, {Jackson},
  {Janssen}, {Jogler}, {J{\'o}hannesson}, {Johnson}, {Johnson}, {Johnson},
  {Johnson}, {Johnston}, {Kamae}, {Kataoka}, {Keith}, {Kerr}, {Kn{\"o}dlseder},
  {Kramer}, {Kuss}, {Lande}, {Larsson}, {Latronico}, {Lemoine-Goumard},
  {Longo}, {Loparco}, {Lovellette}, {Lubrano}, {Lyne}, {Manchester}, {Marelli},
  {Massaro}, {Mayer}, {Mazziotta}, {McEnery}, {McLaughlin}, {Mehault},
  {Michelson}, {Mignani}, {Mitthumsiri}, {Mizuno}, {Moiseev}, {Monzani},
  {Morselli}, {Moskalenko}, {Murgia}, {Nakamori}, {Nemmen}, {Nuss}, {Ohno},
  {Ohsugi}, {Orienti}, {Orlando}, {Ormes}, {Paneque}, {Panetta}, {Parent},
  {Perkins}, {Pesce-Rollins}, {Pierbattista}, {Piron}, {Pivato}, {Pletsch},
  {Porter}, {Possenti}, {Rain{\`o}}, {Rando}, {Ransom}, {Ray}, {Razzano},
  {Rea}, {Reimer}, {Reimer}, {Renault}, {Reposeur}, {Ritz}, {Romani}, {Roth},
  {Rousseau}, {Roy}, {Ruan}, {Sartori}, {Saz Parkinson}, {Scargle}, {Schulz},
  {Sgr{\`o}}, {Shannon}, {Siskind}, {Smith}, {Spandre}, {Spinelli}, {Stappers},
  {Strong}, {Suson}, {Takahashi}, {Thayer}, {Thayer}, {Theureau}, {Thompson},
  {Thorsett}, {Tibaldo}, {Tibolla}, {Tinivella}, {Torres}, {Tosti}, {Troja},
  {Uchiyama}, {Usher}, {Vandenbroucke}, {Vasileiou}, {Venter}, {Vianello},
  {Vitale}, {Wang}, {Weltevrede}, {Winer}, {Wolff}, {Wood}, {Wood}, {Wood}, \&
  {Yang}}]{2013ApJS..208...17A}
{Abdo}, A.~A., {Ajello}, M., {Allafort}, A., {et~al.} 2013, ApJS, 208, 17

\bibitem[{Abdo {et~al.}(2009)}]{Abdo:2009ku}
Abdo, A.~A. {et~al.} 2009, Astrophys. J. Lett., 700, L127, [Erratum:
  Astrophys.J.Lett. 703, L185 (2009), Erratum: Astrophys.J. 703, L185 (2009)]

\bibitem[{{Abdollahi} {et~al.}(2020){Abdollahi}, {Acero}, {Ackermann},
  {Ajello}, {Atwood}, {Axelsson}, {Baldini}, {Ballet}, {Barbiellini},
  {Bastieri}, {Becerra Gonzalez}, {Bellazzini}, {Berretta}, {Bissaldi},
  {Blandford}, {Bloom}, {Bonino}, {Bottacini}, {Brandt}, {Bregeon}, {Bruel},
  {Buehler}, {Burnett}, {Buson}, {Cameron}, {Caputo}, {Caraveo}, {Casandjian},
  {Castro}, {Cavazzuti}, {Charles}, {Chaty}, {Chen}, {Cheung}, {Chiaro},
  {Ciprini}, {Cohen-Tanugi}, {Cominsky}, {Coronado-Bl{\'a}zquez}, {Costantin},
  {Cuoco}, {Cutini}, {D'Ammando}, {DeKlotz}, {de la Torre Luque}, {de Palma},
  {Desai}, {Digel}, {Di Lalla}, {Di Mauro}, {Di Venere}, {Dom{\'\i}nguez},
  {Dumora}, {Fana Dirirsa}, {Fegan}, {Ferrara}, {Franckowiak}, {Fukazawa},
  {Funk}, {Fusco}, {Gargano}, {Gasparrini}, {Giglietto}, {Giommi}, {Giordano},
  {Giroletti}, {Glanzman}, {Green}, {Grenier}, {Griffin}, {Grondin}, {Grove},
  {Guiriec}, {Harding}, {Hayashi}, {Hays}, {Hewitt}, {Horan},
  {J{\'o}hannesson}, {Johnson}, {Kamae}, {Kerr}, {Kocevski}, {Kovac'evic'},
  {Kuss}, {Landriu}, {Larsson}, {Latronico}, {Lemoine-Goumard}, {Li},
  {Liodakis}, {Longo}, {Loparco}, {Lott}, {Lovellette}, {Lubrano}, {Madejski},
  {Maldera}, {Malyshev}, {Manfreda}, {Marchesini}, {Marcotulli},
  {Mart{\'\i}-Devesa}, {Martin}, {Massaro}, {Mazziotta}, {McEnery}, {Mereu},
  {Meyer}, {Michelson}, {Mirabal}, {Mizuno}, {Monzani}, {Morselli},
  {Moskalenko}, {Negro}, {Nuss}, {Ojha}, {Omodei}, {Orienti}, {Orlando},
  {Ormes}, {Palatiello}, {Paliya}, {Paneque}, {Pei}, {Pe{\~n}a-Herazo},
  {Perkins}, {Persic}, {Pesce-Rollins}, {Petrosian}, {Petrov}, {Piron}, {Poon},
  {Porter}, {Principe}, {Rain{\`o}}, {Rando}, {Razzano}, {Razzaque}, {Reimer},
  {Reimer}, {Remy}, {Reposeur}, {Romani}, {Saz Parkinson}, {Schinzel},
  {Serini}, {Sgr{\`o}}, {Siskind}, {Smith}, {Spandre}, {Spinelli}, {Strong},
  {Suson}, {Tajima}, {Takahashi}, {Tak}, {Thayer}, {Thompson}, {Tibaldo},
  {Torres}, {Torresi}, {Valverde}, {Van Klaveren}, {van Zyl}, {Wood},
  {Yassine}, \& {Zaharijas}}]{2020ApJS..247...33A}
{Abdollahi}, S., {Acero}, F., {Ackermann}, M., {et~al.} 2020, ApJS, 247, 33

\bibitem[{{Abdollahi} {et~al.}(2022){Abdollahi}, {Acero}, {Ackermann},
  {Baldini}, {Ballet}, {Barbiellini}, {Bastieri}, {Bellazzini}, {Berenji},
  {Berretta}, {Bissaldi}, {Blandford}, {Bonino}, {Bruel}, {Buson}, {Cameron},
  {Caputo}, {Caraveo}, {Castro}, {Chiaro}, {Cibrario}, {Ciprini},
  {Coronado-Bl{\'a}zquez}, {Crnogorcevic}, {Cutini}, {D'Ammando}, {De Gaetano},
  {Di Lalla}, {Dirirsa}, {Di Venere}, {Dom{\'\i}nguez}, {Fegan}, {Fiori},
  {Fleischhack}, {Franckowiak}, {Fukazawa}, {Fusco}, {Gammaldi}, {Gargano},
  {Gasparrini}, {Giacchino}, {Giglietto}, {Giordano}, {Giroletti}, {Glanzman},
  {Green}, {Grenier}, {Grondin}, {Guiriec}, {Gustafsson}, {Harding}, {Hays},
  {Hewitt}, {Horan}, {Hou}, {J{\'o}hannesson}, {Kayanoki}, {Kerr}, {Kuss},
  {Larsson}, {Latronico}, {Lemoine-Goumard}, {Li}, {Longo}, {Loparco},
  {Lubrano}, {Maldera}, {Malyshev}, {Manfreda}, {Mart{\'\i}-Devesa},
  {Mazziotta}, {Mereu}, {Michelson}, {Mirabal}, {Mitthumsiri}, {Mizuno},
  {Monzani}, {Morselli}, {Moskalenko}, {Nuss}, {Omodei}, {Orienti}, {Orlando},
  {Ormes}, {Paneque}, {Pei}, {Persic}, {Pesce-Rollins}, {Pillera}, {Poon},
  {Porter}, {Principe}, {Rain{\`o}}, {Rando}, {Rani}, {Razzano}, {Razzaque},
  {Reimer}, {Reimer}, {Reposeur}, {S{\'a}nchez-Conde}, {Saz Parkinson},
  {Scotton}, {Serini}, {Sgr{\`o}}, {Siskind}, {Spandre}, {Spinelli}, {Sueoka},
  {Suson}, {Tajima}, {Tak}, {Thayer}, {Torres}, {Troja}, {Valverde},
  {Wadiasingh}, {Wood}, \& {Zaharijas}}]{2022ApJ...933..204A}
{Abdollahi}, S., {Acero}, F., {Ackermann}, M., {et~al.} 2022, ApJ, 933, 204

\bibitem[{{Abeysekara} {et~al.}(2017){Abeysekara}, {Albert}, {Alfaro},
  {Alvarez}, {{\'A}lvarez}, {Arceo}, {Arteaga-Vel{\'a}zquez}, {Avila Rojas},
  {Ayala Solares}, {Barber}, {Bautista-Elivar}, {Becerril}, {Belmont-Moreno},
  {BenZvi}, {Berley}, {Bernal}, {Braun}, {Brisbois}, {Caballero-Mora},
  {Capistr{\'a}n}, {Carrami{\~n}ana}, {Casanova}, {Castillo}, {Cotti},
  {Cotzomi}, {Couti{\~n}o de Le{\'o}n}, {De Le{\'o}n}, {De la Fuente},
  {Dingus}, {DuVernois}, {D{\'\i}az-V{\'e}lez}, {Ellsworth}, {Engel},
  {Enr{\'\i}quez-Rivera}, {Fiorino}, {Fraija}, {Garc{\'\i}a-Gonz{\'a}lez},
  {Garfias}, {Gerhardt}, {Gonz{\'a}lez Mu{\~n}oz}, {Gonz{\'a}lez}, {Goodman},
  {Hampel-Arias}, {Harding}, {Hern{\'a}ndez}, {Hern{\'a}ndez-Almada}, {Hinton},
  {Hona}, {Hui}, {H{\"u}ntemeyer}, {Iriarte}, {Jardin-Blicq}, {Joshi},
  {Kaufmann}, {Kieda}, {Lara}, {Lauer}, {Lee}, {Lennarz}, {Vargas},
  {Linnemann}, {Longinotti}, {Luis Raya}, {Luna-Garc{\'\i}a}, {L{\'o}pez-Coto},
  {Malone}, {Marinelli}, {Martinez}, {Martinez-Castellanos},
  {Mart{\'\i}nez-Castro}, {Mart{\'\i}nez-Huerta}, {Matthews},
  {Miranda-Romagnoli}, {Moreno}, {Mostaf{\'a}}, {Nellen}, {Newbold}, {Nisa},
  {Noriega-Papaqui}, {Pelayo}, {Pretz}, {P{\'e}rez-P{\'e}rez}, {Ren}, {Rho},
  {Rivi{\`e}re}, {Rosa-Gonz{\'a}lez}, {Rosenberg}, {Ruiz-Velasco}, {Salazar},
  {Salesa Greus}, {Sandoval}, {Schneider}, {Schoorlemmer}, {Sinnis}, {Smith},
  {Springer}, {Surajbali}, {Taboada}, {Tibolla}, {Tollefson}, {Torres},
  {Ukwatta}, {Vianello}, {Weisgarber}, {Westerhoff}, {Wisher}, {Wood},
  {Yapici}, {Yodh}, {Younk}, {Zepeda}, {Zhou}, {Guo}, {Hahn}, {Li}, \&
  {Zhang}}]{2017Sci...358..911A}
{Abeysekara}, A.~U., {Albert}, A., {Alfaro}, R., {et~al.} 2017, Science, 358,
  911

\bibitem[{{Abeysekara} {et~al.}(2020){Abeysekara}, {Albert}, {Alfaro}, {Angeles
  Camacho}, {Arteaga-Vel{\'a}zquez}, {Arunbabu}, {Avila Rojas}, {Ayala
  Solares}, {Baghmanyan}, {Belmont-Moreno}, {BenZvi}, {Brisbois},
  {Caballero-Mora}, {Capistr{\'a}n}, {Carrami{\~n}ana}, {Casanova}, {Cotti},
  {Cotzomi}, {Couti{\~n}o de Le{\'o}n}, {De la Fuente}, {de Le{\'o}n},
  {Dichiara}, {Dingus}, {DuVernois}, {D{\'\i}az-V{\'e}lez}, {Ellsworth},
  {Engel}, {Espinoza}, {Fleischhack}, {Fraija}, {Galv{\'a}n-G{\'a}mez},
  {Garcia}, {Garc{\'\i}a-Gonz{\'a}lez}, {Garfias}, {Gonz{\'a}lez}, {Goodman},
  {Harding}, {Hernandez}, {Hinton}, {Hona}, {Huang}, {Hueyotl-Zahuantitla},
  {H{\"u}ntemeyer}, {Iriarte}, {Jardin-Blicq}, {Joshi}, {Kaufmann}, {Kieda},
  {Lara}, {Lee}, {Le{\'o}n Vargas}, {Linnemann}, {Longinotti}, {Luis-Raya},
  {Lundeen}, {L{\'o}pez-Coto}, {Malone}, {Marinelli}, {Martinez},
  {Martinez-Castellanos}, {Mart{\'\i}nez-Castro}, {Mart{\'\i}nez-Huerta},
  {Matthews}, {Miranda-Romagnoli}, {Morales-Soto}, {Moreno}, {Mostaf{\'a}},
  {Nayerhoda}, {Nellen}, {Newbold}, {Nisa}, {Noriega-Papaqui}, {Peisker},
  {P{\'e}rez-P{\'e}rez}, {Pretz}, {Ren}, {Rho}, {Rivi{\`e}re},
  {Rosa-Gonz{\'a}lez}, {Rosenberg}, {Ruiz-Velasco}, {Salesa Greus}, {Sandoval},
  {Schneider}, {Schoorlemmer}, {Sinnis}, {Smith}, {Springer}, {Surajbali},
  {Tabachnick}, {Tanner}, {Tibolla}, {Tollefson}, {Torres}, {Torres-Escobedo},
  {Villase{\~n}or}, {Weisgarber}, {Wood}, {Yapici}, {Zhang}, {Zhou}, \& {HAWC
  Collaboration}}]{2020PhRvL.124b1102A}
{Abeysekara}, A.~U., {Albert}, A., {Alfaro}, R., {et~al.} 2020, PhRvL, 124,
  021102

\bibitem[{{Acero} {et~al.}(2015){Acero}, {Ackermann}, {Ajello}, {Albert},
  {Atwood}, {Axelsson}, {Baldini}, {Ballet}, {Barbiellini}, {Bastieri},
  {Belfiore}, {Bellazzini}, {Bissaldi}, {Blandford}, {Bloom}, {Bogart},
  {Bonino}, {Bottacini}, {Bregeon}, {Britto}, {Bruel}, {Buehler}, {Burnett},
  {Buson}, {Caliandro}, {Cameron}, {Caputo}, {Caragiulo}, {Caraveo},
  {Casandjian}, {Cavazzuti}, {Charles}, {Chaves}, {Chekhtman}, {Cheung},
  {Chiang}, {Chiaro}, {Ciprini}, {Claus}, {Cohen-Tanugi}, {Cominsky}, {Conrad},
  {Cutini}, {D'Ammando}, {de Angelis}, {DeKlotz}, {de Palma}, {Desiante},
  {Digel}, {Di Venere}, {Drell}, {Dubois}, {Dumora}, {Favuzzi}, {Fegan},
  {Ferrara}, {Finke}, {Franckowiak}, {Fukazawa}, {Funk}, {Fusco}, {Gargano},
  {Gasparrini}, {Giebels}, {Giglietto}, {Giommi}, {Giordano}, {Giroletti},
  {Glanzman}, {Godfrey}, {Grenier}, {Grondin}, {Grove}, {Guillemot}, {Guiriec},
  {Hadasch}, {Harding}, {Hays}, {Hewitt}, {Hill}, {Horan}, {Iafrate}, {Jogler},
  {J{\'o}hannesson}, {Johnson}, {Johnson}, {Johnson}, {Johnson}, {Kamae},
  {Kataoka}, {Katsuta}, {Kuss}, {La Mura}, {Landriu}, {Larsson}, {Latronico},
  {Lemoine-Goumard}, {Li}, {Li}, {Longo}, {Loparco}, {Lott}, {Lovellette},
  {Lubrano}, {Madejski}, {Massaro}, {Mayer}, {Mazziotta}, {McEnery},
  {Michelson}, {Mirabal}, {Mizuno}, {Moiseev}, {Mongelli}, {Monzani},
  {Morselli}, {Moskalenko}, {Murgia}, {Nuss}, {Ohno}, {Ohsugi}, {Omodei},
  {Orienti}, {Orlando}, {Ormes}, {Paneque}, {Panetta}, {Perkins},
  {Pesce-Rollins}, {Piron}, {Pivato}, {Porter}, {Racusin}, {Rando}, {Razzano},
  {Razzaque}, {Reimer}, {Reimer}, {Reposeur}, {Rochester}, {Romani},
  {Salvetti}, {S{\'a}nchez-Conde}, {Saz Parkinson}, {Schulz}, {Siskind},
  {Smith}, {Spada}, {Spandre}, {Spinelli}, {Stephens}, {Strong}, {Suson},
  {Takahashi}, {Takahashi}, {Tanaka}, {Thayer}, {Thayer}, {Thompson},
  {Tibaldo}, {Tibolla}, {Torres}, {Torresi}, {Tosti}, {Troja}, {Van Klaveren},
  {Vianello}, {Winer}, {Wood}, {Wood}, {Zimmer}, \& {Fermi-LAT
  Collaboration}}]{2015ApJS..218...23A}
{Acero}, F., {Ackermann}, M., {Ajello}, M., {et~al.} 2015, ApJS, 218, 23

\bibitem[{Acero {et~al.}(2016)Acero, Ackermann, Ajello, Albert, Baldini,
  Ballet, Barbiellini, Bastieri, Bellazzini, Bissaldi, Bloom, Bonino,
  Bottacini, Brandt, Bregeon, Bruel, Buehler, Buson, Caliandro, Cameron,
  Caragiulo, Caraveo, Casandjian, Cavazzuti, Cecchi, Charles, Chekhtman,
  Chiang, Chiaro, Ciprini, Claus, Cohen-Tanugi, Conrad, Cuoco, Cutini,
  D'Ammando, de~Angelis, de~Palma, Desiante, Digel, Venere, Drell, Favuzzi,
  Fegan, Ferrara, Focke, Franckowiak, Funk, Fusco, Gargano, Gasparrini,
  Giglietto, Giordano, Giroletti, Glanzman, Godfrey, Grenier, Guiriec, Hadasch,
  Harding, Hayashi, Hays, Hewitt, Hill, Horan, Hou, Jogler, J{\'{o}}hannesson,
  Kamae, Kuss, Landriu, Larsson, Latronico, Li, Li, Longo, Loparco, Lovellette,
  Lubrano, Maldera, Malyshev, Manfreda, Martin, Mayer, Mazziotta, McEnery,
  Michelson, Mirabal, Mizuno, Monzani, Morselli, Nuss, Ohsugi, Omodei, Orienti,
  Orlando, Ormes, Paneque, Pesce-Rollins, Piron, Pivato, Rain{\`{o}}, Rando,
  Razzano, Razzaque, Reimer, Reimer, Remy, Renault, S{\'{a}}nchez-Conde,
  Schaal, Schulz, Sgr{\`{o}}, Siskind, Spada, Spandre, Spinelli, Strong, Suson,
  Tajima, Takahashi, Thayer, Thompson, Tibaldo, Tinivella, Torres, Tosti,
  Troja, Vianello, Werner, Wood, Wood, Zaharijas, \& Zimmer}]{Acero_2016}
Acero, F., Ackermann, M., Ajello, M., {et~al.} 2016, The Astrophysical Journal
  Supplement Series, 223, 26

\bibitem[{{Acero} {et~al.}(2016){Acero}, {Ackermann}, {Ajello}, {Baldini},
  {Ballet}, {Barbiellini}, {Bastieri}, {Bellazzini}, {Bissaldi}, {Blandford},
  {Bloom}, {Bonino}, {Bottacini}, {Brandt}, {Bregeon}, {Bruel}, {Buehler},
  {Buson}, {Caliandro}, {Cameron}, {Caputo}, {Caragiulo}, {Caraveo},
  {Casandjian}, {Cavazzuti}, {Cecchi}, {Chekhtman}, {Chiang}, {Chiaro},
  {Ciprini}, {Claus}, {Cohen}, {Cohen-Tanugi}, {Cominsky}, {Condon}, {Conrad},
  {Cutini}, {D'Ammando}, {de Angelis}, {de Palma}, {Desiante}, {Digel}, {Di
  Venere}, {Drell}, {Drlica-Wagner}, {Favuzzi}, {Ferrara}, {Franckowiak},
  {Fukazawa}, {Funk}, {Fusco}, {Gargano}, {Gasparrini}, {Giglietto}, {Giommi},
  {Giordano}, {Giroletti}, {Glanzman}, {Godfrey}, {Gomez-Vargas}, {Grenier},
  {Grondin}, {Guillemot}, {Guiriec}, {Gustafsson}, {Hadasch}, {Harding},
  {Hayashida}, {Hays}, {Hewitt}, {Hill}, {Horan}, {Hou}, {Iafrate}, {Jogler},
  {J{\'o}hannesson}, {Johnson}, {Kamae}, {Katagiri}, {Kataoka}, {Katsuta},
  {Kerr}, {Kn{\"o}dlseder}, {Kocevski}, {Kuss}, {Laffon}, {Lande}, {Larsson},
  {Latronico}, {Lemoine-Goumard}, {Li}, {Li}, {Longo}, {Loparco}, {Lovellette},
  {Lubrano}, {Magill}, {Maldera}, {Marelli}, {Mayer}, {Mazziotta}, {Michelson},
  {Mitthumsiri}, {Mizuno}, {Moiseev}, {Monzani}, {Moretti}, {Morselli},
  {Moskalenko}, {Murgia}, {Nemmen}, {Nuss}, {Ohsugi}, {Omodei}, {Orienti},
  {Orlando}, {Ormes}, {Paneque}, {Perkins}, {Pesce-Rollins}, {Petrosian},
  {Piron}, {Pivato}, {Porter}, {Rain{\`o}}, {Rando}, {Razzano}, {Razzaque},
  {Reimer}, {Reimer}, {Renaud}, {Reposeur}, {Rousseau}, {Saz Parkinson},
  {Schmid}, {Schulz}, {Sgr{\`o}}, {Siskind}, {Spada}, {Spandre}, {Spinelli},
  {Strong}, {Suson}, {Tajima}, {Takahashi}, {Tanaka}, {Thayer}, {Thompson},
  {Tibaldo}, {Tibolla}, {Torres}, {Tosti}, {Troja}, {Uchiyama}, {Vianello},
  {Wells}, {Wood}, {Wood}, {Yassine}, {den Hartog}, \&
  {Zimmer}}]{2016ApJS..224....8A}
{Acero}, F., {Ackermann}, M., {Ajello}, M., {et~al.} 2016, ApJS, 224, 8

\bibitem[{{Ackermann} {et~al.}(2013){Ackermann}, {Ajello}, {Allafort},
  {Baldini}, {Ballet}, {Barbiellini}, {Baring}, {Bastieri}, {Bechtol},
  {Bellazzini}, {Blandford}, {Bloom}, {Bonamente}, {Borgland}, {Bottacini},
  {Brandt}, {Bregeon}, {Brigida}, {Bruel}, {Buehler}, {Busetto}, {Buson},
  {Caliandro}, {Cameron}, {Caraveo}, {Casandjian}, {Cecchi}, {{\c{C}}elik},
  {Charles}, {Chaty}, {Chaves}, {Chekhtman}, {Cheung}, {Chiang}, {Chiaro},
  {Cillis}, {Ciprini}, {Claus}, {Cohen-Tanugi}, {Cominsky}, {Conrad}, {Corbel},
  {Cutini}, {D'Ammando}, {de Angelis}, {de Palma}, {Dermer}, {do Couto e
  Silva}, {Drell}, {Drlica-Wagner}, {Falletti}, {Favuzzi}, {Ferrara},
  {Franckowiak}, {Fukazawa}, {Funk}, {Fusco}, {Gargano}, {Germani},
  {Giglietto}, {Giommi}, {Giordano}, {Giroletti}, {Glanzman}, {Godfrey},
  {Grenier}, {Grondin}, {Grove}, {Guiriec}, {Hadasch}, {Hanabata}, {Harding},
  {Hayashida}, {Hayashi}, {Hays}, {Hewitt}, {Hill}, {Hughes}, {Jackson},
  {Jogler}, {J{\'o}hannesson}, {Johnson}, {Kamae}, {Kataoka}, {Katsuta},
  {Kn{\"o}dlseder}, {Kuss}, {Lande}, {Larsson}, {Latronico}, {Lemoine-Goumard},
  {Longo}, {Loparco}, {Lovellette}, {Lubrano}, {Madejski}, {Massaro}, {Mayer},
  {Mazziotta}, {McEnery}, {Mehault}, {Michelson}, {Mignani}, {Mitthumsiri},
  {Mizuno}, {Moiseev}, {Monzani}, {Morselli}, {Moskalenko}, {Murgia},
  {Nakamori}, {Nemmen}, {Nuss}, {Ohno}, {Ohsugi}, {Omodei}, {Orienti},
  {Orlando}, {Ormes}, {Paneque}, {Perkins}, {Pesce-Rollins}, {Piron}, {Pivato},
  {Rain{\`o}}, {Rando}, {Razzano}, {Razzaque}, {Reimer}, {Reimer}, {Ritz},
  {Romoli}, {S{\'a}nchez-Conde}, {Schulz}, {Sgr{\`o}}, {Simeon}, {Siskind},
  {Smith}, {Spandre}, {Spinelli}, {Stecker}, {Strong}, {Suson}, {Tajima},
  {Takahashi}, {Takahashi}, {Tanaka}, {Thayer}, {Thayer}, {Thompson},
  {Thorsett}, {Tibaldo}, {Tibolla}, {Tinivella}, {Troja}, {Uchiyama}, {Usher},
  {Vandenbroucke}, {Vasileiou}, {Vianello}, {Vitale}, {Waite}, {Werner},
  {Winer}, {Wood}, {Wood}, {Yamazaki}, {Yang}, \&
  {Zimmer}}]{2013Sci...339..807A}
{Ackermann}, M., {Ajello}, M., {Allafort}, A., {et~al.} 2013, Science, 339, 807

\bibitem[{{Aharonian} {et~al.}(2008){Aharonian}, {Akhperjanian}, {Bazer-Bachi},
  {Behera}, {Beilicke}, {Benbow}, {Berge}, {Bernl{\"o}hr}, {Boisson}, {Bolz},
  {Borrel}, {Braun}, {Brion}, {Brown}, {B{\"u}hler}, {Bulik}, {B{\"u}sching},
  {Boutelier}, {Carrigan}, {Chadwick}, {Chounet}, {Clapson}, {Coignet},
  {Cornils}, {Costamante}, {Degrange}, {Dickinson}, {Djannati-Ata{\"\i}},
  {Domainko}, {O'C. Drury}, {Dubus}, {Dyks}, {Egberts}, {Emmanoulopoulos},
  {Espigat}, {Farnier}, {Feinstein}, {Fiasson}, {F{\"o}rster}, {Fontaine},
  {Fukui}, {Funk}, {Funk}, {F{\"u}{\ss}ling}, {Gallant}, {Giebels},
  {Glicenstein}, {Gl{\"u}ck}, {Goret}, {Hadjichristidis}, {Hauser}, {Hauser},
  {Heinzelmann}, {Henri}, {Hermann}, {Hinton}, {Hoffmann}, {Hofmann},
  {Holleran}, {Hoppe}, {Horns}, {Jacholkowska}, {de Jager}, {Kendziorra},
  {Kerschhaggl}, {Kh{\'e}lifi}, {Komin}, {Kosack}, {Lamanna}, {Latham}, {Le
  Gallou}, {Lemi{\`e}re}, {Lemoine-Goumard}, {Lenain}, {Lohse}, {Martin},
  {Martineau-Huynh}, {Marcowith}, {Masterson}, {Maurin}, {McComb}, {Moderski},
  {Moriguchi}, {Moulin}, {de Naurois}, {Nedbal}, {Nolan}, {Olive}, {Orford},
  {Osborne}, {Ostrowski}, {Panter}, {Pedaletti}, {Pelletier}, {Petrucci},
  {Pita}, {P{\"u}hlhofer}, {Punch}, {Ranchon}, {Raubenheimer}, {Raue},
  {Rayner}, {Reimer}, {Renaud}, {Ripken}, {Rob}, {Rolland}, {Rosier-Lees},
  {Rowell}, {Rudak}, {Ruppel}, {Sahakian}, {Santangelo}, {Saug{\'e}},
  {Schlenker}, {Schlickeiser}, {Schr{\"o}der}, {Schwanke}, {Schwarzburg},
  {Schwemmer}, {Shalchi}, {Sol}, {Spangler}, {Stawarz}, {Steenkamp},
  {Stegmann}, {Superina}, {Takeuchi}, {Tam}, {Tavernet}, {Terrier}, {van
  Eldik}, {Vasileiadis}, {Venter}, {Vialle}, {Vincent}, {Vivier}, {V{\"o}lk},
  {Volpe}, {Wagner}, \& {Ward}}]{2008A&A...481..401A}
{Aharonian}, F., {Akhperjanian}, A.~G., {Bazer-Bachi}, A.~R., {et~al.} 2008,
  A\&A, 481, 401

\bibitem[{{Aharonian} {et~al.}(2019){Aharonian}, {Yang}, \& {de O{\~n}a
  Wilhelmi}}]{2019NatAs...3..561A}
{Aharonian}, F., {Yang}, R., \& {de O{\~n}a Wilhelmi}, E. 2019, Nature
  Astronomy, 3, 561

\bibitem[{{Aharonian} {et~al.}(2010){Aharonian}, {Kelner}, \&
  {Prosekin}}]{2010PhRvD..82d3002A}
{Aharonian}, F.~A., {Kelner}, S.~R., \& {Prosekin}, A.~Y. 2010, PRD, 82, 043002

\bibitem[{{Albert} {et~al.}(2021){Albert}, {Alfaro}, {Alvarez}, {{\'A}lvarez},
  {Angeles Camacho}, {Arteaga-Vel{\'a}zquez}, {Arunbabu}, {Avila Rojas}, {Ayala
  Solares}, {Baghmanyan}, {Belmont-Moreno}, {BenZvi}, {Brisbois},
  {Caballero-Mora}, {Capistr{\'a}n}, {Carrami{\~n}ana}, {Casanova}, {Cotti},
  {Cotzomi}, {Couti{\~n}o de Le{\'o}n}, {De la Fuente}, {de Le{\'o}n}, {Diaz
  Hernandez}, {Dingus}, {DuVernois}, {Durocher}, {D{\'\i}az-V{\'e}lez},
  {Ellsworth}, {Engel}, {Espinoza}, {Fan}, {Fern{\'a}ndez Alonso}, {Fraija},
  {Galv{\'a}n-G{\'a}mez}, {Garc{\'\i}a-Gonz{\'a}lez}, {Garfias}, {Giacinti},
  {Gonz{\'a}lez}, {Goodman}, {Harding}, {Hernandez}, {Hona}, {Huang},
  {Hueyotl-Zahuantitla}, {H{\"u}ntemeyer}, {Iriarte}, {Jardin-Blicq}, {Joshi},
  {Kieda}, {Lara}, {Lee}, {Lee}, {Le{\'o}n Vargas}, {Linnemann}, {Longinotti},
  {Luis-Raya}, {Lundeen}, {Malone}, {Marandon}, {Martinez},
  {Mart{\'\i}nez-Castro}, {Matthews}, {Miranda-Romagnoli}, {Morales-Soto},
  {Moreno}, {Mostaf{\'a}}, {Nayerhoda}, {Nellen}, {Newbold}, {Nisa},
  {Noriega-Papaqui}, {Olivera-Nieto}, {Omodei}, {Peisker}, {P{\'e}rez Araujo},
  {P{\'e}rez-P{\'e}rez}, {Rho}, {Roh}, {Rosa-Gonz{\'a}lez}, {Ruiz-Velasco},
  {Salazar}, {Salesa Greus}, {Sandoval}, {Schneider}, {Schoorlemmer},
  {Serna-Franco}, {Smith}, {Springer}, {Surajbali}, {Tanner}, {Tollefson},
  {Torres}, {Torres-Escobedo}, {Turner}, {Ure{\~n}a-Mena}, {Villase{\~n}or},
  {Weisgarber}, {Willox}, {Zhou}, \& {HAWC
  Collaboration}}]{2021ApJ...911L..27A}
{Albert}, A., {Alfaro}, R., {Alvarez}, C., {et~al.} 2021, ApJL, 911, L27

\bibitem[{{Albert} {et~al.}(2007){Albert}, {Aliu}, {Anderhub}, {Antoranz},
  {Armada}, {Baixeras}, {Barrio}, {Bartko}, {Bastieri}, {Becker}, {Bednarek},
  {Berger}, {Bigongiari}, {Biland}, {Bock}, {Bordas}, {Bosch-Ramon}, {Bretz},
  {Britvitch}, {Camara}, {Carmona}, {Chilingarian}, {Coarasa}, {Commichau},
  {Contreras}, {Cortina}, {Costado}, {Curtef}, {Danielyan}, {Dazzi}, {De
  Angelis}, {Delgado}, {de los Reyes}, {De Lotto}, {Domingo-Santamar{\'\i}a},
  {Dorner}, {Doro}, {Errando}, {Fagiolini}, {Ferenc}, {Fern{\'a}ndez}, {Firpo},
  {Flix}, {Fonseca}, {Font}, {Fuchs}, {Galante}, {Garc{\'\i}a-L{\'o}pez},
  {Garczarczyk}, {Gaug}, {Giller}, {Goebel}, {Hakobyan}, {Hayashida},
  {Hengstebeck}, {Herrero}, {H{\"o}hne}, {Hose}, {Hsu}, {Jacon}, {Jogler},
  {Kosyra}, {Kranich}, {Kritzer}, {Laille}, {Lindfors}, {Lombardi}, {Longo},
  {L{\'o}pez}, {L{\'o}pez}, {Lorenz}, {Majumdar}, {Maneva}, {Mannheim},
  {Mansutti}, {Mariotti}, {Mart{\'\i}nez}, {Mazin}, {Merck}, {Meucci}, {Meyer},
  {Miranda}, {Mirzoyan}, {Mizobuchi}, {Moralejo}, {Nieto}, {Nilsson},
  {Ninkovic}, {O{\~n}a-Wilhelmi}, {Otte}, {Oya}, {Paneque}, {Panniello},
  {Paoletti}, {Paredes}, {Pasanen}, {Pascoli}, {Pauss}, {Pegna}, {Persic},
  {Peruzzo}, {Piccioli}, {Prandini}, {Puchades}, {Raymers}, {Rhode},
  {Rib{\'o}}, {Rico}, {Rissi}, {Robert}, {R{\"u}gamer}, {Saggion}, {Saito},
  {S{\'a}nchez}, {Sartori}, {Scalzotto}, {Scapin}, {Schmitt}, {Schweizer},
  {Shayduk}, {Shinozaki}, {Shore}, {Sidro}, {Sillanp{\"a}{\"a}}, {Sobczynska},
  {Stamerra}, {Stark}, {Takalo}, {Temnikov}, {Tescaro}, {Teshima}, {Torres},
  {Turini}, {Vankov}, {Vitale}, {Wagner}, {Wibig}, {Wittek}, {Zandanel},
  {Zanin}, \& {Zapatero}}]{2007ApJ...664L..87A}
{Albert}, J., {Aliu}, E., {Anderhub}, H., {et~al.} 2007, ApJL, 664, L87

\bibitem[{{Alispach} {et~al.}(2022){Alispach}, {CTA-LST Project}, {Abe},
  {Aguasca}, {Agudo}, {Antonelli}, {Aramo}, {Armstrong}, {Artero}, {Asano},
  {Ashkar}, {Aubert}, {Baktash}, {Bamba}, {Baquero Larriva}, {Baroncelli},
  {Barres de Almeida}, {Barrio}, {Batkovi{\'c}}, {Becerra Gonzalez},
  {Bernardos}, {Berti}, {Biederbeck}, {Bigongiari}, {Blanch}, {Bonnoli},
  {Bordas}, {Bose}, {Bulgarelli}, {Burelli}, {Buscemi}, {Cardillo}, {Caroff},
  {Carosi}, {Cassol}, {Cerruti}, {Chai}, {Cheng}, {Chikawa}, {Chytka},
  {Contreras}, {Cortina}, {Costantini}, {Dalchenko}, {De Angelis}, {de Bony de
  Lavergne}, {Deleglise}, {Delgado}, {Delgado Mengual}, {Della Volpe},
  {Depaoli}, {Di Pierro}, {Di Venere}, {D{\'\i}az}, {Dominik}, {Dominis
  Prester}, {Donini}, {Dorner}, {Doro}, {Els{\"a}sser}, {Emery}, {Escudero},
  {Fiasson}, {Foffano}, {Fonseca}, {Freixas Coromina}, {Fukami}, {Fukazawa},
  {Garcia}, {Garcia L{\'o}pez}, {Giglietto}, {Giordano}, {Gliwny}, {Godinovic},
  {Green}, {Grespan}, {Gunji}, {Hackfeld}, {Hadasch}, {Hahn}, {Hassan},
  {Hayashi}, {Heckmann}, {Heller}, {Herrera Llorente}, {Hirotani}, {Hoffmann},
  {Horns}, {Houles}, {Hrabovsky}, {Hrupec}, {Hui}, {H{\"u}tten}, {Inada},
  {Inome}, {Iori}, {Ishio}, {Iwamura}, {Jacquemont}, {Jim{\'e}nez
  Mart{\'\i}nez}, {Jouvin}, {Jury{\v{s}}ek}, {Kagaya}, {Karas}, {Katagiri},
  {Kataoka}, {Kerszberg}, {Kobayashi}, {Kong}, {Kubo}, {Kushida}, {Lamanna},
  {Lamastra}, {Le Flour}, {Longo}, {Lopez-Coto}, {L{\'o}pez-Moya},
  {L{\'o}pez-Oramas}, {Luque-Escamilla}, {Majumdar}, {Makariev}, {Mandat},
  {Manganaro}, {Mannheim}, {Mariotti}, {Marquez}, {Marsella}, {Mart{\'\i}},
  {Martinez}, {Mart{\'\i}nez}, {Martinez}, {Marusevec}, {Mas}, {Maurin},
  {Mazin}, {Mestre Guillen}, {Mi{\'c}anovi{\'c}}, {Miceli}, {Miener},
  {Miranda}, {Miranda}, {Mirzoyan}, {Mizuno}, {Molina}, {Montaruli},
  {Monteiro}, {Moralejo}, {Morcuende}, {Moretti}, {Morselli}, {Mrakovcic},
  {Murase}, {Nagai}, {Nakamori}, {Nickel}, {Nieto}, {Nievas}, {Nishijima},
  {Noda}, {Nosek}, {N{\"o}the}, {Nozaki}, {Ohishi}, {Ohtani}, {Oka}, {Okazaki},
  {Okumura}, {Orito}, {Otero-Santos}, {Palatiello}, {Paneque}, {Paoletti},
  {Paredes}, {Pavleti{\'c}}, {Pech}, {Pecimotika}, {Poireau}, {Polo},
  {Prandini}, {Prast}, {Priyadarshi}, {Prouza}, {Rando}, {Rhode}, {Rib{\'o}},
  {Rizi}, {Rugliancich}, {Ruiz}, {Saito}, {Sakurai}, {Sanchez},
  {{\v{S}}ari{\'c}}, {Saturni}, {Scherpenberg}, {Schleicher}, {Schubert},
  {Sch{\"u}ssler}, {Schweizer}, {Seglar Arroyo}, {Shellard}, {Sitarek},
  {Sliusar}, {Spolon}, {Stri{\v{s}}kovi{\'c}}, {Strzys}, {Suda}, {Sunada},
  {Tajima}, {Takahashi}, {Takahashi}, {Takata}, {Takeishi}, {Tam}, {Tanaka},
  {Tateishi}, {Tejedor}, {Temnikov}, {Terada}, {Terzic}, {Teshima},
  {Tluczykont}, {Tokanai}, {Torres}, {Travnicek}, {Truzzi}, {Vacula}, {Vazquez
  Acosta}, {Verguilov}, {Verna}, {Viale}, {Vigorito}, {Vitale}, {Vovk},
  {Vuillaume}, {Walter}, {Will}, {Yamamoto}, {Yamazaki}, {Yoshida},
  {Yoshikoshi}, \& {Zari{\'c}}}]{2022icrc.confE.716A}
{Alispach}, C.~M., {CTA-LST Project}, T., {Abe}, H., {et~al.} 2022, in 37th
  International Cosmic Ray Conference, 716

\bibitem[{{Ambrogi} {et~al.}(2019){Ambrogi}, {Zanin}, {Casanova}, {De O{\~n}a
  Wilhelmi}, {Peron}, \& {Aharonian}}]{2019A&A...623A..86A}
{Ambrogi}, L., {Zanin}, R., {Casanova}, S., {et~al.} 2019, A\&A, 623, A86

\bibitem[{{Amenomori} {et~al.}(2019){Amenomori}, {Bao}, {Bi}, {Chen}, {Chen},
  {Chen}, {Chen}, {Chen}, {Cirennima}, {Cui}, {Danzengluobu}, {Ding}, {Fang},
  {Fang}, {Feng}, {Feng}, {Feng}, {Gao}, {Gou}, {Guo}, {He}, {He}, {Hibino},
  {Hotta}, {Hu}, {Hu}, {Huang}, {Jia}, {Jiang}, {Jin}, {Kajino}, {Kasahara},
  {Katayose}, {Kato}, {Kato}, {Kawata}, {Kozai}, {Labaciren}, {Le}, {Li}, {Li},
  {Li}, {Lin}, {Liu}, {Liu}, {Liu}, {Liu}, {Lou}, {Lu}, {Meng}, {Mitsui},
  {Munakata}, {Nakamura}, {Nanjo}, {Nishizawa}, {Ohnishi}, {Ohta}, {Ozawa},
  {Qian}, {Qu}, {Saito}, {Sakata}, {Sako}, {Sengoku}, {Shao}, {Shibata},
  {Shiomi}, {Sugimoto}, {Takita}, {Tan}, {Tateyama}, {Torii}, {Tsuchiya},
  {Udo}, {Wang}, {Wu}, {Xue}, {Yagisawa}, {Yamamoto}, {Yang}, {Yuan}, {Zhai},
  {Zhang}, {Zhang}, {Zhang}, {Zhang}, {Zhang}, {Zhang}, {Zhang},
  {Zhaxisangzhu}, {Zhou}, \& {Tibet AS {\ensuremath{\gamma}}
  Collaboration}}]{2019PhRvL.123e1101A}
{Amenomori}, M., {Bao}, Y.~W., {Bi}, X.~J., {et~al.} 2019, PhRvL, 123, 051101

\bibitem[{{Ballet} {et~al.}(2020){Ballet}, {Burnett}, {Digel}, \&
  {Lott}}]{2020arXiv200511208B}
{Ballet}, J., {Burnett}, T.~H., {Digel}, S.~W., \& {Lott}, B. 2020, arXiv
  e-prints, arXiv:2005.11208

\bibitem[{Bamba {et~al.}(2010)Bamba, Anada, Dotani, Mori, Yamazaki, Ebisawa, \&
  Vink}]{Bamba_2010}
Bamba, A., Anada, T., Dotani, T., {et~al.} 2010, The Astrophysical Journal,
  719, L116

\bibitem[{{Bell}(1978)}]{1978MNRAS.182..147B}
{Bell}, A.~R. 1978, MNRAS, 182, 147

\bibitem[{{Bell}(2013)}]{2013APh....43...56B}
{Bell}, A.~R. 2013, Astroparticle Physics, 43, 56

\bibitem[{{Berge} {et~al.}(2007){Berge}, {Funk}, \&
  {Hinton}}]{2007A&A...466.1219B}
{Berge}, D., {Funk}, S., \& {Hinton}, J. 2007, A\&A, 466, 1219

\bibitem[{{Bolatto} {et~al.}(2013){Bolatto}, {Wolfire}, \&
  {Leroy}}]{2013ARA&A..51..207B}
{Bolatto}, A.~D., {Wolfire}, M., \& {Leroy}, A.~K. 2013, ARAA, 51, 207

\bibitem[{{Breuhaus} {et~al.}(2021){Breuhaus}, {Hahn}, {Romoli}, {Reville},
  {Giacinti}, {Tuffs}, \& {Hinton}}]{2021ApJ...908L..49B}
{Breuhaus}, M., {Hahn}, J., {Romoli}, C., {et~al.} 2021, ApJL, 908, L49

\bibitem[{{Breuhaus} {et~al.}(2022){Breuhaus}, {Reville}, \&
  {Hinton}}]{2022A&A...660A...8B}
{Breuhaus}, M., {Reville}, B., \& {Hinton}, J.~A. 2022, A\&A, 660, A8

\bibitem[{{Cantat-Gaudin} \& {Anders}(2020)}]{2020A&A...633A..99C}
{Cantat-Gaudin}, T. \& {Anders}, F. 2020, A\&A, 633, A99

\bibitem[{{Cao} {et~al.}(2021{\natexlab{a}}){Cao}, {Aharonian}, {An},
  {Axikegu}, {Bai}, {Bai}, {Bao}, {Bastieri}, {Bi}, {Bi}, {Cai}, {Cai}, {Cao},
  {Chang}, {Chang}, {Chen}, {Chen}, {Chen}, {Chen}, {Chen}, {Chen}, {Chen},
  {Chen}, {Chen}, {Chen}, {Chen}, {Chen}, {Chen}, {Cheng}, {Cheng}, {Cui},
  {Cui}, {Cui}, {Piazzoli}, {Dai}, {Dai}, {Dai}, {Dan-Zeng-Luo-Bu}, {Volpe},
  {Dong}, {Duan}, {Fan}, {Fan}, {Fan}, {Fang}, {Fang}, {Feng}, {Feng}, {Feng},
  {Feng}, {Gao}, {Gao}, {Gao}, {Gao}, {Gao}, {Ge}, {Geng}, {Gong}, {Gou}, {Gu},
  {Guo}, {Guo}, {Guo}, {Guo}, {Guo}, {Han}, {He}, {He}, {He}, {He}, {He}, {He},
  {Heller}, {Hor}, {Hou}, {Hu}, {Hu}, {Hu}, {Hu}, {Huang}, {Huang}, {Huang},
  {Huang}, {Huang}, {Huang}, {Ji}, {Ji}, {Jia}, {Jiang}, {Jiang}, {Jin}, {Ke},
  {Kuleshov}, {Levochkin}, {Li}, {Li}, {Li}, {Li}, {Li}, {Li}, {Li}, {Li},
  {Li}, {Li}, {Li}, {Li}, {Li}, {Li}, {Li}, {Li}, {Li}, {Liang}, {Liang},
  {Lin}, {Liu}, {Liu}, {Liu}, {Liu}, {Liu}, {Liu}, {Liu}, {Liu}, {Liu}, {Liu},
  {Liu}, {Liu}, {Liu}, {Liu}, {Liu}, {Liu}, {Long}, {Lu}, {Lv}, {Ma}, {Ma},
  {Ma}, {Mao}, {Masood}, {Min}, {Mitthumsiri}, {Montaruli}, {Nan}, {Pang},
  {Pattarakijwanich}, {Pei}, {Qi}, {Qi}, {Qiao}, {Qin}, {Ruffolo}, {Rulev},
  {S{\'a}iz}, {Shao}, {Shchegolev}, {Sheng}, {Shi}, {Song}, {Stenkin},
  {Stepanov}, {Su}, {Sun}, {Sun}, {Sun}, {Tam}, {Tang}, {Tian}, {Wang}, {Wang},
  {Wang}, {Wang}, {Wang}, {Wang}, {Wang}, {Wang}, {Wang}, {Wang}, {Wang},
  {Wang}, {Wang}, {Wang}, {Wang}, {Wang}, {Wang}, {Wang}, {Wang}, {Wang},
  {Wang}, {Wang}, {Wei}, {Wei}, {Wei}, {Wen}, {Wu}, {Wu}, {Wu}, {Wu}, {Wu},
  {Xi}, {Xia}, {Xia}, {Xiang}, {Xiao}, {Xiao}, {Xiao}, {Xin}, {Xin}, {Xing},
  {Xu}, {Xu}, {Xue}, {Yan}, {Yan}, {Yang}, {Yang}, {Yang}, {Yang}, {Yang},
  {Yang}, {Yang}, {Yao}, {Yao}, {Ye}, {Yin}, {Yin}, {You}, {You}, {Yu}, {Yuan},
  {Zeng}, {Zeng}, {Zeng}, {Zeng}, {Zha}, {Zhai}, {Zhang}, {Zhang}, {Zhang},
  {Zhang}, {Zhang}, {Zhang}, {Zhang}, {Zhang}, {Zhang}, {Zhang}, {Zhang},
  {Zhang}, {Zhang}, {Zhang}, {Zhang}, {Zhang}, {Zhang}, {Zhang}, {Zhang},
  {Zhao}, {Zhao}, {Zhao}, {Zhao}, {Zhao}, {Zheng}, {Zheng}, {Zhou}, {Zhou},
  {Zhou}, {Zhou}, {Zhou}, {Zhou}, {Zhu}, {Zhu}, {Zhu}, {Zhu}, \&
  {Zuo}}]{2021ApJ...919L..22C}
{Cao}, Z., {Aharonian}, F., {An}, Q., {et~al.} 2021{\natexlab{a}}, ApJL, 919,
  L22

\bibitem[{{Cao} {et~al.}(2021{\natexlab{b}}){Cao}, {Aharonian}, {An},
  {Axikegu}, {Bai}, {Bao}, {Bastieri}, {Bi}, {Bi}, {Cai}, {Cai}, {Cao},
  {Chang}, {Chang}, {Chang}, {Chen}, {Chen}, {Chen}, {Chen}, {Chen}, {Chen},
  {Chen}, {Chen}, {Chen}, {Chen}, {Chen}, {Chen}, {Chen}, {Cheng}, {Cheng},
  {Cui}, {Cui}, {Cui}, {Dai}, {Dai}, {Dai}, {Danzengluobu}, {della Volpe},
  {D'Ettorre Piazzoli}, {Dong}, {Fan}, {Fan}, {Fan}, {Fang}, {Fang}, {Feng},
  {Feng}, {Feng}, {Feng}, {Gao}, {Gao}, {Gao}, {Gao}, {Ge}, {Geng}, {Gong},
  {Gou}, {Gu}, {Guo}, {Guo}, {Guo}, {Guo}, {Han}, {He}, {He}, {He}, {He}, {He},
  {He}, {Heller}, {Hor}, {Hou}, {Hou}, {Hu}, {Hu}, {Hu}, {Hu}, {Huang},
  {Huang}, {Huang}, {Huang}, {Huang}, {Ji}, {Ji}, {Jia}, {Jiang}, {Jiang},
  {Jin}, {Kuleshov}, {Levochkin}, {Li}, {Li}, {Li}, {Li}, {Li}, {Li}, {Li},
  {Li}, {Li}, {Li}, {Li}, {Li}, {Li}, {Li}, {Li}, {Li}, {Li}, {Liang}, {Liang},
  {Lin}, {Liu}, {Liu}, {Liu}, {Liu}, {Liu}, {Liu}, {Liu}, {Liu}, {Liu}, {Liu},
  {Liu}, {Liu}, {Liu}, {Liu}, {Liu}, {Long}, {Lu}, {Lv}, {Ma}, {Ma}, {Ma},
  {Mao}, {Masood}, {Mitthumsiri}, {Montaruli}, {Nan}, {Pang},
  {Pattarakijwanich}, {Pei}, {Qi}, {Ruffolo}, {Rulev}, {S{\'a}iz}, {Shao},
  {Shchegolev}, {Sheng}, {Shi}, {Song}, {Stenkin}, {Stepanov}, {Sun}, {Sun},
  {Sun}, {Tam}, {Tang}, {Tian}, {Wang}, {Wang}, {Wang}, {Wang}, {Wang}, {Wang},
  {Wang}, {Wang}, {Wang}, {Wang}, {Wang}, {Wang}, {Wang}, {Wang}, {Wang},
  {Wang}, {Wang}, {Wang}, {Wang}, {Wang}, {Wang}, {Wei}, {Wei}, {Wei}, {Wen},
  {Wu}, {Wu}, {Wu}, {Wu}, {Wu}, {Xi}, {Xia}, {Xia}, {Xiang}, {Xiao}, {Xiao},
  {Xin}, {Xin}, {Xing}, {Xu}, {Xu}, {Xue}, {Yan}, {Yang}, {Yang}, {Yang},
  {Yang}, {Yang}, {Yang}, {Yang}, {Yao}, {Yao}, {Ye}, {Yin}, {Yin}, {You},
  {You}, {Yu}, {Yuan}, {Zeng}, {Zeng}, {Zeng}, {Zeng}, {Zha}, {Zhai}, {Zhang},
  {Zhang}, {Zhang}, {Zhang}, {Zhang}, {Zhang}, {Zhang}, {Zhang}, {Zhang},
  {Zhang}, {Zhang}, {Zhang}, {Zhang}, {Zhang}, {Zhang}, {Zhang}, {Zhang},
  {Zhang}, {Zhang}, {Zhao}, {Zhao}, {Zhao}, {Zhao}, {Zhao}, {Zheng}, {Zheng},
  {Zhou}, {Zhou}, {Zhou}, {Zhou}, {Zhou}, {Zhou}, {Zhu}, {Zhu}, {Zhu}, {Zhu},
  \& {Zuo}}]{2021Natur.594...33C}
{Cao}, Z., {Aharonian}, F.~A., {An}, Q., {et~al.} 2021{\natexlab{b}}, Nature,
  594, 33

\bibitem[{{Celli} {et~al.}(2019){Celli}, {Morlino}, {Gabici}, \&
  {Aharonian}}]{celli2019}
{Celli}, S., {Morlino}, G., {Gabici}, S., \& {Aharonian}, F.~A. 2019, MNRAS,
  490, 4317

\bibitem[{{Condon} {et~al.}(1998){Condon}, {Cotton}, {Greisen}, {Yin},
  {Perley}, {Taylor}, \& {Broderick}}]{1998AJ....115.1693C}
{Condon}, J.~J., {Cotton}, W.~D., {Greisen}, E.~W., {et~al.} 1998, AJ, 115,
  1693

\bibitem[{{CTA Consortium} {et~al.}(2019){CTA Consortium}, {Acharya}, {Agudo},
  {Al Samarai}, {Alfaro}, {Alfaro}, {Alispach}, {Alves Batista}, {Amans},
  {Amato}, {Ambrosi}, {Antolini}, {Antonelli}, {Aramo}, {Araya}, {Armstrong},
  {Arqueros}, {Arrabito}, {Asano}, {Ashley}, {Backes}, {Balazs}, {Balbo},
  {Ballester}, {Ballet}, {Bamba}, {Barkov}, {Barres de Almeida}, {Barrio},
  {Bastieri}, {Becherini}, {Belfiore}, {Benbow}, {Berge}, {Bernardini},
  {Bernardini}, {Bernardos}, {Bernl{\"o}hr}, {Bertucci}, {Biasuzzi},
  {Bigongiari}, {Biland}, {Bissaldi}, {Biteau}, {Blanch}, {Blazek}, {Boisson},
  {Bolmont}, {Bonanno}, {Bonardi}, {Bonavolont{\`a}}, {Bonnoli}, {Bosnjak},
  {B{\"o}ttcher}, {Braiding}, {Bregeon}, {Brill}, {Brown}, {Brun}, {Brunetti},
  {Buanes}, {Buckley}, {Bugaev}, {B{\"u}hler}, {Bulgarelli}, {Bulik}, {Burton},
  {Burtovoi}, {Busetto}, {Canestrari}, {Capalbi}, {Capitanio}, {Caproni},
  {Caraveo}, {C{\'a}rdenas}, {Carlile}, {Carosi}, {Carqu{\'\i}n}, {Carr},
  {Casanova}, {Cascone}, {Catalani}, {Catalano}, {Cauz}, {Cerruti}, {Chadwick},
  {Chaty}, {Chaves}, {Chen}, {Chen}, {Chernyakova}, {Chikawa}, {Christov},
  {Chudoba}, {Cie{\'s}lar}, {Coco}, {Colafrancesco}, {Colin}, {Conforti},
  {Connaughton}, {Conrad}, {Contreras}, {Cortina}, {Costa}, {Costantini},
  {Cotter}, {Covino}, {Crocker}, {Cuadra}, {Cuevas}, {Cumani}, {D'A{\`\i}},
  {D'Ammando}, {D'Avanzo}, {D'Urso}, {Daniel}, {Davids}, {Dawson}, {Dazzi}, {De
  Angelis}, {de C{\'a}ssia dos Anjos}, {De Cesare}, {De Franco}, {de Gouveia
  Dal Pino}, {de la Calle}, {de los Reyes Lopez}, {De Lotto}, {De Luca}, {De
  Lucia}, {de Naurois}, {de O{\~n}a Wilhelmi}, {De Palma}, {De Persio}, {de
  Souza}, {Deil}, {Del Santo}, {Delgado}, {della Volpe}, {Di Girolamo}, {Di
  Pierro}, {Di Venere}, {D{\'\i}az}, {Dib}, {Diebold}, {Djannati-Ata{\"\i}},
  {Dom{\'\i}nguez}, {Dominis Prester}, {Dorner}, {Doro}, {Drass}, {Dravins},
  {Dubus}, {Dwarkadas}, {Ebr}, {Eckner}, {Egberts}, {Einecke}, {Ekoume},
  {Els{\"a}sser}, {Ernenwein}, {Espinoza}, {Evoli}, {Fairbairn},
  {Falceta-Goncalves}, {Falcone}, {Farnier}, {Fasola}, {Fedorova}, {Fegan},
  {Fernandez-Alonso}, {Fern{\'a}ndez-Barral}, {Ferrand}, {Fesquet},
  {Filipovic}, {Fioretti}, {Fontaine}, {Fornasa}, {Fortson}, {Freixas
  Coromina}, {Fruck}, {Fujita}, {Fukazawa}, {Funk}, {F{\"u}{\ss}ling},
  {Gabici}, {Gadola}, {Gallant}, {Garcia}, {Garcia L{\'o}pez}, {Garczarczyk},
  {Gaskins}, {Gasparetto}, {Gaug}, {Gerard}, {Giavitto}, {Giglietto}, {Giommi},
  {Giordano}, {Giro}, {Giroletti}, {Giuliani}, {Glicenstein}, {Gnatyk},
  {Godinovic}, {Goldoni}, {G{\'o}mez-Vargas}, {Gonz{\'a}lez}, {Gonz{\'a}lez},
  {G{\"o}tz}, {Graham}, {Grandi}, {Granot}, {Green}, {Greenshaw}, {Griffiths},
  {Gunji}, {Hadasch}, {Hara}, {Hardcastle}, {Hassan}, {Hayashi}, {Hayashida},
  {Heller}, {Helo}, {Hermann}, {Hinton}, {Hnatyk}, {Hofmann}, {Holder},
  {Horan}, {H{\"o}randel}, {Horns}, {Horvath}, {Hovatta}, {Hrabovsky},
  {Hrupec}, {Humensky}, {H{\"u}tten}, {Iarlori}, {Inada}, {Inome}, {Inoue},
  {Inoue}, {Inoue}, {Iocco}, {Ioka}, {Iori}, {Ishio}, {Iwamura}, {Jamrozy},
  {Janecek}, {Jankowsky}, {Jean}, {Jung-Richardt}, {Jurysek}, {Kaaret},
  {Karkar}, {Katagiri}, {Katz}, {Kawanaka}, {Kazanas}, {Kh{\'e}lifi}, {Kieda},
  {Kimeswenger}, {Kimura}, {Kisaka}, {Knapp}, {Kn{\"o}dlseder}, {Koch},
  {Kohri}, {Komin}, {Kosack}, {Kraus}, {Krause}, {Krau{\ss}}, {Kubo}, {Kukec
  Mezek}, {Kuroda}, {Kushida}, {La Palombara}, {Lamanna}, {Lang}, {Lapington},
  {Le Blanc}, {Leach}, {Lees}, {Lefaucheur}, {Leigui de Oliveira}, {Lenain},
  {Lico}, {Limon}, {Lindfors}, {Lohse}, {Lombardi}, {Longo}, {L{\'o}pez},
  {L{\'o}pez-Coto}, {Lu}, {Lucarelli}, {Luque-Escamilla}, {Lyard}, {Maccarone},
  {Maier}, {Majumdar}, {Malaguti}, {Mandat}, {Maneva}, {Manganaro}, {Mangano},
  {Marcowith}, {Mar{\'\i}n}, {Markoff}, {Mart{\'\i}}, {Martin},
  {Mart{\'\i}nez}, {Mart{\'\i}nez}, {Masetti}, {Masuda}, {Maurin}, {Maxted},
  {Mazin}, {Medina}, {Melandri}, {Mereghetti}, {Meyer}, {Minaya}, {Mirabal},
  {Mirzoyan}, {Mitchell}, {Mizuno}, {Moderski}, {Mohammed}, {Mohrmann},
  {Montaruli}, {Moralejo}, {Morcuende-Parrilla}, {Mori}, {Morlino}, {Morris},
  {Morselli}, {Moulin}, {Mukherjee}, {Mundell}, {Murach}, {Muraishi}, {Murase},
  {Nagai}, {Nagataki}, {Nagayoshi}, {Naito}, {Nakamori}, {Nakamura}, {Niemiec},
  {Nieto}, {Niko{\l}ajuk}, {Nishijima}, {Noda}, {Nosek}, {Novosyadlyj},
  {Nozaki}, {O'Brien}, {Oakes}, {Ohira}, {Ohishi}, {Ohm}, {Okazaki}, {Okumura},
  {Ong}, {Orienti}, {Orito}, {Osborne}, {Ostrowski}, {Otte}, {Oya}, {Padovani},
  {Paizis}, {Palatiello}, {Palatka}, {Paoletti}, {Paredes}, {Pareschi},
  {Parsons}, {Pe'er}, {Pech}, {Pedaletti}, {Perri}, {Persic}, {Petrashyk},
  {Petrucci}, {Petruk}, {Peyaud}, {Pfeifer}, {Piano}, {Pisarski}, {Pita},
  {Pohl}, {Polo}, {Pozo}, {Prandini}, {Prast}, {Principe}, {Prokhorov},
  {Prokoph}, {Prouza}, {P{\"u}hlhofer}, {Punch}, {P{\"u}rckhauer}, {Queiroz},
  {Quirrenbach}, {Rain{\`o}}, {Razzaque}, {Reimer}, {Reimer}, {Reisenegger},
  {Renaud}, {Rezaeian}, {Rhode}, {Ribeiro}, {Rib{\'o}}, {Richtler}, {Rico},
  {Rieger}, {Riquelme}, {Rivoire}, {Rizi}, {Rodriguez}, {Rodriguez Fernandez},
  {Rodr{\'\i}guez V{\'a}zquez}, {Rojas}, {Romano}, {Romeo}, {Rosado}, {Rovero},
  {Rowell}, {Rudak}, {Rugliancich}, {Rulten}, {Sadeh}, {Safi-Harb}, {Saito},
  {Sakaki}, {Sakurai}, {Salina}, {S{\'a}nchez-Conde}, {Sandaker}, {Sandoval},
  {Sangiorgi}, {Sanguillon}, {Sano}, {Santander}, {Sarkar}, {Satalecka},
  {Saturni}, {Schioppa}, {Schlenstedt}, {Schneider}, {Schoorlemmer},
  {Schovanek}, {Schulz}, {Schussler}, {Schwanke}, {Sciacca}, {Scuderi},
  {Seitenzahl}, {Semikoz}, {Sergijenko}, {Servillat}, {Shalchi}, {Shellard},
  {Sidoli}, {Siejkowski}, {Sillanp{\"a}{\"a}}, {Sironi}, {Sitarek}, {Sliusar},
  {Slowikowska}, {Sol}, {Stamerra}, {Stani{\v{c}}}, {Starling}, {Stawarz},
  {Stefanik}, {Stephan}, {Stolarczyk}, {Stratta}, {Straumann}, {Suomijarvi},
  {Supanitsky}, {Tagliaferri}, {Tajima}, {Tavani}, {Tavecchio}, {Tavernet},
  {Tayabaly}, {Tejedor}, {Temnikov}, {Terada}, {Terrier}, {Terzic}, {Teshima},
  {Testa}, {Thoudam}, {Tian}, {Tibaldo}, {Tluczykont}, {Todero Peixoto},
  {Tokanai}, {Tomastik}, {Tonev}, {Tornikoski}, {Torres}, {Torresi}, {Tosti},
  {Tothill}, {Tovmassian}, {Travnicek}, {Trichard}, {Trifoglio}, {Troyano
  Pujadas}, {Tsujimoto}, {Umana}, {Vagelli}, {Vagnetti}, {Valentino},
  {Vallania}, {Valore}, {van Eldik}, {Vandenbroucke}, {Varner}, {Vasileiadis},
  {Vassiliev}, {V{\'a}zquez Acosta}, {Vecchi}, {Vega}, {Vercellone}, {Veres},
  {Vergani}, {Verzi}, {Vettolani}, {Viana}, {Vigorito}, {Villanueva}, {Voelk},
  {Vollhardt}, {Vorobiov}, {Vrastil}, {Vuillaume}, {Wagner}, {Wagner},
  {Walter}, {Ward}, {Warren}, {Watson}, {Werner}, {White}, {White},
  {Wierzcholska}, {Wilcox}, {Will}, {Williams}, {Wischnewski}, {Wood},
  {Yamamoto}, {Yamazaki}, {Yanagita}, {Yang}, {Yoshida}, {Yoshiike},
  {Yoshikoshi}, {Zacharias}, {Zaharijas}, {Zampieri}, {Zandanel}, {Zanin},
  {Zavrtanik}, {Zavrtanik}, {Zdziarski}, {Zech}, {Zechlin}, {Zhdanov},
  {Ziegler}, \& {Zorn}}]{2019scta.book.....C}
{CTA Consortium}, {Acharya}, B.~S., {Agudo}, I., {et~al.} 2019, {Science with
  the Cherenkov Telescope Array}

\bibitem[{{Cui} {et~al.}(2018){Cui}, {Yeung}, {Tam}, \&
  {P{\"u}hlhofer}}]{2018ApJ...860...69C}
{Cui}, Y., {Yeung}, P. K.~H., {Tam}, P.~H.~T., \& {P{\"u}hlhofer}, G. 2018,
  ApJ, 860, 69

\bibitem[{{Dame} {et~al.}(2001){Dame}, {Hartmann}, \&
  {Thaddeus}}]{2001ApJ...547..792D}
{Dame}, T.~M., {Hartmann}, D., \& {Thaddeus}, P. 2001, ApJ, 547, 792

\bibitem[{{de Bruyn} {et~al.}(2000){de Bruyn}, {Miley}, {Rengelink}, {Tang},
  {Bremer}, {Rottgering}, {Raimond}, {Bremer}, \&
  {Fullagar}}]{2000yCat.8062....0D}
{de Bruyn}, G., {Miley}, G., {Rengelink}, R., {et~al.} 2000, VizieR Online Data
  Catalog, VIII/62

\bibitem[{{de Naurois}(2021)}]{2021Univ....7..421D}
{de Naurois}, M. 2021, Universe, 7, 421

\bibitem[{{de O{\~n}a Wilhelmi} {et~al.}(2022){de O{\~n}a Wilhelmi},
  {L{\'o}pez-Coto}, {Amato}, \& {Aharonian}}]{2022ApJ...930L...2D}
{de O{\~n}a Wilhelmi}, E., {L{\'o}pez-Coto}, R., {Amato}, E., \& {Aharonian},
  F. 2022, ApJL, 930, L2

\bibitem[{{Deil} {et~al.}(2017){Deil}, {Zanin}, {Lefaucheur}, {Boisson},
  {Khelifi}, {Terrier}, {Wood}, {Mohrmann}, {Chakraborty}, {Watson},
  {Lopez-Coto}, {Klepser}, {Cerruti}, {Lenain}, {Acero}, {Djannati-Ata{\"\i}},
  {Pita}, {Bosnjak}, {Trichard}, {Vuillaume}, {Donath}, {Consortium}, {King},
  {Jouvin}, {Owen}, {Sipocz}, {Lennarz}, {Voruganti}, {Spir-Jacob}, {Ruiz}, \&
  {Arribas}}]{2017ICRC...35..766D}
{Deil}, C., {Zanin}, R., {Lefaucheur}, J., {et~al.} 2017, in International
  Cosmic Ray Conference, Vol. 301, 35th International Cosmic Ray Conference
  (ICRC2017), 766

\bibitem[{{Dickey} \& {Lockman}(1990)}]{1990ARA&A..28..215D}
{Dickey}, J.~M. \& {Lockman}, F.~J. 1990, ARAA, 28, 215

\bibitem[{{Eckert} {et~al.}(2017){Eckert}, {Ettori}, {Pointecouteau},
  {Molendi}, {Paltani}, \& {Tchernin}}]{eckert17}
{Eckert}, D., {Ettori}, S., {Pointecouteau}, E., {et~al.} 2017, Astronomische
  Nachrichten, 338, 293

\bibitem[{{Eckert} {et~al.}(2020){Eckert}, {Finoguenov}, {Ghirardini},
  {Grandis}, {Kaefer}, {Sanders}, \& {Ramos-Ceja}}]{eckert20}
{Eckert}, D., {Finoguenov}, A., {Ghirardini}, V., {et~al.} 2020, The Open
  Journal of Astrophysics, 3, 12

\bibitem[{{Faherty} {et~al.}(2007){Faherty}, {Walter}, \&
  {Anderson}}]{2007Ap&SS.308..225F}
{Faherty}, J., {Walter}, F.~M., \& {Anderson}, J. 2007, ApSS, 308, 225

\bibitem[{{Fermi-LAT collaboration} {et~al.}(2022){Fermi-LAT collaboration},
  {:}, {Abdollahi}, {Acero}, {Baldini}, {Ballet}, {Bastieri}, {Bellazzini},
  {Berenji}, {Berretta}, {Bissaldi}, {Blandford}, {Bloom}, {Bonino}, {Brill},
  {Britto}, {Bruel}, {Burnett}, {Buson}, {Cameron}, {Caputo}, {Caraveo},
  {Castro}, {Chaty}, {Cheung}, {Chiaro}, {Cibrario}, {Ciprini},
  {Coronado-Blazquez}, {Crnogorcevic}, {Cutini}, {D'Ammando}, {De Gaetano},
  {Digel}, {Di Lalla}, {Dirirsa}, {Di Venere}, {Dominguez}, {Fallah Ramazani},
  {Fegan}, {Ferrara}, {Fiori}, {Fleischhack}, {Franckowiak}, {Fukazawa},
  {Funk}, {Fusco}, {Galanti}, {Gammaldi}, {Gargano}, {Garrappa}, {Gasparrini},
  {Giacchino}, {Giglietto}, {Giordano}, {Giroletti}, {Glanzman}, {Green},
  {Grenier}, {Grondin}, {Guillemot}, {Guiriec}, {Gustafsson}, {Harding},
  {Hays}, {Hewitt}, {Horan}, {Hou}, {Johannesson}, {Karwin}, {Kayanoki},
  {Kerr}, {Kuss}, {Landriu}, {Larsson}, {Latronico}, {Lemoine-Goumard}, {Li},
  {Liodakis}, {Longo}, {Loparco}, {Lott}, {Lubrano}, {Maldera}, {Malyshev},
  {Manfreda}, {Marti-Devesa}, {Mazziotta}, {Mereu}, {Meyer}, {Michelson},
  {Mirabal}, {Mitthumsiri}, {Mizuno}, {Moiseev}, {Monzani}, {Morselli},
  {Moskalenko}, {Negro}, {Nuss}, {Omodei}, {Orienti}, {Orlando}, {Paneque},
  {Pei}, {Perkins}, {Persic}, {Pesce-Rollins}, {Petrosian}, {Pillera}, {Poon},
  {Porter}, {Principe}, {Raino}, {Rando}, {Rani}, {Razzano}, {Razzaque},
  {Reimer}, {Reimer}, {Reposeur}, {Sanchez-Conde}, {Saz Parkinson}, {Scotton},
  {Serini}, {Sgro}, {Siskind}, {Smith}, {Spandre}, {Spinelli}, {Sueoka},
  {Suson}, {Tajima}, {Tak}, {Thayer}, {Thompson}, {Torres}, {Troja},
  {Valverde}, {Wood}, \& {Zaharijas}}]{2022arXiv220111184F}
{Fermi-LAT collaboration}, {:}, {Abdollahi}, S., {et~al.} 2022, arXiv e-prints,
  arXiv:2201.11184

\bibitem[{Fomin {et~al.}(1994)Fomin, Stepanian, Lamb, Lewis, Punch, \&
  Weekes}]{FOMIN1994137}
Fomin, V., Stepanian, A., Lamb, R., {et~al.} 1994, Astroparticle Physics, 2,
  137

\bibitem[{Fruck {et~al.}(2022)}]{Fruck:2022igg}
Fruck, C. {et~al.} 2022, Mon. Not. Roy. Astron. Soc., 515, 4520

\bibitem[{{Gabici} {et~al.}(2019){Gabici}, {Evoli}, {Gaggero}, {Lipari},
  {Mertsch}, {Orlando}, {Strong}, \& {Vittino}}]{2019IJMPD..2830022G}
{Gabici}, S., {Evoli}, C., {Gaggero}, D., {et~al.} 2019, International Journal
  of Modern Physics D, 28, 1930022

\bibitem[{{Ghirardini} {et~al.}(2019){Ghirardini}, {Eckert}, {Ettori},
  {Pointecouteau}, {Molendi}, {Gaspari}, {Rossetti}, {De Grandi}, {Roncarelli},
  {Bourdin}, {Mazzotta}, {Rasia}, \& {Vazza}}]{ghirardini19}
{Ghirardini}, V., {Eckert}, D., {Ettori}, S., {et~al.} 2019, A\&A, 621, A41

\bibitem[{{Grant} {et~al.}(2019){Grant}, {Ackermann}, {Karle}, \&
  {Kowalski}}]{2019BAAS...51g.288G}
{Grant}, D., {Ackermann}, M., {Karle}, A., \& {Kowalski}, M. 2019, in Bulletin
  of the American Astronomical Society, Vol.~51, 288

\bibitem[{{H.~E.~S.~S. Collaboration} {et~al.}(2018{\natexlab{a}}){H.~E.~S.~S.
  Collaboration}, {Abdalla}, {Abramowski}, {Aharonian}, {Ait Benkhali},
  {Akhperjanian}, {Andersson}, {Ang{\"u}ner}, {Arrieta}, {Aubert}, {Backes},
  {Balzer}, {Barnard}, {Becherini}, {Becker Tjus}, {Berge}, {Bernhard},
  {Bernl{\"o}hr}, {Blackwell}, {B{\"o}ttcher}, {Boisson}, {Bolmont}, {Bordas},
  {Bregeon}, {Brun}, {Brun}, {Bryan}, {Bulik}, {Capasso}, {Carr}, {Carrigan},
  {Casanova}, {Cerruti}, {Chakraborty}, {Chalme-Calvet}, {Chaves}, {Chen},
  {Chevalier}, {Chr{\'e}tien}, {Colafrancesco}, {Cologna}, {Condon}, {Conrad},
  {Couturier}, {Cui}, {Davids}, {Degrange}, {Deil}, {Devin}, {deWilt},
  {Dirson}, {Djannati-Ata{\"\i}}, {Domainko}, {Donath}, {Drury}, {Dubus},
  {Dutson}, {Dyks}, {Edwards}, {Egberts}, {Eger}, {Ernenwein}, {Eschbach},
  {Farnier}, {Fegan}, {Fernandes}, {Fiasson}, {Fontaine}, {F{\"o}rster},
  {Funk}, {F{\"u}{\ss}ling}, {Gabici}, {Gajdus}, {Gallant}, {Garrigoux},
  {Giavitto}, {Giebels}, {Glicenstein}, {Gottschall}, {Goyal}, {Grondin},
  {Hadasch}, {Hahn}, {Haupt}, {Hawkes}, {Heinzelmann}, {Henri}, {Hermann},
  {Hervet}, {Hillert}, {Hinton}, {Hofmann}, {Hoischen}, {Holler}, {Horns},
  {Ivascenko}, {Jacholkowska}, {Jamrozy}, {Janiak}, {Jankowsky}, {Jankowsky},
  {Jingo}, {Jogler}, {Jouvin}, {Jung-Richardt}, {Kastendieck},
  {Katarzy{\'n}ski}, {Katz}, {Kerszberg}, {Kh{\'e}lifi}, {Kieffer}, {King},
  {Klepser}, {Klochkov}, {Klu{\'z}niak}, {Kolitzus}, {Komin}, {Kosack},
  {Krakau}, {Kraus}, {Krayzel}, {Kr{\"u}ger}, {Laffon}, {Lamanna}, {Lau},
  {Lees}, {Lefaucheur}, {Lefranc}, {Lemi{\`e}re}, {Lemoine-Goumard}, {Lenain},
  {Leser}, {Lohse}, {Lorentz}, {Liu}, {L{\'o}pez-Coto}, {Lypova}, {Marandon},
  {Marcowith}, {Mariaud}, {Marx}, {Maurin}, {Maxted}, {Mayer}, {Meintjes},
  {Meyer}, {Mitchell}, {Moderski}, {Mohamed}, {Mohrmann}, {Mor{\r{a}}},
  {Moulin}, {Murach}, {de Naurois}, {Niederwanger}, {Niemiec}, {Oakes},
  {O'Brien}, {Odaka}, {{\"O}ttl}, {Ohm}, {de O{\~n}a Wilhelmi}, {Ostrowski},
  {Oya}, {Padovani}, {Panter}, {Parsons}, {Paz Arribas}, {Pekeur}, {Pelletier},
  {Perennes}, {Petrucci}, {Peyaud}, {Pita}, {Poon}, {Prokhorov}, {Prokoph},
  {P{\"u}hlhofer}, {Punch}, {Quirrenbach}, {Raab}, {Reimer}, {Reimer},
  {Renaud}, {de los Reyes}, {Rieger}, {Romoli}, {Rosier-Lees}, {Rowell},
  {Rudak}, {Rulten}, {Sahakian}, {Salek}, {Sanchez}, {Santangelo}, {Sasaki},
  {Schlickeiser}, {Sch{\"u}ssler}, {Schulz}, {Schwanke}, {Schwemmer},
  {Settimo}, {Seyffert}, {Shafi}, {Shilon}, {Simoni}, {Sol}, {Spanier},
  {Spengler}, {Spies}, {Stawarz}, {Steenkamp}, {Stegmann}, {Stinzing}, {Stycz},
  {Sushch}, {Tavernet}, {Tavernier}, {Taylor}, {Terrier}, {Tibaldo}, {Tiziani},
  {Tluczykont}, {Trichard}, {Tuffs}, {Uchiyama}, {Valerius}, {van der Walt},
  {van Eldik}, {van Soelen}, {Vasileiadis}, {Veh}, {Venter}, {Viana},
  {Vincent}, {Vink}, {Voisin}, {V{\"o}lk}, {Vuillaume}, {Wadiasingh}, {Wagner},
  {Wagner}, {Wagner}, {White}, {Wierzcholska}, {Willmann}, {W{\"o}rnlein},
  {Wouters}, {Yang}, {Zabalza}, {Zaborov}, {Zacharias}, {Zdziarski}, {Zech},
  {Zefi}, {Ziegler}, \& {{\.Z}ywucka}}]{2018A&A...612A...2H}
{H.~E.~S.~S. Collaboration}, {Abdalla}, H., {Abramowski}, A., {et~al.}
  2018{\natexlab{a}}, A\&A, 612, A2

\bibitem[{{H.~E.~S.~S. Collaboration} {et~al.}(2018{\natexlab{b}}){H.~E.~S.~S.
  Collaboration}, {Abdalla}, {Abramowski}, {Aharonian}, {Ait Benkhali},
  {Akhperjanian}, {Andersson}, {Ang{\"u}ner}, {Arrieta}, {Aubert}, {Backes},
  {Balzer}, {Barnard}, {Becherini}, {Becker Tjus}, {Berge}, {Bernhard},
  {Bernl{\"o}hr}, {Blackwell}, {B{\"o}ttcher}, {Boisson}, {Bolmont}, {Bordas},
  {Bregeon}, {Brun}, {Brun}, {Bryan}, {Bulik}, {Capasso}, {Carr}, {Casanova},
  {Cerruti}, {Chakraborty}, {Chalme-Calvet}, {Chaves}, {Chen}, {Chevalier},
  {Chr{\'e}tien}, {Colafrancesco}, {Cologna}, {Condon}, {Conrad}, {Cui},
  {Davids}, {Decock}, {Degrange}, {Deil}, {Devin}, {deWilt}, {Dirson},
  {Djannati-Ata{\"\i}}, {Domainko}, {Donath}, {Drury}, {Dubus}, {Dutson},
  {Dyks}, {Edwards}, {Egberts}, {Eger}, {Ernenwein}, {Eschbach}, {Farnier},
  {Fegan}, {Fernandes}, {Fiasson}, {Fontaine}, {F{\"o}rster}, {Funk},
  {F{\"u}{\ss}ling}, {Gabici}, {Gajdus}, {Gallant}, {Garrigoux}, {Giavitto},
  {Giebels}, {Glicenstein}, {Gottschall}, {Goyal}, {Grondin}, {Hadasch},
  {Hahn}, {Haupt}, {Hawkes}, {Heinzelmann}, {Henri}, {Hermann}, {Hervet},
  {Hinton}, {Hofmann}, {Hoischen}, {Holler}, {Horns}, {Ivascenko},
  {Jacholkowska}, {Jamrozy}, {Janiak}, {Jankowsky}, {Jankowsky}, {Jingo},
  {Jogler}, {Jouvin}, {Jung-Richardt}, {Kastendieck}, {Katarzy{\'n}ski},
  {Katz}, {Kerszberg}, {Kh{\'e}lifi}, {Kieffer}, {King}, {Klepser}, {Klochkov},
  {Klu{\'z}niak}, {Kolitzus}, {Komin}, {Kosack}, {Krakau}, {Kraus}, {Krayzel},
  {Kr{\"u}ger}, {Laffon}, {Lamanna}, {Lau}, {Lees}, {Lefaucheur}, {Lefranc},
  {Lemi{\`e}re}, {Lemoine-Goumard}, {Lenain}, {Leser}, {Lohse}, {Lorentz},
  {Liu}, {L{\'o}pez-Coto}, {Lypova}, {Marandon}, {Marcowith}, {Mariaud},
  {Marx}, {Maurin}, {Maxted}, {Mayer}, {Meintjes}, {Meyer}, {Mitchell},
  {Moderski}, {Mohamed}, {Mohrmann}, {Mor{\r{a}}}, {Moulin}, {Murach}, {de
  Naurois}, {Niederwanger}, {Niemiec}, {Oakes}, {O'Brien}, {Odaka}, {{\"O}ttl},
  {Ohm}, {Ostrowski}, {Oya}, {Padovani}, {Panter}, {Parsons}, {Pekeur},
  {Pelletier}, {Perennes}, {Petrucci}, {Peyaud}, {Piel}, {Pita}, {Poon},
  {Prokhorov}, {Prokoph}, {P{\"u}hlhofer}, {Punch}, {Quirrenbach}, {Raab},
  {Reimer}, {Reimer}, {Renaud}, {de los Reyes}, {Rieger}, {Romoli},
  {Rosier-Lees}, {Rowell}, {Rudak}, {Rulten}, {Sahakian}, {Salek}, {Sanchez},
  {Santangelo}, {Sasaki}, {Schlickeiser}, {Sch{\"u}ssler}, {Schulz},
  {Schwanke}, {Schwemmer}, {Settimo}, {Seyffert}, {Shafi}, {Shilon}, {Simoni},
  {Sol}, {Spanier}, {Spengler}, {Spies}, {Stawarz}, {Steenkamp}, {Stegmann},
  {Stinzing}, {Stycz}, {Sushch}, {Tavernet}, {Tavernier}, {Taylor}, {Terrier},
  {Tibaldo}, {Tiziani}, {Tluczykont}, {Trichard}, {Tuffs}, {Uchiyama}, {van der
  Walt}, {van Eldik}, {van Rensburg}, {van Soelen}, {Vasileiadis}, {Veh},
  {Venter}, {Viana}, {Vincent}, {Vink}, {Voisin}, {V{\"o}lk}, {Vuillaume},
  {Wadiasingh}, {Wagner}, {Wagner}, {Wagner}, {White}, {Wierzcholska},
  {Willmann}, {W{\"o}rnlein}, {Wouters}, {Yang}, {Zabalza}, {Zaborov},
  {Zacharias}, {Zdziarski}, {Zech}, {Zefi}, {Ziegler}, {{\.Z}ywucka},
  {Fermi-LAT Collaboration}, \& {Katsuta}}]{2018A&A...612A...5H}
{H.~E.~S.~S. Collaboration}, {Abdalla}, H., {Abramowski}, A., {et~al.}
  2018{\natexlab{b}}, A\&A, 612, A5

\bibitem[{{H.~E.~S.~S. Collaboration} {et~al.}(2018{\natexlab{c}}){H.~E.~S.~S.
  Collaboration}, {Abdalla}, {Abramowski}, {Aharonian}, {Ait Benkhali},
  {Ang{\"u}ner}, {Arakawa}, {Arrieta}, {Aubert}, {Backes}, {Balzer}, {Barnard},
  {Becherini}, {Becker Tjus}, {Berge}, {Bernhard}, {Bernl{\"o}hr}, {Blackwell},
  {B{\"o}ttcher}, {Boisson}, {Bolmont}, {Bonnefoy}, {Bordas}, {Bregeon},
  {Brun}, {Brun}, {Bryan}, {B{\"u}chele}, {Bulik}, {Capasso}, {Carrigan},
  {Caroff}, {Carosi}, {Casanova}, {Cerruti}, {Chakraborty}, {Chaves}, {Chen},
  {Chevalier}, {Colafrancesco}, {Condon}, {Conrad}, {Davids}, {Decock}, {Deil},
  {Devin}, {deWilt}, {Dirson}, {Djannati-Ata{\"\i}}, {Domainko}, {Donath},
  {Drury}, {Dutson}, {Dyks}, {Edwards}, {Egberts}, {Eger}, {Emery},
  {Ernenwein}, {Eschbach}, {Farnier}, {Fegan}, {Fernandes}, {Fiasson},
  {Fontaine}, {F{\"o}rster}, {Funk}, {F{\"u}{\ss}ling}, {Gabici}, {Gallant},
  {Garrigoux}, {Gast}, {Gat{\'e}}, {Giavitto}, {Giebels}, {Glawion},
  {Glicenstein}, {Gottschall}, {Grondin}, {Hahn}, {Haupt}, {Hawkes},
  {Heinzelmann}, {Henri}, {Hermann}, {Hinton}, {Hofmann}, {Hoischen}, {Holch},
  {Holler}, {Horns}, {Ivascenko}, {Iwasaki}, {Jacholkowska}, {Jamrozy},
  {Jankowsky}, {Jankowsky}, {Jingo}, {Jouvin}, {Jung-Richardt}, {Kastendieck},
  {Katarzy{\'n}ski}, {Katsuragawa}, {Katz}, {Kerszberg}, {Khangulyan},
  {Kh{\'e}lifi}, {King}, {Klepser}, {Klochkov}, {Klu{\'z}niak}, {Komin},
  {Kosack}, {Krakau}, {Kraus}, {Kr{\"u}ger}, {Laffon}, {Lamanna}, {Lau},
  {Lees}, {Lefaucheur}, {Lemi{\`e}re}, {Lemoine-Goumard}, {Lenain}, {Leser},
  {Lohse}, {Lorentz}, {Liu}, {L{\'o}pez-Coto}, {Lypova}, {Marandon},
  {Malyshev}, {Marcowith}, {Mariaud}, {Marx}, {Maurin}, {Maxted}, {Mayer},
  {Meintjes}, {Meyer}, {Mitchell}, {Moderski}, {Mohamed}, {Mohrmann},
  {Mor{\r{a}}}, {Moulin}, {Murach}, {Nakashima}, {de Naurois}, {Ndiyavala},
  {Niederwanger}, {Niemiec}, {Oakes}, {O'Brien}, {Odaka}, {Ohm}, {Ostrowski},
  {Oya}, {Padovani}, {Panter}, {Parsons}, {Paz Arribas}, {Pekeur}, {Pelletier},
  {Perennes}, {Petrucci}, {Peyaud}, {Piel}, {Pita}, {Poireau}, {Poon},
  {Prokhorov}, {Prokoph}, {P{\"u}hlhofer}, {Punch}, {Quirrenbach}, {Raab},
  {Rauth}, {Reimer}, {Reimer}, {Renaud}, {de los Reyes}, {Rieger}, {Rinchiuso},
  {Romoli}, {Rowell}, {Rudak}, {Rulten}, {Safi-Harb}, {Sahakian}, {Saito},
  {Sanchez}, {Santangelo}, {Sasaki}, {Schandri}, {Schlickeiser},
  {Sch{\"u}ssler}, {Schulz}, {Schwanke}, {Schwemmer}, {Seglar-Arroyo},
  {Settimo}, {Seyffert}, {Shafi}, {Shilon}, {Shiningayamwe}, {Simoni}, {Sol},
  {Spanier}, {Spir-Jacob}, {Stawarz}, {Steenkamp}, {Stegmann}, {Steppa},
  {Sushch}, {Takahashi}, {Tavernet}, {Tavernier}, {Taylor}, {Terrier},
  {Tibaldo}, {Tiziani}, {Tluczykont}, {Trichard}, {Tsirou}, {Tsuji}, {Tuffs},
  {Uchiyama}, {van der Walt}, {van Eldik}, {van Rensburg}, {van Soelen},
  {Vasileiadis}, {Veh}, {Venter}, {Viana}, {Vincent}, {Vink}, {Voisin},
  {V{\"o}lk}, {Vuillaume}, {Wadiasingh}, {Wagner}, {Wagner}, {Wagner}, {White},
  {Wierzcholska}, {Willmann}, {W{\"o}rnlein}, {Wouters}, {Yang}, {Zaborov},
  {Zacharias}, {Zanin}, {Zdziarski}, {Zech}, {Zefi}, {Ziegler}, {Zorn}, \&
  {{\.Z}ywucka}}]{2018A&A...612A...1H}
{H.~E.~S.~S. Collaboration}, {Abdalla}, H., {Abramowski}, A., {et~al.}
  2018{\natexlab{c}}, A\&A, 612, A1

\bibitem[{{HI4PI Collaboration} {et~al.}(2016){HI4PI Collaboration}, {Ben
  Bekhti}, {Fl{\"o}er}, {Keller}, {Kerp}, {Lenz}, {Winkel}, {Bailin},
  {Calabretta}, {Dedes}, {Ford}, {Gibson}, {Haud}, {Janowiecki}, {Kalberla},
  {Lockman}, {McClure-Griffiths}, {Murphy}, {Nakanishi}, {Pisano}, \&
  {Staveley-Smith}}]{2016A&A...594A.116H}
{HI4PI Collaboration}, {Ben Bekhti}, N., {Fl{\"o}er}, L., {et~al.} 2016, A\&A,
  594, A116

\bibitem[{{Hillas}(1985)}]{1985ICRC....3..445H}
{Hillas}, A.~M. 1985, in International Cosmic Ray Conference, Vol.~3, 19th
  International Cosmic Ray Conference (ICRC19), Volume 3, 445

\bibitem[{{Hinton} \& {Hofmann}(2009)}]{2009ARA&A..47..523H}
{Hinton}, J.~A. \& {Hofmann}, W. 2009, ARAA, 47, 523

\bibitem[{{Jogler} \& {Funk}(2016)}]{2016ApJ...816..100J}
{Jogler}, T. \& {Funk}, S. 2016, ApJ, 816, 100

\bibitem[{Jury\v{s}ek {et~al.}(2021)Jury\v{s}ek, Lyard, \&
  Walter}]{Jurysek:2021iig}
Jury\v{s}ek, J., Lyard, E., \& Walter, R. 2021, in {31st annual conference
  on~Astronomical Data Analysis Software and Systems}

\bibitem[{{Kafexhiu} {et~al.}(2014){Kafexhiu}, {Aharonian}, {Taylor}, \&
  {Vila}}]{2014PhRvD..90l3014K}
{Kafexhiu}, E., {Aharonian}, F., {Taylor}, A.~M., \& {Vila}, G.~S. 2014, PRD,
  90, 123014

\bibitem[{Kar \& Gupta(2022)}]{Kar_2022}
Kar, A. \& Gupta, N. 2022, The Astrophysical Journal, 926, 110

\bibitem[{{Kelner} {et~al.}(2006){Kelner}, {Aharonian}, \&
  {Bugayov}}]{kelner_2006}
{Kelner}, S.~R., {Aharonian}, F.~A., \& {Bugayov}, V.~V. 2006, PRD, 74, 034018

\bibitem[{{Khangulyan} {et~al.}(2014){Khangulyan}, {Aharonian}, \&
  {Kelner}}]{2014ApJ...783..100K}
{Khangulyan}, D., {Aharonian}, F.~A., \& {Kelner}, S.~R. 2014, ApJ, 783, 100

\bibitem[{{Kharchenko} {et~al.}(2016){Kharchenko}, {Piskunov}, {Schilbach},
  {R{\"o}ser}, \& {Scholz}}]{2016A&A...585A.101K}
{Kharchenko}, N.~V., {Piskunov}, A.~E., {Schilbach}, E., {R{\"o}ser}, S., \&
  {Scholz}, R.~D. 2016, A\&A, 585, A101

\bibitem[{{Kronberger} {et~al.}(2006){Kronberger}, {Teutsch}, {Alessi},
  {Steine}, {Ferrero}, {Graczewski}, {Juchert}, {Patchick}, {Riddle},
  {Saloranta}, {Schoenball}, \& {Watson}}]{2006A&A...447..921K}
{Kronberger}, M., {Teutsch}, P., {Alessi}, B., {et~al.} 2006, A\&A, 447, 921

\bibitem[{Leung {et~al.}(2014)Leung, Takata, Ng, Kong, Tam, Hui, \&
  Cheng}]{Leung_2014}
Leung, G. C.~K., Takata, J., Ng, C.~W., {et~al.} 2014, The Astrophysical
  Journal, 797, L13

\bibitem[{{Li} {et~al.}(2021){Li}, {Liu}, {de O{\~n}a Wilhelmi}, {Torres},
  {Liu}, {Kerr}, {B{\"u}hler}, {Su}, {He}, \& {Xiao}}]{2021ApJ...913L..33L}
{Li}, J., {Liu}, R.-Y., {de O{\~n}a Wilhelmi}, E., {et~al.} 2021, ApJL, 913,
  L33

\bibitem[{{Li} \& {Ma}(1983)}]{1983ApJ...272..317L}
{Li}, T.~P. \& {Ma}, Y.~Q. 1983, ApJ, 272, 317

\bibitem[{{Linden} {et~al.}(2017){Linden}, {Auchettl}, {Bramante}, {Cholis},
  {Fang}, {Hooper}, {Karwal}, \& {Li}}]{2017PhRvD..96j3016L}
{Linden}, T., {Auchettl}, K., {Bramante}, J., {et~al.} 2017, PRD, 96, 103016

\bibitem[{Liu {et~al.}(2019)Liu, Ge, Sun, \& Wang}]{Liu_2019}
Liu, R.-Y., Ge, C., Sun, X.-N., \& Wang, X.-Y. 2019, The Astrophysical Journal,
  875, 149

\bibitem[{{L{\'o}pez-Coto} {et~al.}(2022){L{\'o}pez-Coto}, {de O{\~n}a
  Wilhelmi}, {Aharonian}, {Amato}, \& {Hinton}}]{2022NatAs...6..199L}
{L{\'o}pez-Coto}, R., {de O{\~n}a Wilhelmi}, E., {Aharonian}, F., {Amato}, E.,
  \& {Hinton}, J. 2022, Nature Astronomy, 6, 199

\bibitem[{Lopez-Coto {et~al.}(2022)Lopez-Coto, Vuillaume, Moralejo, Cassol,
  Nöthe, Morcuende, Priyadarshi, Bernardos, Nozaki, Gliwny, Ruiz, Garcia,
  Dalchenko, yrenier, Saha, Jacquemont, Neise, Alispach, Pillera,
  Andres-Baquero, Sitarek, aaguasca, Takahashi, sn621, \&
  yiwamura}]{ruben_lopez_coto_2022_6458862}
Lopez-Coto, R., Vuillaume, T., Moralejo, A., {et~al.} 2022,
  cta-observatory/cta-lstchain: v0.9.6 - 2022-04-13

\bibitem[{L\'opez-Coto {et~al.}(2021)}]{CTALSTproject:2021mfp}
L\'opez-Coto, R. {et~al.} 2021, in {37th International Cosmic Ray Conference}

\bibitem[{{MAGIC Collaboration} {et~al.}(2020){MAGIC Collaboration}, {Acciari},
  {Ansoldi}, {Antonelli}, {Arbet Engels}, {Asano}, {Baack}, {Babi{\'c}},
  {Baquero}, {Barres de Almeida}, {Barrio}, {Becerra Gonz{\'a}lez}, {Bednarek},
  {Bellizzi}, {Bernardini}, {Bernardos}, {Berti}, {Besenrieder},
  {Bhattacharyya}, {Bigongiari}, {Biland}, {Blanch}, {Bonnoli},
  {Bo{\v{s}}njak}, {Busetto}, {Carosi}, {Ceribella}, {Cerruti}, {Chai},
  {Chilingarian}, {Cikota}, {Colak}, {Colombo}, {Contreras}, {Cortina},
  {Covino}, {D'Amico}, {D'Elia}, {da Vela}, {Dazzi}, {de Angelis}, {de Lotto},
  {Delfino}, {Delgado}, {Delgado Mendez}, {Depaoli}, {di Girolamo}, {di
  Pierro}, {di Venere}, {Do Souto Espi{\~n}eira}, {Dominis Prester}, {Donini},
  {Dorner}, {Doro}, {Elsaesser}, {Fallah Ramazani}, {Fattorini}, {Ferrara},
  {Foffano}, {Fonseca}, {Font}, {Fruck}, {Fukami}, {Garc{\'\i}a L{\'o}pez},
  {Garczarczyk}, {Gasparyan}, {Gaug}, {Giglietto}, {Giordano}, {Gliwny},
  {Godinovi{\'c}}, {Green}, {Green}, {Hadasch}, {Hahn}, {Heckmann}, {Herrera},
  {Hoang}, {Hrupec}, {H{\"u}tten}, {Inada}, {Inoue}, {Ishio}, {Iwamura},
  {Jormanainen}, {Jouvin}, {Kajiwara}, {Karjalainen}, {Kerszberg}, {Kobayashi},
  {Kubo}, {Kushida}, {Lamastra}, {Lelas}, {Leone}, {Lindfors}, {Lombardi},
  {Longo}, {L{\'o}pez-Coto}, {L{\'o}pez-Moya}, {L{\'o}pez-Oramas}, {Loporchio},
  {Machado de Oliveira Fraga}, {Maggio}, {Majumdar}, {Makariev}, {Mallamaci},
  {Maneva}, {Manganaro}, {Mannheim}, {Maraschi}, {Mariotti}, {Mart{\'\i}nez},
  {Mazin}, {Mender}, {Mi{\'c}anovi{\'c}}, {Miceli}, {Miener}, {Minev},
  {Miranda}, {Mirzoyan}, {Molina}, {Moralejo}, {Morcuende}, {Moreno},
  {Moretti}, {Munar-Adrover}, {Neustroev}, {Nigro}, {Nilsson}, {Ninci},
  {Nishijima}, {Noda}, {Nozaki}, {Ohtani}, {Oka}, {Otero-Santos}, {Palatiello},
  {Paneque}, {Paoletti}, {Paredes}, {Pavleti{\'c}}, {Pe{\~n}il}, {Perennes},
  {Persic}, {Prada Moroni}, {Prandini}, {Priyadarshi}, {Puljak}, {Rhode},
  {Rib{\'o}}, {Rico}, {Righi}, {Rugliancich}, {Saha}, {Sahakyan}, {Saito},
  {Sakurai}, {Satalecka}, {Saturni}, {Schleicher}, {Schmidt}, {Schweizer},
  {Sitarek}, {{\v{S}}nidari{\'c}}, {Sobczynska}, {Spolon}, {Stamerra}, {Strom},
  {Strzys}, {Suda}, {Suri{\'c}}, {Takahashi}, {Tavecchio}, {Temnikov},
  {Terzi{\'c}}, {Teshima}, {Torres-Alb{\`a}}, {Tosti}, {Truzzi}, {Tutone}, {van
  Scherpenberg}, {Vanzo}, {Vazquez Acosta}, {Ventura}, {Verguilov}, {Vigorito},
  {Vitale}, {Vovk}, {Will}, {Zari{\'c}}, {Hirotani}, \& {Saz
  Parkinson}}]{2020A&A...643L..14M}
{MAGIC Collaboration}, {Acciari}, V.~A., {Ansoldi}, S., {et~al.} 2020, A\&A,
  643, L14

\bibitem[{Manchester {et~al.}(2005)Manchester, Hobbs, Teoh, \&
  Hobbs}]{Manchester_2005}
Manchester, R.~N., Hobbs, G.~B., Teoh, A., \& Hobbs, M. 2005, The Astronomical
  Journal, 129, 1993

\bibitem[{{Martin} {et~al.}(2014){Martin}, {Torres}, {Cillis}, \& {de O{\~n}a
  Wilhelmi}}]{2014MNRAS.443..138M}
{Martin}, J., {Torres}, D.~F., {Cillis}, A., \& {de O{\~n}a Wilhelmi}, E. 2014,
  MNRAS, 443, 138

\bibitem[{{Miville-Desch{\^e}nes} {et~al.}(2017){Miville-Desch{\^e}nes},
  {Murray}, \& {Lee}}]{2017ApJ...834...57M}
{Miville-Desch{\^e}nes}, M.-A., {Murray}, N., \& {Lee}, E.~J. 2017, ApJ, 834,
  57

\bibitem[{{Moderski} {et~al.}(2005){Moderski}, {Sikora}, {Coppi}, \&
  {Aharonian}}]{2005MNRAS.363..954M}
{Moderski}, R., {Sikora}, M., {Coppi}, P.~S., \& {Aharonian}, F. 2005, MNRAS,
  363, 954

\bibitem[{{Popescu} {et~al.}(2017){Popescu}, {Yang}, {Tuffs}, {Natale},
  {Rushton}, \& {Aharonian}}]{2017MNRAS.470.2539P}
{Popescu}, C.~C., {Yang}, R., {Tuffs}, R.~J., {et~al.} 2017, MNRAS, 470, 2539

\bibitem[{{Roman-Duval} {et~al.}(2009){Roman-Duval}, {Jackson}, {Heyer},
  {Johnson}, {Rathborne}, {Shah}, \& {Simon}}]{2009ApJ...699.1153R}
{Roman-Duval}, J., {Jackson}, J.~M., {Heyer}, M., {et~al.} 2009, ApJ, 699, 1153

\bibitem[{{Russeil} {et~al.}(2017){Russeil}, {Zavagno}, {M{\`e}ge}, {Poulin},
  {Molinari}, \& {Cambresy}}]{2017A&A...601L...5R}
{Russeil}, D., {Zavagno}, A., {M{\`e}ge}, P., {et~al.} 2017, A\&A, 601, L5

\bibitem[{{Saz Parkinson} {et~al.}(2016){Saz Parkinson}, {Xu}, {Yu},
  {Salvetti}, {Marelli}, \& {Falcone}}]{2016ApJ...820....8S}
{Saz Parkinson}, P.~M., {Xu}, H., {Yu}, P.~L.~H., {et~al.} 2016, ApJ, 820, 8

\bibitem[{{Sudoh} {et~al.}(2019){Sudoh}, {Linden}, \&
  {Beacom}}]{2019PhRvD.100d3016S}
{Sudoh}, T., {Linden}, T., \& {Beacom}, J.~F. 2019, PRD, 100, 043016

\bibitem[{{Taylor} {et~al.}(2003){Taylor}, {Gibson}, {Peracaula}, {Martin},
  {Landecker}, {Brunt}, {Dewdney}, {Dougherty}, {Gray}, {Higgs}, {Kerton},
  {Knee}, {Kothes}, {Purton}, {Uyaniker}, {Wallace}, {Willis}, \&
  {Durand}}]{2003AJ....125.3145T}
{Taylor}, A.~R., {Gibson}, S.~J., {Peracaula}, M., {et~al.} 2003, AJ, 125, 3145

\bibitem[{{Tibet AS{\ensuremath{\gamma}} Collaboration} {et~al.}(2021){Tibet
  AS{\ensuremath{\gamma}} Collaboration}, {Amenomori}, {Bao}, {Bi}, {Chen},
  {Chen}, {Chen}, {Chen}, {Chen}, {Cirennima}, {Danzengluobu}, {Fang}, {Fang},
  {Feng}, {Feng}, {Feng}, {Gao}, {Gou}, {Guo}, {Guo}, {He}, {He}, {Hibino},
  {Hotta}, {Hu}, {Hu}, {Huang}, {Jia}, {Jiang}, {Jin}, {Kasahara}, {Katayose},
  {Kato}, {Kato}, {Kawata}, {Kihara}, {Ko}, {Kozai}, {Labaciren}, {Li}, {Li},
  {Li}, {Lin}, {Liu}, {Liu}, {Liu}, {Liu}, {Liu}, {Lou}, {Lu}, {Meng},
  {Munakata}, {Nakada}, {Nakamura}, {Nanjo}, {Nishizawa}, {Ohnishi}, {Ohura},
  {Ozawa}, {Qian}, {Qu}, {Saito}, {Sakata}, {Sako}, {Shao}, {Shibata},
  {Shiomi}, {Sugimoto}, {Takano}, {Takita}, {Tan}, {Tateyama}, {Torii},
  {Tsuchiya}, {Udo}, {Wang}, {Wu}, {Xue}, {Yamamoto}, {Yang}, {Yokoe}, {Yuan},
  {Zhai}, {Zhang}, {Zhang}, {Zhang}, {Zhang}, {Zhang}, {Zhang}, {Zhang},
  {Zhao}, \& {Zhaxisangzhu}}]{2021NatAs...5..460T}
{Tibet AS{\ensuremath{\gamma}} Collaboration}, {Amenomori}, M., {Bao}, Y.~W.,
  {et~al.} 2021, Nature Astronomy, 5, 460

\bibitem[{{Tramacere}(2020)}]{2020ascl.soft09001T}
{Tramacere}, A. 2020, {JetSeT: Numerical modeling and SED fitting tool for
  relativistic jets}, Astrophysics Source Code Library, record ascl:2009.001

\bibitem[{{Tramacere} {et~al.}(2009){Tramacere}, {Giommi}, {Perri},
  {Verrecchia}, \& {Tosti}}]{Tramacere2009}
{Tramacere}, A., {Giommi}, P., {Perri}, M., {Verrecchia}, F., \& {Tosti}, G.
  2009, A\&A, 501, 879

\bibitem[{{Tramacere} {et~al.}(2011){Tramacere}, {Massaro}, \&
  {Taylor}}]{Tramacere2011}
{Tramacere}, A., {Massaro}, E., \& {Taylor}, A.~M. 2011, ApJ, 739, 66

\bibitem[{{Vannoni} {et~al.}(2009){Vannoni}, {Gabici}, \&
  {Aharonian}}]{2009A&A...497...17V}
{Vannoni}, G., {Gabici}, S., \& {Aharonian}, F.~A. 2009, A\&A, 497, 17

\bibitem[{{Wood} {et~al.}(2017){Wood}, {Caputo}, {Charles}, {Di Mauro},
  {Magill}, {Perkins}, \& {Fermi-LAT Collaboration}}]{2017ICRC...35..824W}
{Wood}, M., {Caputo}, R., {Charles}, E., {et~al.} 2017, in International Cosmic
  Ray Conference, Vol. 301, 35th International Cosmic Ray Conference
  (ICRC2017), 824

\bibitem[{Yang {et~al.}(2016)Yang, Aharonian, \& Evoli}]{PhysRevD.93.123007}
Yang, R., Aharonian, F., \& Evoli, C. 2016, Phys. Rev. D, 93, 123007

\bibitem[{{Yuan} {et~al.}(2012){Yuan}, {Liu}, \& {Bi}}]{2012ApJ...761..133Y}
{Yuan}, Q., {Liu}, S., \& {Bi}, X. 2012, ApJ, 761, 133

\bibitem[{{Zabalza}(2015)}]{2015ICRC...34..922Z}
{Zabalza}, V. 2015, in International Cosmic Ray Conference, Vol.~34, 34th
  International Cosmic Ray Conference (ICRC2015), 922

\bibitem[{{Zeng} {et~al.}(2019){Zeng}, {Xin}, \& {Liu}}]{2019ApJ...874...50Z}
{Zeng}, H., {Xin}, Y., \& {Liu}, S. 2019, ApJ, 874, 50

\bibitem[{{Zhu} {et~al.}(2018){Zhu}, {Zhang}, \& {Fang}}]{2018A&A...609A.110Z}
{Zhu}, B.-T., {Zhang}, L., \& {Fang}, J. 2018, A\&A, 609, A110

\end{thebibliography}

\begin{appendix}
\section{Two-dimensional LST-1 analysis}\label{sec_appendix}

Here, we present the results of a first attempt to build significance and excess sky maps. At present,  LST-1 does not have a background model that could be used for background-count predictions. Nevertheless, a tool\footnote{\url{https://github.com/mdebony/acceptance_modelisation/tree/main/acceptance_modelisation}} has been developed for the creation of an acceptance model  from real data that can be used for radial corrections in \texttt{Gammapy} background models. In our case, the background method adopted for the 2D analysis is the \textit{ring}-background technique: in this method the OFF region is defined as a ring around a trial source position, and is used to provide a background estimate (\citet{2007A&A...466.1219B}).
As opposed to the \textit{reflected-region}-background model, in which we use the same offset between the ON and OFF regions and the pointing direction, with this method the detector acceptance cannot be assumed to be constant within the ring, because it covers areas with different offsets from the pointing position (\citet{2007A&A...466.1219B}).\par
As mentioned, the radial acceptance model (which assumes the instrument response to be symmetrical under rotations around the pointing direction) is extracted directly from the data set under analysis on a run-by-run basis and is then projected onto the sky in order to evaluate the background for the whole data set (\citet{2021Univ....7..421D}). We excluded the three putative $\gamma$-ray sources (LHAASO J2108+5157, 4FGL J2108.0+5155  and  HS) in order to avoid contamination of the acceptance (\citet{2021Univ....7..421D}). 

As expected, we noticed that the background acceptance rapidly decreases with energy; on the other hand, it does not significantly change with the offset from the center of the FoV at the high energies we are looking at, because showers from high-energy gamma rays have a larger Cherenkov photon density on the ground than the ones produced by low-energy gamma rays, and the telescope is triggered by showers with larger impact parameters and hence large angular offsets (\citet{2007A&A...466.1219B}). 

Finally, in order to take into account the acceptance dependence on the zenith angle variation across the FoV, we divided the sample into four zenith bins and stacked them together according to the average run zenith angle. 

\par
\begin{figure*}
\centering

\resizebox{\hsize}{!}
    {\includegraphics[width=0.33\hsize]{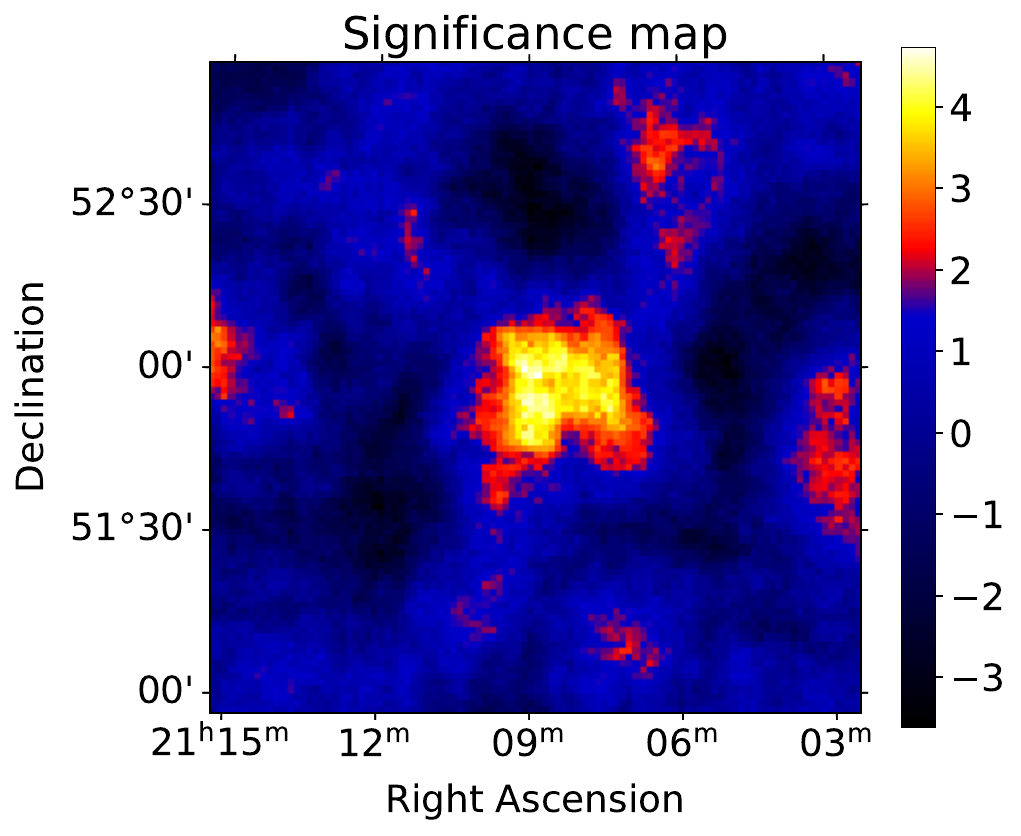}
    \includegraphics[width=0.33\hsize]{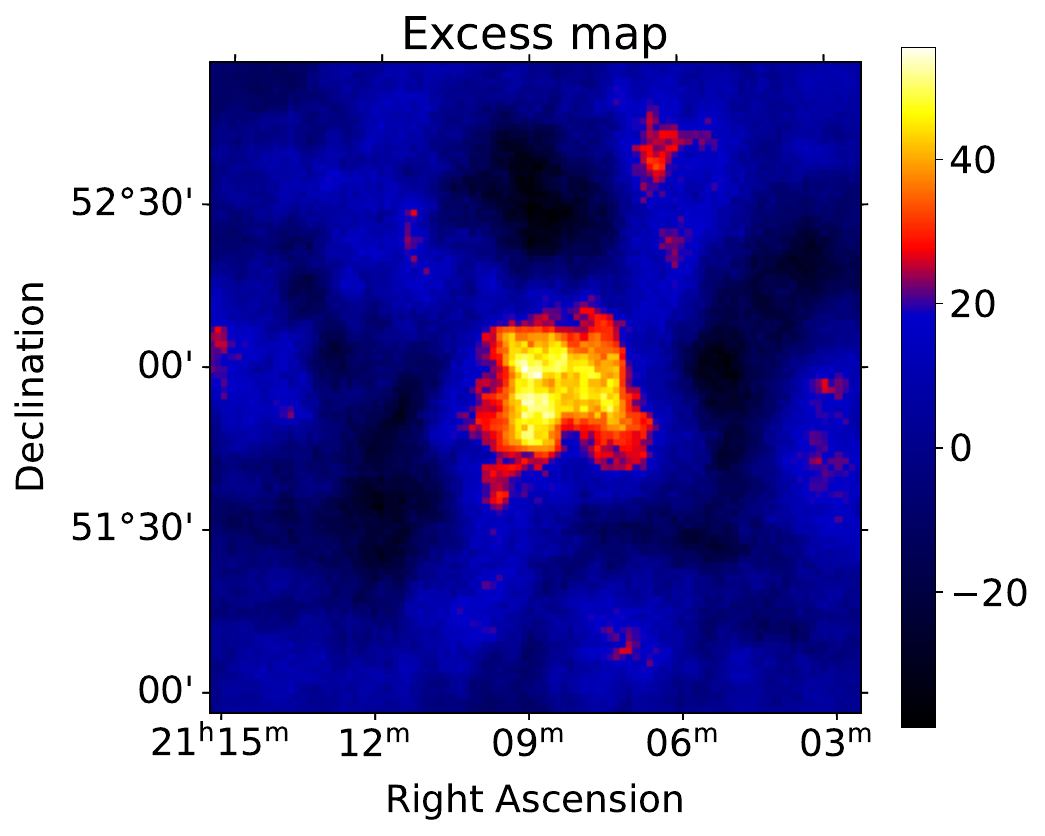}
    \includegraphics[width=0.33\hsize]{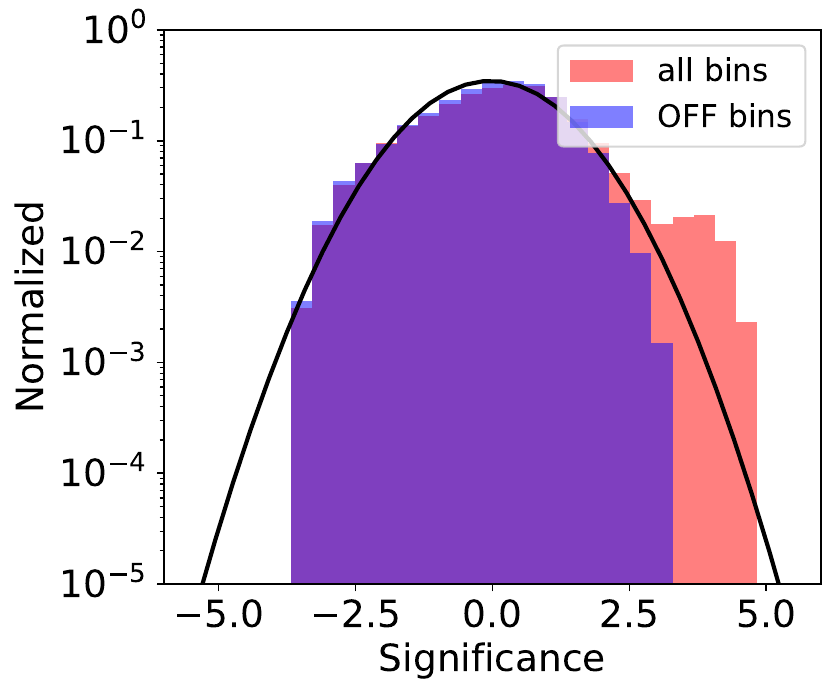}}
  \caption{Statistical significance (\textit{left}) and excess (\textit{center}) maps in a region of $2^\circ \times 2^\circ$ around LHAASO J2108+5157 in the range of energy between 3 TeV and 100 TeV. The map was produced using the \texttt{Gammapy} tool for the \textit{ring}-background model: the OFF region is defined as a $0.3^\circ$ radius ring, with a width of $0.3^\circ$, around a $0.2^\circ$ radius circular ON region. The correlation radius for the building of the maps was fixed at $0.2^\circ$ according to the value of the PSF ($68\%$ containment) in this range of energy, averaged for the different zenith bins used for the production of the IRFs. The same exclusion region was used as that adopted to build the acceptance model. \textit{Right}: Significance distribution for the off bins (purple) and for all bins (pink). The black line represents a Gaussian fit of the background, providing the mean value $\mu=-0.03$, and standard deviation $\sigma=1.15$.} 
     \label{fig_sig_dist_zd_yes_mask}
\end{figure*}

In Figure \ref{fig_sig_dist_zd_yes_mask}, the significance and excess maps show an excess of signal above 3 TeV up to significance values above 4$\sigma$ in the proximity of LHAASO J2018+5157. The Gaussian fit over the significance distribution for the off bins (see Figure \ref{fig_sig_dist_zd_yes_mask}) gives an average value consistent with 0, but the distribution is slightly asymmetric, meaning that the background modeling for our dataset is not perfect and could be improved. \par  
An LST-1 background modeling program is still under development, but it is already possible to use it to extract the radial acceptance from real data and use it for radial corrections on the \textit{ring}-background tool implemented in gammapy. Although the obtained background model is not found to be perfect for our dataset, we find significance values in the skymap that are reasonably consistent with those achieved through the 1D signal extraction above 3 TeV under the point-like-source assumption (see Figure~\ref{fig_lhaaso_theta2_onoff}).

\end{appendix}

\end{document}